\documentclass{article}

\usepackage[english]{babel}
\usepackage{booktabs}
\usepackage{bbm}
\usepackage[
  a4paper,
  verbose=true,
  top=2.5cm,
  bottom=2.5cm,
  left=2.5cm,
  right=2.5cm,
  headheight=14pt,
  headsep=25pt,
  footskip=30pt
]{geometry}
\usepackage{siunitx}
\usepackage[bb=px]{mathalpha}
\usepackage{natbib}
\usepackage{arxiv}
\usepackage{nameref}
\usepackage{etoolbox}
\usepackage[utf8]{inputenc} 
\usepackage[T1]{fontenc}    
\usepackage{url}            
\usepackage{booktabs}       
\usepackage{amsfonts}       
\usepackage{nicefrac}       
\usepackage{microtype}      
\usepackage{lipsum}
\usepackage[pdftex]{graphicx}
\graphicspath{ {./images/} }

\usepackage{amsmath}
\usepackage{graphicx}
\usepackage[colorlinks=false, allcolors=black, hidelinks]{hyperref}
\usepackage[table]{xcolor}
\usepackage{doi}
\usepackage[acronym]{glossaries}
\makeglossaries
\usepackage{colortbl}
\newcommand{\lightmidrule}{\arrayrulecolor{lightgray}\midrule\arrayrulecolor{black}}
\usepackage{pdflscape}
\usepackage{longtable}

\usepackage{subcaption}
\usepackage{graphicx}
\usepackage{tabularx}
\usepackage{multirow} 
\usepackage{natbib}
\usepackage{amsmath,amssymb}

\usepackage{svg}
\usepackage{diagbox}
\usepackage{makecell}
\usepackage{tablefootnote}
\usepackage{threeparttable}
\usepackage{tikz}

\usepackage{amssymb}
\usepackage{pifont}
\newacronym{dqn}{DQN}{Deep Q-Network}

\newacronym{smape}{SMAPE}{symmetric mean percentage error}

\newacronym{mae}{MAE}{mean absolute error}

\newacronym{rmse}{RMSE}{root mean squared error}

\newacronym{rl}{RL}{Reinforcement learning}

\newacronym{ig}{IG}{Integrated Gradients}

\newacronym{al}{AL}{Active Learning}

\newacronym{csr}{CSR}{complete spatial randomness}

\newacronym{mape}{MAPE}{mean absolute percentage error}

\newacronym{td}{TD}{Temporal Difference}

\newacronym{logo}{LOGO}{Leave-One-Group-Out}

\newacronym{tslp}{TSLP}{Traffic Sensor Location Problem}

\title{Sensor Placement for Urban Traffic Interpolation: \\A Data-Driven Evaluation to Inform Policy}

\author{
  Silke K. Kaiser\textsuperscript{a,b,*} \\
  \textsuperscript{a}Data Science Lab, Hertie School, Berlin, Germany \\
  \textsuperscript{b}Centre for Sustainability, Hertie School, Berlin, Germany \\
  \textsuperscript{*}Corresponding author: \texttt{s.kaiser@phd.hertie-school.org}
}

\begin{document}
\date{}
\maketitle

\begin{abstract}

Data on citywide street-segment traffic volumes are essential for urban planning and sustainable mobility management. Yet such data are available only for a limited subset of streets due to the high costs of sensor deployment and maintenance. Traffic volumes on the remaining network are therefore interpolated based on existing sensor measurements. However, current sensor locations are often determined by administrative priorities rather than by data-driven optimization, leading to biased coverage and reduced estimation performance. 
This study provides a large-scale, real-world benchmarking of easily implementable, data-driven strategies for optimizing the placement of permanent and temporary traffic sensors, using segment-level data from Berlin (Strava bicycle counts) and Manhattan (taxi counts). 
It compares spatial placement strategies based on network centrality, spatial coverage, feature coverage, and active learning. In addition, the study examines temporal deployment schemes for temporary sensors.
The findings highlight that spatial placement strategies that emphasize even spatial coverage and employ active learning achieve the lowest prediction errors. With only 10 sensors, they reduce  the mean absolute error by over 60\% in Berlin and 70\% in Manhattan compared to alternatives. Temporal deployment choices further improve performance: distributing measurements evenly across weekdays reduces error by an additional 7\% in Berlin and 21\% in Manhattan. Together, these spatial and temporal principles allow temporary deployments to closely approximate the performance of optimally placed permanent deployments. From a policy perspective, the results indicate that cities can substantially improve data usefulness by adopting data-driven sensor placement strategies, while retaining flexibility in choosing between temporary and permanent deployments.

\end{abstract}

\section{Introduction}

Traffic volume measures the number of cyclists, motorists, or pedestrians passing a location. When available at high spatial and temporal resolution across an entire city, traffic volume data support a wide range of applications; for example, targeted infrastructure investments, adaptive traffic management, improved public transport planning, or accurate estimation of emissions and environmental impacts \citep{leduc_road_2008,liu_spatial-temporal_2019, zheng_urban_2014}. However, obtaining this level of detailed, citywide coverage remains a substantial challenge. 

In practice, urban traffic volume data have historically been collected through physical sensors such as inductive loops, radar counters, or cameras \citep{leduc_road_2008}. While these technologies provide high-quality observations, they also entail significant operational challenges: they require continuous power supply, data transmission, processing infrastructure, and regular maintenance, all of which contribute to substantial cost \citep{gagliardi_optimal_2024}. As a result, network-wide deployment of such sensors is economically and logistically infeasible, and most cities monitor only a small fraction of their street network \citep{gagliardi_optimal_2024,leduc_road_2008}. Moreover, sensor locations are often determined by planning or administrative priorities, such as high-usage sites or locations near planned investments, which can limit the representativeness of the collected data \citep{claes_bicycle_2016, turner_strategic_2012}. 
At the same time, sensor data remain critical for estimating citywide traffic volumes. The strategic placement of a limited number of sensors is therefore a key determinant of reliable traffic data and thus should be evidence-based. Accordingly, this study identifies and compares placement strategies that maximize citywide interpolation performance.

Despite major advances in modeling frameworks and data sources, traffic volume interpolation still depends critically on ground-truth data from physical sensors. 
Citywide estimation of link-level traffic volumes remains a central challenge in urban transportation research for both motorized traffic \citep{xing_urban_2024, yu_review_2016} and bicycle traffic \citep{bhowmick_systematic_2023}. A wide range of methodological approaches has been proposed to address this problem, spanning from classical linear regression and distance- and kernel-based statistical interpolation methods \citep{ cover_nearest_1967,lu_adaptive_2008,shan_urban_2013,zhou_kernelized_2012} to more advanced machine learning techniques \citep{kaiser_counting_2025,miah_estimation_2023, sekula_estimating_2018}. Recent developments further extend these approaches by employing graph neural networks to explicitly model spatial and temporal dependencies \citep{dai_dynamic_2023, kaiser_spatio-temporal_2025}.
All of these approaches are trained and validated on data from physical traffic sensors, often in combination with auxiliary datasets. Auxiliary data include for example weather conditions \citep{koesdwiady_improving_2016, nosal_incorporating_2014}, infrastructure and points of interest \citep{askari_taxi_2020, fazio_bike_2021, strauss_spatial_2013}, GPS-based trajectory data  \citep{brown_spatial_2022, hochmair_estimating_2019, zhan_citywide_2017} and satellite imagery \citep{ganji_methodology_2020, mccord_estimating_2003}. 
 Yet, across most studies, the most critical input remains the observed traffic counts from existing physical sensors. These measurements are typically used either as features for model calibration or as ground-truth data for validation \citep{cai_traffic_2023, dai_dynamic_2023, yao_spatiotemporal_2023, zhu_dual_2025}, as they are often the only reliable source of directly observed link-level traffic volumes.
 
Given the critical role of sensor data, the performance of traffic estimation models is inherently constrained by the quality and coverage of these data, which, in turn, are strongly determined by sensor placement. This has motivated a growing body of research on the \acrlong{tslp} (\acrshort{tslp}), which studies how a limited sensor budget should be allocated to most effectively capture network-wide traffic dynamics \citep{owais_traffic_2022}.
The \acrshort{tslp} is commonly divided into two main categories \citep{gentili_locating_2012}. The first, known as the flow observability problem, seeks sensor placements that ensure a unique solution to the system of network flow equations, such that traffic flows on all network links are uniquely determined. The literature on flow observability typically focuses on small or simulated networks rather than real-world settings \citep{agarwal_dynamic_2016,contreras_observability_2016, shao_optimization_2021}. The second problem, flow estimation, seeks sensor placements that allow the best possible traffic volume estimation with limited sensor coverage. As real-world sensor coverage is typically sparse, this work aligns with this line of research.
 
In this context, both the location and timing of sensor deployment become critical design factors. A substantial body of research has addressed the spatial dimension of sensor placement: Information-theoretic measures have been used to identify sensor locations that maximize the informational value of observed flows \citep{ivanchev_information_2016}, while clustering-based approaches group network segments with similar characteristics to reduce redundancy and improve representativeness \citep{kianfar_optimizing_2010}. More recent studies frame the problem as a submodular or control-theoretic optimization task, enabling near-optimal or dynamically adaptive sensor configurations under computational and budgetary constraints \citep{ li_submodularity_2023,mehr_submodular_2018, nugroho_where_2022}. 
Regarding the temporal dimension, several studies highlight the benefits of strategically scheduling or mobilizing sensors to capture dynamic traffic conditions. For instance, dynamic optimization models for mobile sensor placement on freeways \citep{sun_dynamic_2021} and mobile sensor routing approaches based on hybrid swarm heuristics \citep{zhu_mobile_2014} have demonstrated promising performance in enhancing network surveillance. 
However, the majority of these studies rely solely on simulated motorized traffic, linearized dynamics, or small-scale test networks, which limits the transferability of their insights to dense urban environments where network structure is more heterogeneous and sensor coverage is typically more sparse \citep{kaiser_spatio-temporal_2025}. In addition, despite their theoretical appeal, many such methods remain computationally demanding and require detailed traffic models or extensive prior data, which constrain their practical applicability for transport agencies. 
Some studies have proposed more pragmatic sensor placement strategies that emphasize ease of implementation and interpretability, including network centrality measures \citep{paluch_optimizing_2020,senturk_connectivity_2014}, spatial dispersion objectives \citep{bao_sensor_2016, robinson_optimizing_2022}, and active learning approaches \citep{muttreja_active_2006,singh_active_2006, yang_active_2024}, yet methods are typically evaluated in isolation and have not been systematically compared.

Despite this rich body of work, clear, empirically grounded guidance for real-world sensor deployment remains limited. To date, most strategies have been evaluated under heterogeneous data, modeling choices, and resource assumptions, which complicate direct comparison of their relative performance. This lack of systematic, real-world benchmarking hinders the translation of methodological advances into actionable deployment decisions for cities, particularly across different urban contexts and transport modes.

This study addresses the question of how a limited number of traffic sensors should be placed in space and time to maximize the performance of citywide traffic volume interpolation, by systematically benchmarking alternative placement strategies within a unified evaluation framework. The analysis validates and compares strategies for both spatial and temporal sensor placement, and further assesses how interpolation performance differs when sensors are deployed permanently (long-term, continuous recording) versus temporarily (short-term, intermittent deployment). The spatial placement strategies include network centrality, feature coverage, spatial coverage, and an active learning approach. For the temporal dimension, the evaluation considers revisiting versus rotating locations as well as the temporal distribution of sampled days. Citywide interpolation is performed using XGBoost, a model previously shown to perform well in traffic volume estimation \citep{kaiser_counting_2025}.
This analysis is based on real-world, segment-level traffic volume data from two distinct urban contexts: Strava-based bicycle counts in Berlin and taxi trip data in Manhattan, New York City \citep{kaiser_spatio-temporal_2025}.
Although these datasets only represent subsets of overall traffic activity and are therefore subject to sampling bias, they are strongly correlated with total cycling and motorized traffic volumes (correlation coefficients of 0.61 and 0.78, respectively).
Evaluating two cities and two transport modes with data of varying bias thus provides a stringent test of the robustness and generalizability of the proposed strategies. 

In summary, the main contributions of this paper are the following:
\begin{itemize}
     \item The study addresses the lack of unified benchmarking by providing a novel, large-scale, real-world comparison of spatial and temporal sensor placement strategies for urban traffic volume interpolation. 
     \item Under optimal placement, the analysis quantifies the loss in interpolation performance when using temporary instead of permanent sensors, thereby providing direct evidence for a key deployment trade-off faced by cities. 
     \item By evaluating readily deployable placement strategies across two cities and two transport modes, the study provides clear, empirically grounded, and transferable guidance for real-world sensor deployment decisions.
\end{itemize}

\section{Methods \label{sec:methods}}
The objective of this study is to identify sensor placement strategies that maximize the informational value of collected traffic data for accurate citywide traffic volume interpolation. In urban practice, both permanent and temporary sensors are widely used. Because permanent sensors record traffic volumes continuously, their placement requires optimization only in the spatial dimension. Temporary sensors, by contrast, introduce an additional temporal design dimension, as decisions must be made not only about where but also when to deploy them. Accordingly, the study (i) evaluates permanent sensor placement strategies (methods in Section~\ref{sec:methods_sensor_placement_strategies_spatially} and results in Section~\ref{sec:results_subsection1}), (ii) optimizes the temporal allocation of temporary sensors (Sections~\ref{sec:methods_sensor_placement_strategies_temporally} and~\ref{sec:results_subsection2}), and (iii) concludes by comparing interpolation performance under temporary and permanent sensor deployments (Sections~\ref{sec:methods_comparison} and~\ref{sec:results_subsection3}).

\subsection{Experimental Setup and Notation}

The urban street network and its associated traffic observations are formally represented as a set of street segments $S$, indexed by $i \in\{1,\dots,N\}$, where each street segment corresponds to the portion of a street between two intersections. Traffic volume on each street segment is observed over $J$ time steps, indexed by $j \in \{1,\dots, J\}$. Each segment $i$ at time step $j$ is described by a feature vector $x_{i,j} \in \mathbb{R}^d$, where $d$ is the number of features. The feature set comprises spatially varying (e.g., speed limits), temporally varying (e.g., weather conditions), and spatio-temporally varying attributes (e.g., the number of inhabitants in the area).

For model training and evaluation, the set of street segments $S$ is split into disjoint training, validation, and test subsets. The validation set $S_{\text{val}}$ and the test set $S_{\text{test}}$ contain each 15\% of the segments. These subsets are excluded from training and are not considered for sensor placement. The remaining segments form the training set $S_{\text{train}} = S \setminus (S_{\text{val}} \cup S_{\text{test}})$. 

Sensor placement is formulated as the iterative selection of up to $K$ sensors from the training network, where $K$ denotes the sensor budget. $S_{\text{selected}}$ denotes the set of placed sensors and $S_{\text{candidate}} = S_{\text{train}} \setminus S_{\text{selected}}$ the remaining candidate locations, with $|S_{selected}|=K$ and $S_{selected} \subseteq S_{train}$, $S_{candidate} \subseteq S_{train}$. For permanent sensor placement, each sensor location can be selected at most once, and once selected, it provides continuous observations across all time steps $J$.
For temporary placement, placement is defined jointly over space and time. Here, a sensor observation corresponds to a segment–time window pair $(i, [j, j+h))$, where $i \in S_{\text{train}}$ and $[j, j+h)$ denotes a contiguous interval of $h$ time steps within the total time horizon $J$ (i.e. $j \in J, j+h \leq |J|$). The same segment may be selected multiple times for different, non-overlapping intervals.
All sensor placement strategies evaluated in this study can place sensors either from scratch or extend an existing sensor network. 
If there are existing sensors, they are included in $S_{\text{selected}}$ at the start of the optimization, and their number contributes to the total sensor budget $K$. 
Hence, the sensor budget $K$ always denotes the final total number of sensors after placement, including both pre-existing and newly added ones.

Model performance is evaluated primarily using the \acrlong{mae} (\acrshort{mae}). Results using the \acrlong{rmse} (\acrshort{rmse}) are included in the Appendix. MAE is preferred because the ground-truth distributions are strongly right-skewed (Figure~\ref{fig:city_maps}); unlike RMSE, it is less sensitive to extreme outliers and therefore provides a more robust measure of typical interpolation performance. Percentage-based error measures are avoided because the ground-truth distributions contain a large proportion of zero observations (47.8\% for Berlin and 23.6\% for Manhattan). Formal definitions of error metrics are provided in Appendix \ref{app:error_metrices}.

For the interpolation model, which predicts citywide traffic volumes based on the sensed training data, the study uses XGBoost. For each sensor configuration, the model is trained on observations from $S_{\text{selected}}$. During model development and strategy selection, performance is evaluated on the validation set $S_{\text{val}}$. All results reported in this paper are based exclusively on the held-out test set $S_{\text{test}}$. XGBoost is selected because of its strong performance on tabular data and its demonstrated effectiveness for traffic volume interpolation in urban settings \citep{kaiser_counting_2025}. All numerical features are standardized, and categorical features are one-hot encoded.

\subsection{Spatial Sensor Placement Strategies \label{sec:methods_sensor_placement_strategies_spatially}} 
This study evaluates a suite of placement strategies for selecting sensor locations. The strategies can be broadly grouped into four categories: 
(i) network centrality-based placement strategies, which rely on graph-theoretic measures of structural importance; 
(ii) feature-based placement strategies, which aim to ensure diversity and representativeness in the feature space;
(iii) spatial-based placement strategies, which aim to ensure diversity and representativeness in the geographic domain; and 
(iv) an active learning approach, which iteratively reduces model uncertainty through data-driven sensor selection.

The spatial placement strategies are evaluated with sensor budgets $K \in \{10, 25, 50, 75, 100\}$, representing realistic stages of permanent sensor deployment that balance gains in spatial coverage against installation and maintenance costs. The existing deployments, with 34 sensors in Berlin and 8 in Manhattan, fall within this range and are included as additional configurations. This enables direct comparison between existing and simulated sensor placements.

The sensor placement strategies considered here differ in their selection mechanism; the individual strategies are described in detail below. Some strategies follow a simple ranking approach, in which each candidate segment is assigned a precomputed score that is independent of other sensor locations, and the $K$ segments with the highest scores are selected directly (i.e., the network-centrality–based strategies). Other strategies are implemented in a greedy, iterative fashion (including the feature-, spatial-, and learning-based methods). For those, starting from an existing set of sensors $S_{\text{selected}}^{(0)}$, one additional sensor is added at each iteration by maximizing (or minimizing) the respective placement strategy criterion over the remaining candidate locations $S_{\text{candidate}}^{(t)} = S_{\text{train}} \setminus S_{\text{selected}}^{(t)}$, where $t = 0,1,\dots, K - |S_{\text{selected}}^{(0)}|$ indexes the greedy selection steps. 
If the goal is to extend the existing sensor network, $S_{\text{selected}}^{(0)}$ corresponds to the existing sensors; otherwise, when designing a network from scratch, one initial sensor location is selected at random, so that $S_{\text{selected}}^{(0)}$ contains a single randomly chosen segment. 
The selection process continues until $|S_{\text{selected}}| = K$.

For notational clarity across all placement strategies, $i$ indexes any street segment, while $v$ and $u$ denote specific segments within the selected set, and $s, t$ refer to origin–destination pairs used in network-based measures such as betweenness and closeness.

\paragraph{Network centrality-based placement strategies.}
Two widely used network centrality measures for sensor placement are considered: betweenness and closeness centrality. These measures identify structurally important street segments and have been shown to improve the efficiency and robustness of sensor placement 
\citep{diao_sensor_2023, ivanchev_information_2016, jain_node_2013, senturk_connectivity_2014, zhao_understanding_2016}.
The underlying rationale is that central locations tend to capture key flows or interactions within the network, making them valuable candidates for monitoring and control. To compute these measures, the street network is represented as an undirected graph in which each street segment $i \in S$ corresponds to a node. An edge connects two nodes if the corresponding street segments share a common intersection. 

Betweenness centrality quantifies the extent to which a node lies on the shortest paths between all other pairs of nodes \citep{bloch_centrality_2023}. Intuitively, nodes with high betweenness scores represent key corridors through which much of the traffic flows, such as bridges or arterial streets connecting otherwise weakly connected parts of a city.
According to the betweenness-based placement strategy, $K$ sensors are placed on the segments with the highest betweenness scores: 

\begin{equation}
H_{\text{betweenness}} \;=\;
\operatorname*{arg\,max}_{\substack{S_{selected}}}
\sum_{v \in S_{selected}}
\sum_{\substack{s, t \in S \\ s \ne t \ne v}}
\frac{\sigma_{st}(v)}{\sigma_{st}},
\end{equation}

where $\sigma_{st}$ denotes the number of shortest paths between nodes $s$ and $t$ and $\sigma_{st}(v)$ the number of those paths that pass through node $v$.

Closeness centrality, in contrast, measures how near a node is to all other nodes in the network \citep{bloch_centrality_2023}. 
Nodes with high closeness scores minimize the average shortest-path distance to all other nodes. Intuitively, those are segments located in dense urban cores from which many other streets can be reached via short paths. As a result, these locations enable efficient coverage of spatial variations in traffic flow. 
According to this placement strategy, $K$ sensors are placed on the segments with the highest closeness scores:

\begin{equation}
    H_{\text{closeness}} \;=\;
    \operatorname*{arg\,max}_{S_{selected}}
    \sum_{v \in S_{selected}}
    \frac{N - 1}{\sum_{\substack{u \in S \\ u \ne v}} d(v, u)},
\end{equation}

where $d(v,u)$ denotes the shortest-path distance between nodes $v$ and $u$.

\paragraph{Feature-based placement strategies.}
Three feature-based sensor placement strategies are considered: feature diversity, feature redundancy, and feature coverage, which differ in their balance of representativeness and information overlap in feature space. Feature-based placement strategies aim to ensure that the selected sensors collectively capture the diversity and representativeness of the feature space, an objective that has been shown to improve machine learning performance \citep{gong_diversity_2019}. Because sensor locations are selected independently of time, only time-invariant features $d' \subset d$ can be used for feature-based placement. Accordingly, feature vectors in this context are denoted by $x_i \in \mathbb{R}^{d'}$, omitting the time index. All numerical features used to compute the feature-based placement strategies are standardized, and categorical features are one-hot encoded.
Because feature-based strategies depend highly on the feature set used to compute them, the analysis evaluates each strategy using two alternative sets of features to control for potential feature-selection effects: one using all available time-invariant features, and another using a selected set of infrastructure-related features, comprised of the number of car lanes, street type (e.g., residential, motorway link), street surface (e.g., asphalt, cobblestone), and maximum permitted speed (which are all time invariant). Additional feature selections were also tested; however, the corresponding results are reported only in the Appendix \ref{appx:permanent_further_strategies}, as they yielded inferior performance.

Feature diversity seeks to select sensors whose feature vectors are as dissimilar as possible, thereby maximizing coverage of the feature space. 
Diversity is measured using the Euclidean distance between feature vectors, a standard and interpretable choice in diversity optimization and subset selection problems \citep{gong_diversity_2019, parreno_measuring_2021}. 
For a selected subset $S_{\text{selected}}$, feature diversity is quantified as the mean pairwise Euclidean distance between feature vectors $x_v,x_u \in \mathbb{R}^{d'}$. Higher values indicate more distinct sensors that capture a broader range of features, while lower values imply greater similarity among selected sites; accordingly, the feature diversity objective is maximized:
\begin{equation}
H_{\text{feature-diversity}} \;=\; \operatorname*{arg\,max}_{\substack{S_{\text{selected}}}} \frac{2}{K(K-1)} \sum_{\substack{v,u\in S_{selected} \\ v<u}} \left\lVert x_v - x_u \right\rVert_2.
\end{equation}

Feature redundancy measures the degree of similarity among the selected sensors in feature space. It is quantified as the mean pairwise cosine similarity between the feature vectors of $S_{selected}$, a widely used similarity measure in machine learning \citep{vaswani_attention_2017,wang_feature_2015}. Higher values indicate greater overlap (i.e., more redundant information), whereas lower values indicate greater complementarity. Accordingly, the redundancy-based placement strategy selects $K$ sensors by minimizing this quantity:

\begin{equation}
    H_\text{Redundancy}
    \;=\; \operatorname*{arg\,min}_{\substack{S_{\text{selected}}}}
    \frac{2}{K(K-1)} \sum_{\substack{v,u\in S_{selected} \\ v<u}}
    \frac{x_v^\top x_u}{\lVert x_v\rVert_2\,\lVert x_u\rVert_2}.
\end{equation}

Feature coverage quantifies how well the selected sensors span the range of values across all feature dimensions. It is computed as the mean variance of each feature over $S_{selected}$ following the principle that higher variance implies greater coverage and informativeness \citep{krause_near-optimal_2008}. Accordingly, the coverage-based placement strategy maximizes this quantity:

\begin{equation}
H_{\text{coverage}} \;=\;
\operatorname*{arg\,max}_{\substack{S_{\text{selected}}}}
\frac{1}{d'} \sum_{p=1}^{d'} \mathrm{Var}\!\big(\{x_{v p} : v \in S_{\text{selected}}\}\big),
\end{equation}
 
where $x_{v p}$ denotes the value of feature dimension $p$ of the time-invariant feature vector $x_v$ for segment $v$.

\paragraph{Spatial-based placement strategies.}
Two spatial-based sensor placement strategies are considered: spatial dispersion and Voronoi area inequality, which aim to distribute sensors evenly across the study area to enhance representativeness and reduce spatial bias. 
Recent studies in transport and environmental monitoring show that well-dispersed or uniformly distributed sensor networks improve estimation accuracy and that an equitable sensor distribution enhances representativeness and fairness in urban monitoring  \citep{bao_sensor_2016,robinson_optimizing_2022, zied_abozied_spatial_2025}. Because both spatial measures operate on point locations, while street segments are line geometries, the analysis represents each segment by its midpoint and defines $m_v \in \mathbb{R}^2$ as the geographic coordinates of the midpoint of street segment $v$.

Spatial dispersion is assessed using the Clark–Evans Aggregation Index $R$ \citep{clark_distance_1954}, a classical measure of spatial ordering in point patterns. The index is defined as the ratio between the observed mean nearest-neighbor distance among points $\bar r_{\text{obs}}$ and its expected value $\bar r_{\exp}$ under \acrlong{csr} (\acrshort{csr}), with:
\[
\bar r_{\text{obs}} = \frac{1}{K} \sum_{v \in S_{\text{selected}}}
\min_{\substack{u \in S_{\text{selected}} \\ u \neq v}} \lVert m_v - m_u\rVert_2,
\]
\[
\bar r_{\exp} = \frac{1}{2\sqrt{K/A}},
\]
\[
R = \frac{\bar r_{\text{obs}}}{\bar r_{\exp}},
\]
where $A$ denotes the study area, defined by the city boundary. $R>1$ indicates over-dispersion (points more evenly spaced than random), $R<1$ indicates clustering, and $R\approx1$ corresponds to a random distribution. Accordingly, the objective is to maximize $R$. Intuitively, this means the objective favors sensor configurations in which locations are, on average, as far apart as possible, thereby promoting uniform spatial coverage. Because $K$ and $A$ are fixed in the experiments, maximizing $R$ is equivalent to maximizing $\bar r_{\text{obs}}$. The spatial dispersion placement strategy is therefore defined as:

\begin{equation}
H_{\text{dispersion}} =
\operatorname*{arg\,max}_{\substack{S_{\text{selected}}}}
\frac{1}{K}\sum_{v \in S_{\text{selected}}}
\min_{\substack{u \in S_{\text{selected}} \\ u \neq v}} \lVert m_v - m_u\rVert_2.
\end{equation}

The Voronoi area inequality strategy assesses the spatial uniformity of sensor locations by comparing the areas of Voronoi cells constructed around the selected sensors \citep{voronoi_nouvelles_1908}. 
The study area is partitioned into Voronoi cells using the midpoints $m_v$ of the selected street segments. Each Voronoi cell spans the region of the study area that is closer to its corresponding midpoint than to any other; the area of this region, denoted $A_v$, represents the spatial zone of influence of sensor $v$.
Uniformly distributed sensors yield Voronoi cells of similar size, whereas clustered sensors produce many small cells in dense regions and large cells in sparse ones.
To quantify spatial uniformity, the analysis computes the Gini coefficient $G$ of the Voronoi cell areas, following \citet{patrignani_optimizing_2020}:
\begin{equation}
    G = \frac{\sum_{v=1}^{K}\sum_{u=1}^{K} |A_{v} - A_{u}|}{2K^2\bar{A}},
\end{equation}
where $\bar{A}$ denotes the mean cell area.
Lower $G$ values indicate a more even spatial distribution (i.e., more uniform coverage), while higher values correspond to stronger clustering.
The Voronoi area inequality strategy, therefore, selects the sensor configuration that minimizes the Gini coefficient:
\begin{equation}
H_{\text{Voronoi}} =
\operatorname*{arg\,min}_{\substack{S_{\text{selected}}}}
G(S_{\text{selected}}).
\end{equation}

\paragraph{Active learning.}
Beyond rule-based placement strategies, this work implements an \acrlong{al} (\acrshort{al}) approach designed to be simple and easily replicable in practice. Active learning frames sensor placement as a sequential data-acquisition problem aimed at reducing predictive uncertainty \citep{settles_active_2009}. This paradigm has been successfully applied in various domains, including environmental monitoring \citep{singh_active_2006, xie_physics-constrained_2024}, network energy efficiency \citep{muttreja_active_2006}, and structural health monitoring \citep{yang_active_2024}. 

Conceptually, active learning follows a simple iterative procedure. At iteration $t$, an interpolation model is trained on the currently labeled sensor set $S_{\text{selected}}^{(t)}$ and used to predict traffic volumes at the candidate locations $S_{\text{candidate}}^{(t)}$. Based on these predictions, the method identifies locations where model uncertainty is highest, reflecting limited information. The candidate location with the highest uncertainty is then selected as the next sensor location. 

Specifically, in our context, active learning is implemented as follows. Because the interpolation model, XGBoost, does not natively provide predictive uncertainty estimates, uncertainty is approximated using an ensemble-based approach. At each iteration $t$, a fixed ensemble of $M$ independent interpolation models $\{f_{\theta_m}^{(t)}\}_{m=1}^{M}$ are trained on bootstrap resamples of the labeled data $\{(x_v, y_v) \mid v \in S_{\text{selected}}^{(t)}\}$. Each model produces predictions $\hat{y}_{v}^{(m)} = f_{\theta_m}^{(t)}(x_v)$ for all candidate locations $v \in S_{\text{candidate}}^{(t)}$. For each candidate location $v$, the ensemble mean prediction and predictive variance are computed as
\begin{align}
    \hat{\mu}_v^{(t)} &= \frac{1}{M} \sum_{m=1}^{M} f_{\theta_m}^{(t)}(x_v), \\
    \hat{\sigma}_v^{2\,(t)} &= \frac{1}{M-1} \sum_{m=1}^{M} \left(f_{\theta_m}^{(t)}(x_v) - \hat{\mu}_v^{(t)}\right)^2.
\end{align}
The predictive variance $\hat{\sigma}_v^{2\,(t)}$ serves as an estimate of uncertainty, and thus, the next sensor is selected by greedily choosing the candidate location with the highest predictive variance: 
\begin{equation}
    v^* = \arg\max_{v \in S_{\text{candidate}}^{(t)}} \hat{\sigma}_v^{2\,(t)}.
\end{equation}
The selected location is then added to the labeled set,
\[
S_{\text{selected}}^{(t+1)} = S_{\text{selected}}^{(t)} \cup \{v^*\}.
\]
Subsequently, the model is retrained, and the procedure repeats until $K$ sensors are placed.

\paragraph{Baselines.}
To contextualize the performance of the proposed sensor placement strategies, this study compares them against three baseline configurations: random sensor placement, the existing real-world deployments, and an all-training-data scenario.

A randomized sampling approach is used to characterize the range of performance achievable, following prior work \citep{paluch_optimizing_2020}. For each sensor budget $K$, the analysis generates 1,000 random sets of $K$ sensor locations. The interpolation model is trained and evaluated separately for each set, yielding 1,000 prediction errors. The minimum, median, and maximum of these errors are reported, representing best-case, typical, and worst-case sensor placements, respectively.

The existing sensor networks in Berlin and Manhattan are included as a baseline to benchmark the interpolation performance of the evaluated placement strategies against the status quo. For this baseline, the existing real-world sensor locations are used as $S_{\text{selected}}$, and the corresponding observations are extracted from the respective datasets (Strava for Berlin and taxi trips for Manhattan). This baseline is available only for $K=34$ in Berlin and $K=8$ in Manhattan, reflecting the fixed number of existing sensors.

In the all-training-data scenario, the interpolation model is trained on data from all segments in the training set, such that $S_{selected}=S_{\text{train}}$. This represents an idealized scenario in which every segment, except those reserved for validation and testing, is equipped with a sensor. The resulting error reflects a best-case reference for citywide interpolation performance under the given model and feature set.

\subsection{Temporal Deployment Schemes for Temporary Sensors \label{sec:methods_sensor_placement_strategies_temporally}}

When deploying sensors temporally, cities not only need to choose the locations where to deploy them, but also the time window during which to deploy them. Temporary sensor deployments differ greatly across cities. To systematically evaluate how temporal design choices affect interpolation performance, the analysis examines the three key dimensions of temporal sampling: deployment duration per location, revisiting versus rotating locations, and the temporal distribution of sampled days. To isolate the effects of temporal placement strategies, across all temporal simulations, spatial locations are predetermined by a given spatial placement strategy, and the same set of randomly selected days is employed.

To ensure comparability across datasets, all temporal simulations are standardized to the day level. For each sampled date, all observations recorded on that day are included. This harmonizes the hourly taxi counts in Manhattan with the daily bicycle counts in Berlin. Because the Manhattan dataset spans only two months, individual calendar days are reused when necessary to construct larger simulated deployment budgets. Finally, to reflect realistic large-scale temporary monitoring programs observed in practice, the analysis considers deployment scenarios ranging from sparse sampling to complete coverage of the training network.

\paragraph{Deployment duration per location.}
The deployment duration dimension specifies the number of consecutive days a sensor remains active at a given location. For instance, traffic volume on a street segment may be observed for one day, two days, or longer continuous periods. This design choice has been examined by \citet{kaiser_counting_2025}, who find that single-day deployments consistently outperform deployments spanning three or seven consecutive days. The superior performance of shorter deployments is attributed to their broader temporal coverage and the resulting increase in the dispersion of training data across time. Informed by these findings, all subsequent experiments in this study adopt a non-consecutive single-day sampling scheme, such that each temporary deployment corresponds to an isolated day rather than a continuous multi-day observation window.

\paragraph{Revisiting versus rotating locations.}
This temporal design dimension examines whether a fixed observation budget should be concentrated by repeatedly revisiting the same locations or dispersed by rotating across a larger set of locations. For instance, with a total budget of four observation days, sensors may be deployed at a single location on four non-consecutive days, at two locations on two days each, or at four distinct locations on one day each. 
To evaluate the relative merits of revisiting versus rotating locations, the analysis simulates temporary deployment schemes in which each location is observed either once or revisited multiple times (2, 5, or 10 observations per location), while holding the total observation budget constant. 
Concretely, for a given deployment budget of $D$ days, the procedure first draws $D$ dates uniformly at random from the study period. Under a fully rotating strategy, these dates are paired with the first $D$ distinct locations picked according to a placement strategy, each observed once. Under revisiting strategies, the same set of dates is instead allocated to a smaller number of locations, such that each selected location is observed repeatedly while other candidate locations remain unobserved. This design ensures that all strategies are directly comparable under an identical total observation budget, differing only in how observations are distributed across space.

\paragraph{Temporal distribution of sampled days.}
Finally, the evaluation examines the temporal distribution of sampled days, specifically whether the choice of specific weekdays affects interpolation performance. To isolate effects, the design uses non-consecutive single-day sampling and collects one observation per location. The only variation across scenarios is the temporal allocation of observation days: either restricting all measurements to a single weekday or distributing them evenly across the week.
Sampled days are drawn from the same randomly chosen calendar weeks. For example, if seven locations are sampled, each is observed on a different Thursday for the Thursday-only strategy, whereas the evenly distributed strategy observes the same seven locations in the same weeks but assigns each location to a different weekday.

\subsection{Comparison of temporary and permanent deployment \label{sec:methods_comparison}}
The analysis compares interpolation performance under permanent and temporary sensor deployments, holding the number of observed locations constant. In both cases, sensor locations are selected using the best-performing spatial placement strategy identified in the experiments. For temporary deployments, sensors are additionally allocated in time according to the optimal temporal placement strategies identified in the experiments. Results are reported up to the point at which each street segment in the training dataset is observed by a sensor. This level of coverage reflects realistic large-scale temporary monitoring programs and is achievable for temporary sensors, but it represents an unrealistic scenario for permanent deployments, as discussed earlier. This scenario is included solely to enable a direct comparison between permanent and temporary sensing strategies.

\section{Data \label{sec:datasection}}

This study uses two street-segment–level traffic volume datasets with temporal coverage and rich auxiliary features: Strava bicycle volumes for Berlin and taxi volumes for Manhattan, New York City. The datasets are published in \cite{kaiser_data_2025}, with detailed descriptions provided in \cite{kaiser_spatio-temporal_2025}. 

\begin{figure}[!ht]
    \centering
        \begin{subfigure}{0.79\textwidth}
        \centering
        \includegraphics[width=\textwidth]{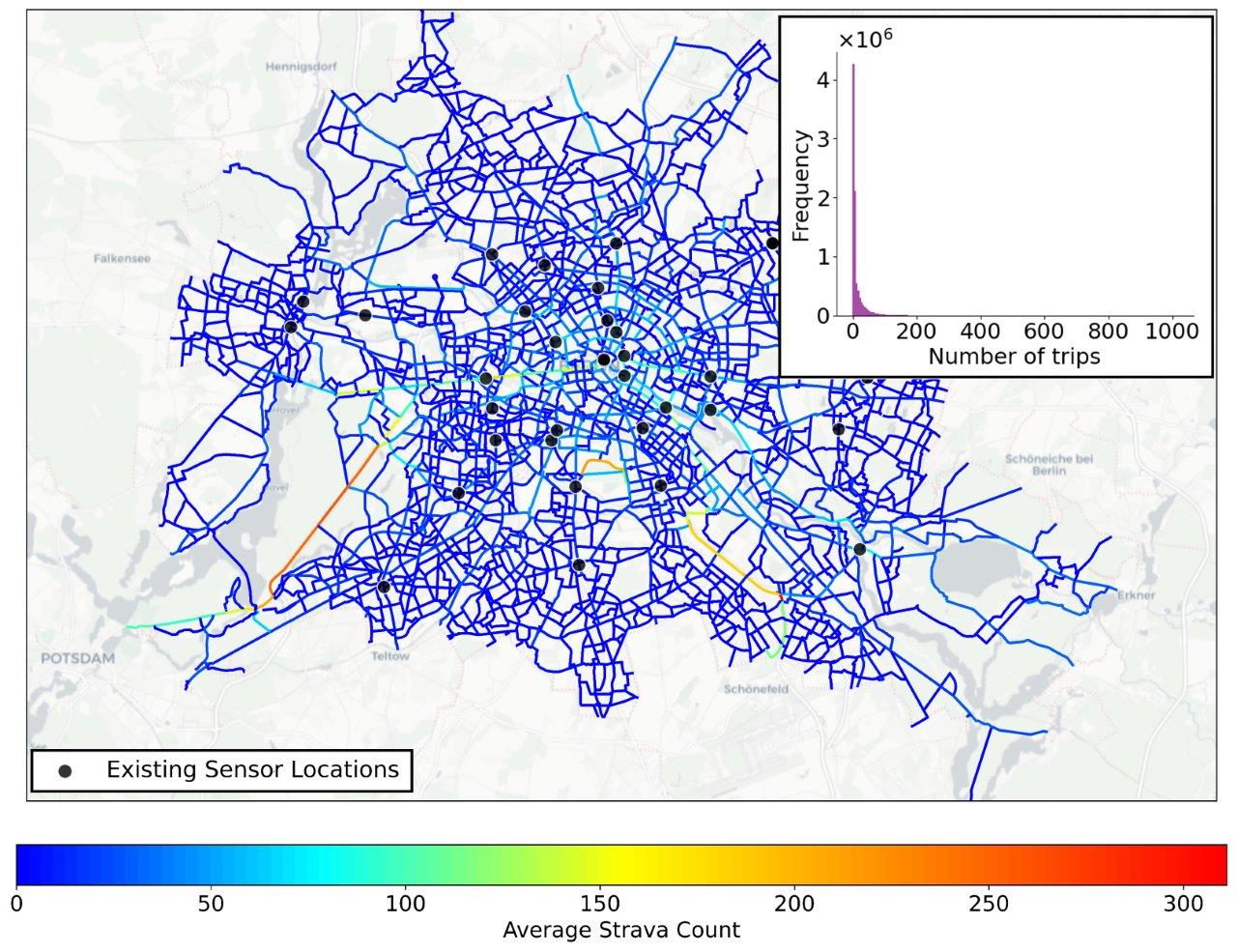}
        \caption{Cycling network in Berlin.}
        \label{fig:network_compraison_berlin}
    \end{subfigure}
    
     \vspace{0.5em}
     
    \begin{subfigure}{0.79\textwidth}
        \centering
        \includegraphics[width=\textwidth]{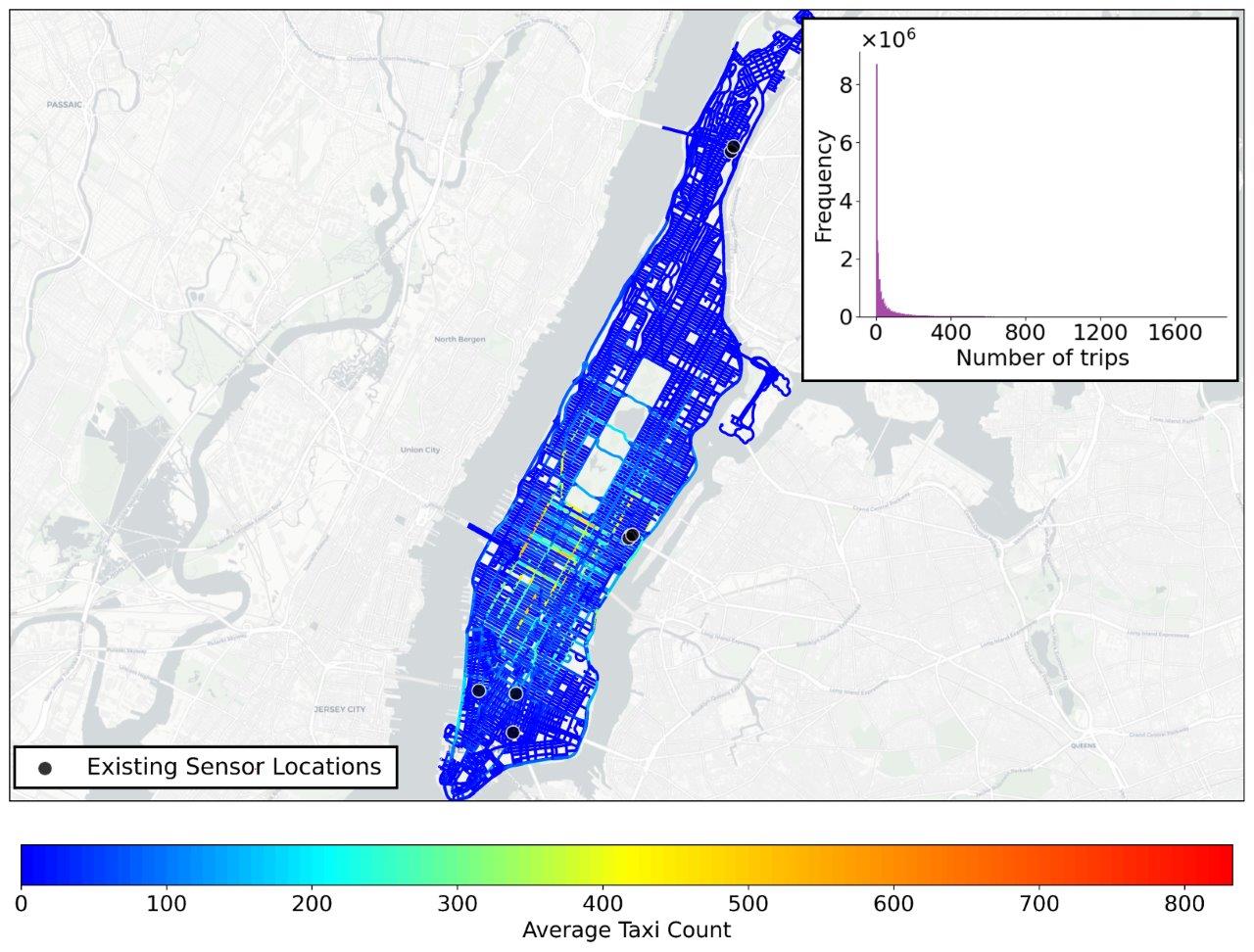}
        \caption{Street network in Manhattan, NYC.}
        \label{fig:network_compraison_ny}
    \end{subfigure}
    \caption[Street networks, sensor locations, and traffic volume]{Street networks and sensor locations in the two study areas. The Berlin cycling network consists of 4,958 street segments and 34 bicycle counting stations (21 permanent and 13 temporary). The Manhattan street network comprises 8,156 street segments and 8 temporary traffic sensors. Dots mark sensor locations; in Manhattan, several sensors are clustered spatially, with four nearly overlapping in Midtown. For visualization purposes, street segments are colored by average traffic volumes over the respective observation periods, although the analysis uses daily Strava counts for Berlin and hourly taxi counts for Manhattan. Insets show the distributions of daily and hourly traffic volumes, respectively.}
    \label{fig:city_maps}
\end{figure}

The Berlin dataset captures bicycle traffic volumes recorded by Strava users \citep{strava_metro_strava_2024} across 4,958 street segments in the city's dedicated cycling network \citep{senate_department_for_the_environment_mobility_consumer_and_climate_protection_berlin_radverkehrsnetz_2024}. This network includes both major roads and smaller streets, providing a comprehensive view of urban bicycle movement patterns. The dataset covers daily measurements from 2019 to 2023. The Manhattan dataset records the number of taxis passing each street segment in Manhattan \citep{new_york_taxi_and_limousine_commission_new_2016}, encompassing the entire public street network of 8,156 street segments. The dataset records hourly counts for January and February 2016. Both networks are depicted in Figure \ref{fig:city_maps}. 
To focus the analysis on typical traffic behavior, observations deviating by more than 3 standard deviations from a segment’s mean are excluded. This removes only a small share of observations (0.7\% in Berlin and 0.4\% in New York). Because traffic volumes in both cities are strongly right-skewed, the filter removes only high outliers associated with exceptional events, such as large bicycle demonstrations or unusually high taxi demand following major public events. Applying the filter at the segment level preserves meaningful heterogeneity across the network, including consistently high-traffic streets. The resulting filtered distributions are shown in the upper right corners of Figures~\ref{fig:network_compraison_berlin} and~\ref{fig:network_compraison_ny}.

Both datasets comprise a rich set of auxiliary features describing the physical, functional, and temporal characteristics of each street segment. In total, the Berlin dataset comprises 98 features per street segment, while the Manhattan dataset includes 79. The features are briefly outlined here; for a detailed list, see \cite{kaiser_data_2025} and \cite{kaiser_spatio-temporal_2025}. The features include street infrastructure details, as infrastructure design directly affects perceived safety and traffic behavior \citep{costa_unraveling_2024, kononov_relationships_2008,yang_towards_2019} (e.g., speed limits, road classification, number of lanes, and the presence of bicycle lanes or parking facilities). 
Structural properties of the street network are captured through standard graph-theoretic connectivity measures, which characterize the relative importance of each street segment within the overall network (e.g., centrality, betweenness, degree). These features reflect how network structure can influence movement patterns \citep{hillier_network_2005}.
Local activity and land-use intensity are represented by counts of nearby points of interest (e.g., shops, educational facilities, or transportation hubs), which have been shown to shape both taxi and bicycle demand \citep{askari_taxi_2020, fazio_bike_2021}. To account for environmental conditions, the dataset includes weather variables (e.g., temperature, precipitation, and sunshine duration) that modulate mode choice and daily traffic intensity \citep{koesdwiady_improving_2016, miranda-moreno_weather_2011}.
The Berlin dataset additionally contains socioeconomic indicators, which are known correlates of cycling participation \citep{goel_cycling_2022} (e.g., population density, age structure, and gender composition). The Berlin dataset also incorporates motorized traffic volumes measured at varying radii around each segment, as surrounding traffic conditions are strong predictors of cycling activity \citep{kaiser_counting_2025}.

Both cities have existing traffic sensor networks and pursue different strategies for temporal sensor deployment. In Berlin, 34 sensors (21 permanent and 13 temporary) record cyclist volumes during the study period (2019–2023) \citep{senate_department_for_the_environment_mobility_consumer_and_climate_protection_berlin_radverkehrszahlstellen_2024}. The permanent sensors are installed at fixed locations and record traffic volumes continuously. The temporary sensors are deployed recurrently at the same locations for one day per month, typically on a Wednesday in the first or second week, yielding twelve observations per site and year. Additionally, Berlin deploys further temporary sensors across the entire street network, yielding more than 10,000 daily observations at approximately 2,500 locations over the study period. However, these measurements cannot be matched reliably to the cycling network used in this study due to limited spatial referencing. Therefore, the study abstains from using them. These additional deployments are nevertheless described here to document existing monitoring practices and to provide context about the scale and structure of temporary sensor deployments observed in practice, which inform the range of deployment scenarios considered in this study.
In Manhattan, 8 temporary sensors measured overall motorized traffic volumes during the study period (January and February 2016) \citep{nyc_open_data_automated_2024}. Manhattan conducts these traffic counts over continuous one-week periods, with sensors rotating across locations rather than revisiting the same sites. The sites of these existing sensors are depicted in Figure~\ref{fig:city_maps}. 
Because physical sensors measure total traffic flows (i.e., all cyclists in Berlin, respectively all vehicles in Manhattan), whereas Strava and taxi data capture only subsets of these flows, the two data sources are not directly comparable. Thus, the analysis uses existing sensors solely to define observation locations. Computations based on existing sensors accordingly employ Strava or taxi counts at these locations, rather than the original sensor measurements.

Both datasets represent non-random subsamples of overall traffic volumes and therefore exhibit inherent biases. Strava rides are disproportionately recorded by young, male, and performance-oriented cyclists \citep{kaiser_spatio-temporal_2025}, while taxi trips in Manhattan tend to be shorter and reflect different travel purposes than private car traffic. Despite these limitations, both datasets are strongly correlated with overall cycling (0.61) and motorized traffic volumes (0.78) and are widely considered to provide meaningful insights into urban mobility patterns \citep{kaiser_counting_2025, lee_strava_2021}. Moreover, no alternative data currently offer comparable spatial and temporal coverage across entire city networks.

Given these known biases, explicit debiasing would in principle be desirable. Such debiasing aims to estimate true cycling or motorized traffic volumes, as measured by physical sensors, using the biased Strava and taxi data, potentially in combination with additional information \citep{kaiser_counting_2025, roy_correcting_2019}. Such a debiasing approach is implemented and documented in Appendix~\ref{apx:debiasing_data}. However, the resulting errors are large relative to the observed traffic levels. Moreover, debiasing based on available physical sensors would anchor the correction to those sensors, thereby reintroducing the very biases that this study seeks to overcome when optimizing sensor placement. For these reasons, no debiasing is applied in the main analysis; instead, the known dataset-specific biases are acknowledged and discussed when interpreting the results.

\section{Results \label{sec:results}}

The results section consists of three parts. The first part reports the interpolation performance of alternative spatial placement strategies for permanent sensors. The second part examines different temporal placement strategies for temporary sensors. The final part compares the interpolation performance achieved under optimal permanent and temporary sensor deployments.

\subsection{Spatial Placement Strategies for Permanent Sensors \label{sec:results_subsection1}}

Table~\ref{Table:results_1_permanent} reports the interpolation error for Berlin and Manhattan across spatial placement strategies. Sensor selection follows the strategies described in Section~\ref{sec:methods_sensor_placement_strategies_spatially}, and the table also includes the three corresponding baseline configurations (existing deployment, all-training-data, and random placement). Random placement is included solely as a scale reference to illustrate the range of prediction errors that can arise from uninformed sensor placement. In particular, the minimum random error constitutes an ex post oracle lower bound: it is observed only after evaluating many random configurations and cannot be realized through any ex ante sensor placement decision. 

The results indicate that the spatial sensor placement strategies (Voronoi area inequality and, more prominently, spatial dispersion) as well as active learning, consistently perform best across sensor budgets and cities. These approaches yield substantially lower errors than network-centrality and feature-based strategies, particularly at small sensor budgets. For example, with only 10 sensors, spatial dispersion in Berlin achieves an error 61.7\% lower than the worst-performing strategy (\acrshort{mae} 12.2 vs. 31.3), while in Manhattan, active learning reduces the error by a factor of 74.5\% (43.1 vs. 169.2).
Two additional aspects are noteworthy. First, feature-based strategies occasionally perform comparably to spatial or learning-based approaches. But they exhibit high variability and occasionally very large errors, rendering them unreliable. Second, while Voronoi-based placement performs competitively in Berlin, its advantages are weaker in Manhattan, where spatial dispersion and active learning consistently outperform it. This indicates that among the spatial sensor placement strategies, spatial dispersion is preferable over Voronoi area inequality.

The best-performing strategies also demonstrate clear advantages relative to the included baselines. 
In both cities, spatial dispersion and active learning outperform the median random placement across most sensor budgets, which represents typical uninformed sensor placement.
Relative to existing deployments, targeted placement yields marked performance gains: in Berlin, prediction error is reduced by roughly half when replacing the existing network with a well-placed configuration. In Manhattan, active learning also improves performance, though the gains are more modest, indicating that existing sensors are already relatively well placed. 
Finally, a performance gap remains between the best 100-sensor configurations and the idealized all-training-data benchmark, as expected given the much larger size of the all-training-data sensor set (3,470 sensors in Berlin and 5,708 in Manhattan).

\newcolumntype{C}[1]{>{\centering\arraybackslash}p{#1}}
\newcolumntype{R}[1]{>{\raggedleft\arraybackslash}p{#1}}
\begin{table*}[ht]
\centering
\caption[Prediction results under different placement strategies]{MAE prediction results of city-wide traffic using permanent sensor deployments under different sensor budgets in Berlin and Manhattan. For each column, the best value is shown in bold, followed by the second–, third–, and fourth–best values shaded from dark to light gray.}
\label{Table:results_1_permanent}
\definecolor{rankSecond}{gray}{0.70}
\definecolor{rankThird}{gray}{0.82}
\definecolor{rankFourth}{gray}{0.92}
\scriptsize
\begin{subtable}{\textwidth}
\centering
\begin{tabular}{ l|
    rrrrrrr|rrrrrrr}
    \toprule
    City & \multicolumn{7}{c|}{Berlin} & \multicolumn{7}{c}{Manhattan}\\
    Sensor budget (K) &10 & 25 & 34$^{*}$ & 50 & 75 & 100 & all$^{**}$ & 8$^{*}$ & 10 & 25 & 50 & 75 & 100 & all$^{**}$ \\
    \midrule
Betweenness &28.7 & 19.4 & 30.3 & 31.2 & 30.8 & 38.2 & -- & 125.4 & 118.3 & 91.5 & 57.9 & 59.8 & 54.3 & -- \\
Closeness &19.9 & 32.7 & 40.9 & 24.2 & 26.9 & 29.3 & -- & 98.2 & 104.0 & 110.8 & 59.2 & 53.8 & 51.1 & -- \\
\lightmidrule
Feature div. (all) &31.3 & 62.9 & 31.7 & 33.8 & 29.1 & 28.5 & -- & 146.8 & 169.2 & 109.2 & 93.6 & 107.3 & 105.2 & -- \\
Feature div. (infr. sel.) &27.9 & 16.0 & 15.0 & 14.0 & 13.6 & 14.7 & -- & 47.6 & \cellcolor{rankThird}{47.8} & \cellcolor{rankFourth}{47.2} & 49.0 & 45.3 & 44.4 & -- \\
Redundancy (all) &\cellcolor{rankFourth}{13.3} & 14.4 & 13.2 & 13.3 & 12.4 & \cellcolor{rankFourth}{12.1} & -- & 88.5 & 85.0 & 53.3 & 45.1 & 48.6 & 43.4 & -- \\
Redundancy (infr. sel.) &17.1 & 23.6 & 18.9 & 17.0 & 14.6 & 14.6 & -- & \cellcolor{rankThird}{43.9} & 58.8 & 49.2 & \cellcolor{rankThird}{42.9} & \cellcolor{rankThird}{41.7} & \cellcolor{rankFourth}{41.7} & -- \\
Coverage (all) &30.2 & 63.1 & 36.8 & 37.3 & 33.3 & 27.7 & -- & 167.0 & 99.9 & 121.2 & 90.4 & 82.7 & 60.5 & -- \\
Coverage (infr. sel.) &19.9 & 14.8 & 14.1 & 13.6 & 13.4 & 13.0 & -- & 54.6 & 55.8 & 49.5 & 46.1 & 43.2 & 41.9 & -- \\
\lightmidrule
Voronoi &15.3 & \cellcolor{rankSecond}{11.6} & \cellcolor{rankThird}{12.0} & \cellcolor{rankFourth}{12.1} & \cellcolor{rankSecond}{11.5} & 14.0 & -- & 66.3 & 62.1 & 48.0 & 47.4 & 43.0 & \cellcolor{rankThird}{41.0} & -- \\
Spatial dispersion &\cellcolor{rankSecond}{12.2} & \cellcolor{rankThird}{12.0} & \cellcolor{rankSecond}{11.5} & \cellcolor{rankSecond}{11.2} & \cellcolor{rankFourth}{11.8} & \cellcolor{rankThird}{11.9} & -- & 48.1 & \cellcolor{rankFourth}{48.0} & \cellcolor{rankThird}{45.8} & \cellcolor{rankSecond}{41.3} & \cellcolor{rankSecond}{40.9} & 43.3 & -- \\
\lightmidrule
Active learning &\cellcolor{rankThird}{12.5} & 13.7 & \cellcolor{rankFourth}{12.7} & \cellcolor{rankThird}{11.9} & \cellcolor{rankThird}{11.6} & \cellcolor{rankSecond}{11.3} & -- & \cellcolor{rankSecond}{42.3} & \cellcolor{rankSecond}{43.1} & \cellcolor{rankSecond}{44.1} & 44.7 & \cellcolor{rankFourth}{42.1} & \cellcolor{rankSecond}{40.6} & -- \\
\midrule
Random (median) &14.5 & \cellcolor{rankFourth}{13.6} & 13.4 & 13.0 & 12.7 & 12.4 & -- & 48.7 & 49.1 & \cellcolor{rankFourth}{47.3} & \cellcolor{rankFourth}{44.6} & 43.1 & 41.8 & -- \\
Random (min) &\textbf{11.6} & \textbf{11.2} & \textbf{11.1} & \textbf{10.9} & \textbf{10.7} & \textbf{10.9} & -- & \textbf{39.4} & \textbf{40.2} & \textbf{38.3} & \textbf{36.1} & \textbf{35.4} & \textbf{35.5} & -- \\
Random (max) &40.6 & 28.9 & 23.4 & 20.1 & 19.1 & 17.4 & -- & 216.4 & 275.2 & 109.6 & 75.2 & 55.7 & 54.5 & -- \\
\midrule
All training data &-- & -- & -- & -- & -- & -- & \textbf{9.0} & -- & -- & -- & -- & -- & -- & \textbf{28.1} \\
Existing &-- & -- & 21.7 & -- & -- & -- & -- & \cellcolor{rankThird}{44.0} & -- & -- & -- & -- & -- & -- \\
    \bottomrule
\end{tabular}
\end{subtable}

\vspace{1em} 

{\raggedright \scriptsize $^{*}$ For both cities, the table includes a column corresponding to the number of existing sensors: 34 in Berlin and 8 in Manhattan.\par}
{\raggedright \scriptsize $^{**}$ 'all' denotes the case where all segments in the training data are equipped with sensors: 3,470 in Berlin and 5,708 in Manhattan.\par}
\end{table*}

The results further clarify how sensor budget interacts with placement quality. 
Poor placement can outweigh the benefits of larger sensor budgets: under the random baseline, the worst configurations with 100 sensors yield higher errors than the median configurations with only 10 sensors in both cities (e.g., in Berlin, \acrshort{mae} 17.4 vs. 14.5; in Manhattan, 54.5 vs. 48.7). 
Furthermore, as the number of sensors increases, performance differences between placement strategies systematically narrow, reflecting that once spatial coverage becomes sufficiently dense, marginal placement decisions matter less. Nevertheless, even at 100 sensors, spatial and learning-based strategies continue to outperform centrality- and feature-driven approaches. 
Lastly, it should also be noted that prediction error generally decreases as the number of sensors increases, though not strictly monotonically. Occasional irregularities appear, suggesting that newly added sensors sometimes introduce local biases in model fitting.

For completeness and robustness, several additional analyses are reported in the Appendix. Appendices~\ref{appx:permanent_further_strategies},~\ref{apx:extending_existing_sensor_network} and~\ref{appx:permanent_rmse} jointly assess the sensitivity of the results to alternative experimental choices; namely, additional feature-based placement strategies, extending existing sensor networks rather than placing sensors from scratch, and evaluating performance using \acrshort{rmse} instead of \acrshort{mae}. Across all extensions, the interpretation patterns hold.

\subsection{Temporal Placement Strategies for Temporal Sensors \label{sec:results_subsection2}}

\begin{figure}[ht!]
    \centering
    \begin{subfigure}{0.49\linewidth}
        \includegraphics[width=\linewidth]{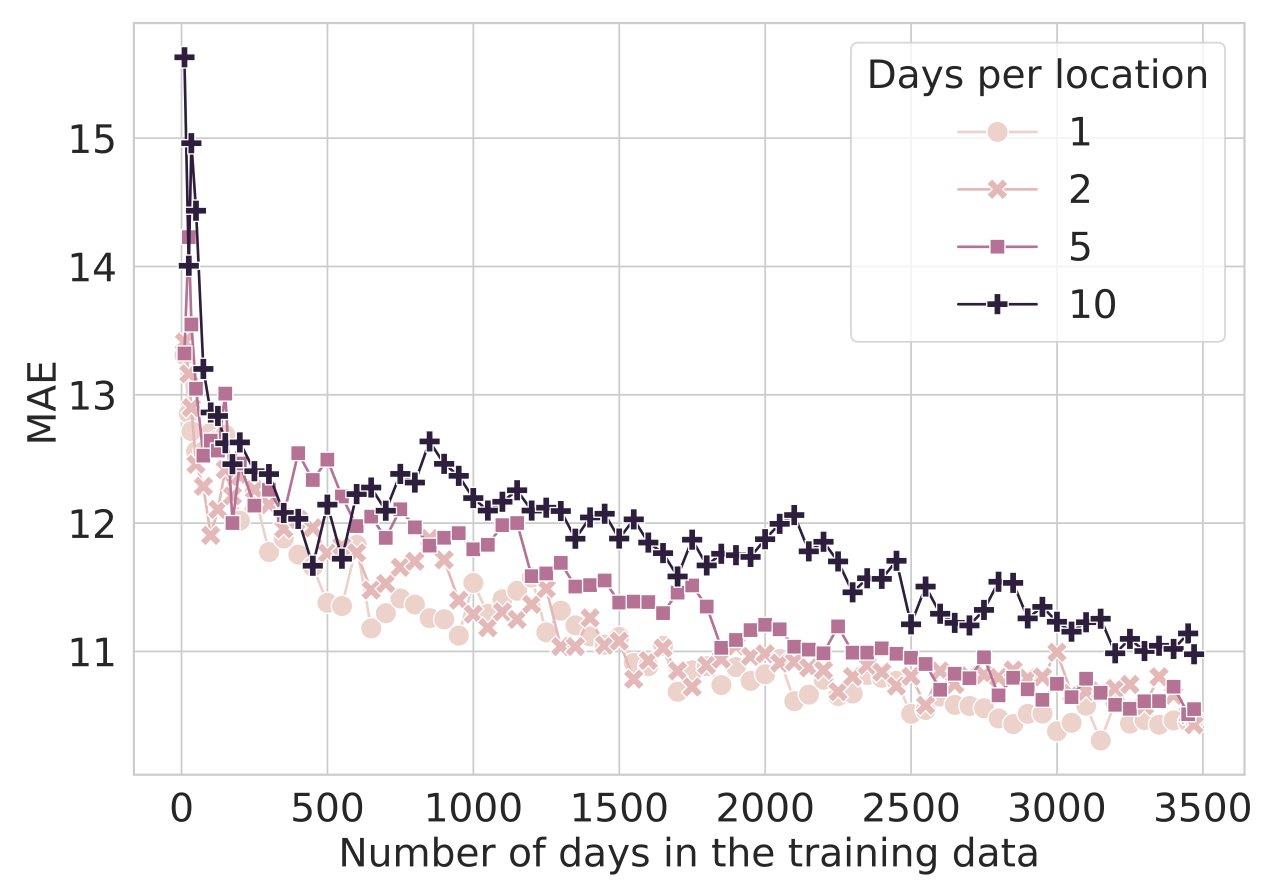}
        \caption{Berlin — Days per location}
        \label{fig:location_berlin_mae_evaluation}
    \end{subfigure}
    \hfill
            \begin{subfigure}{0.49\linewidth}
        \includegraphics[width=\linewidth]{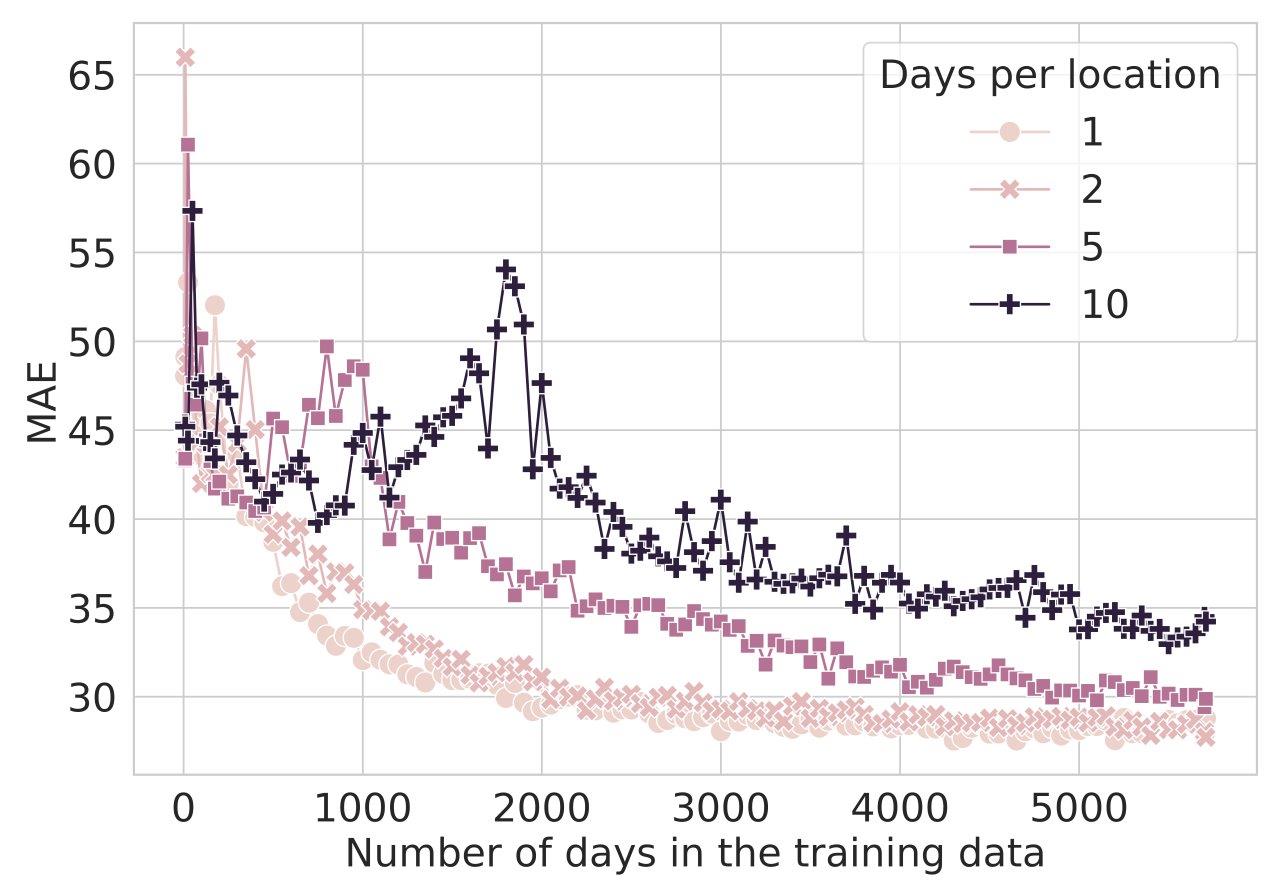}
        \caption{Manhattan — Days per location}
        \label{fig:newyork_location_mae_evaluation}
    \end{subfigure}

    \vspace{0.5cm}
    
        \begin{subfigure}{0.49\linewidth}
        \includegraphics[width=\linewidth]{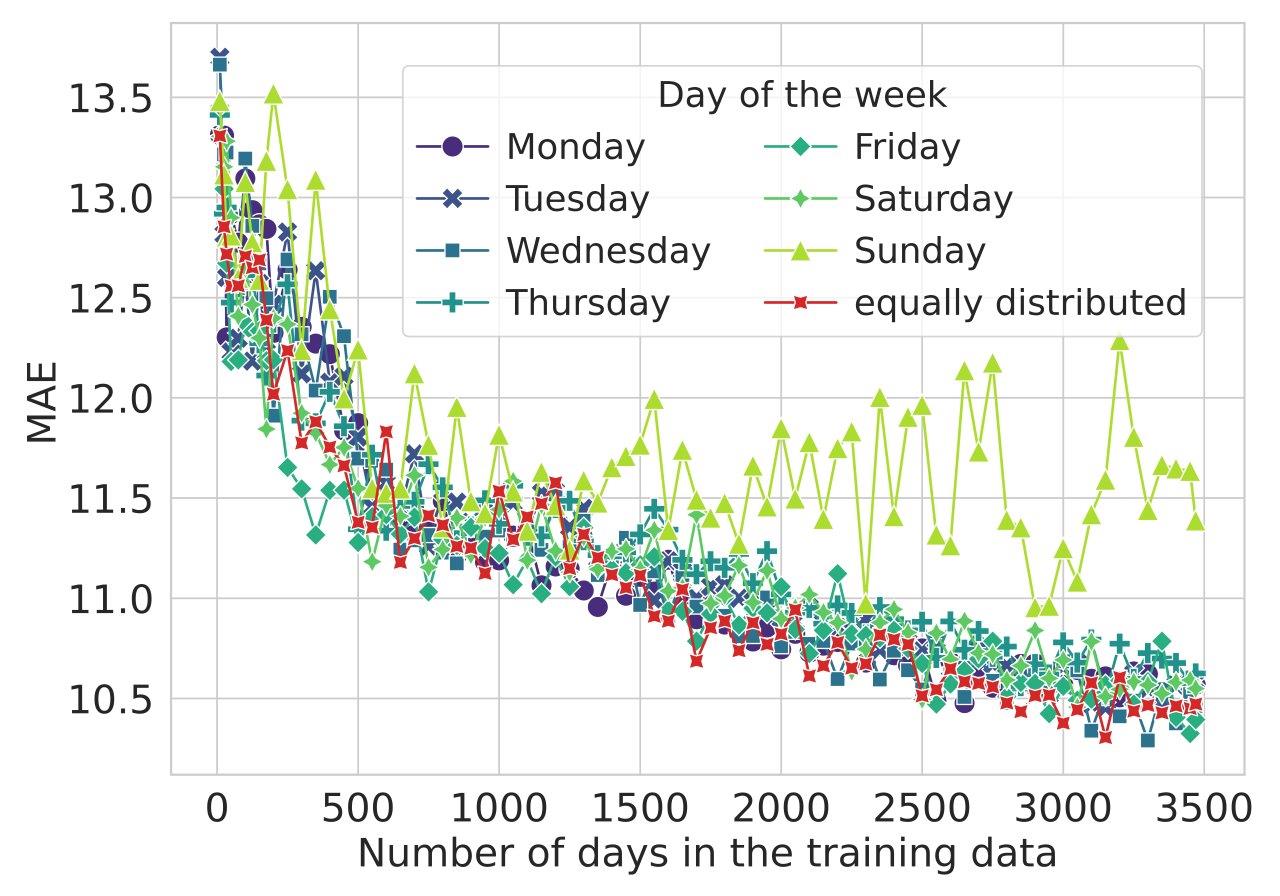}
        \caption{Berlin — Day-of-week effects}
        \label{fig:berlin_week_mae_evaluation}
    \end{subfigure}
    \hfill
    \begin{subfigure}{0.49\linewidth}
        \includegraphics[width=\linewidth]{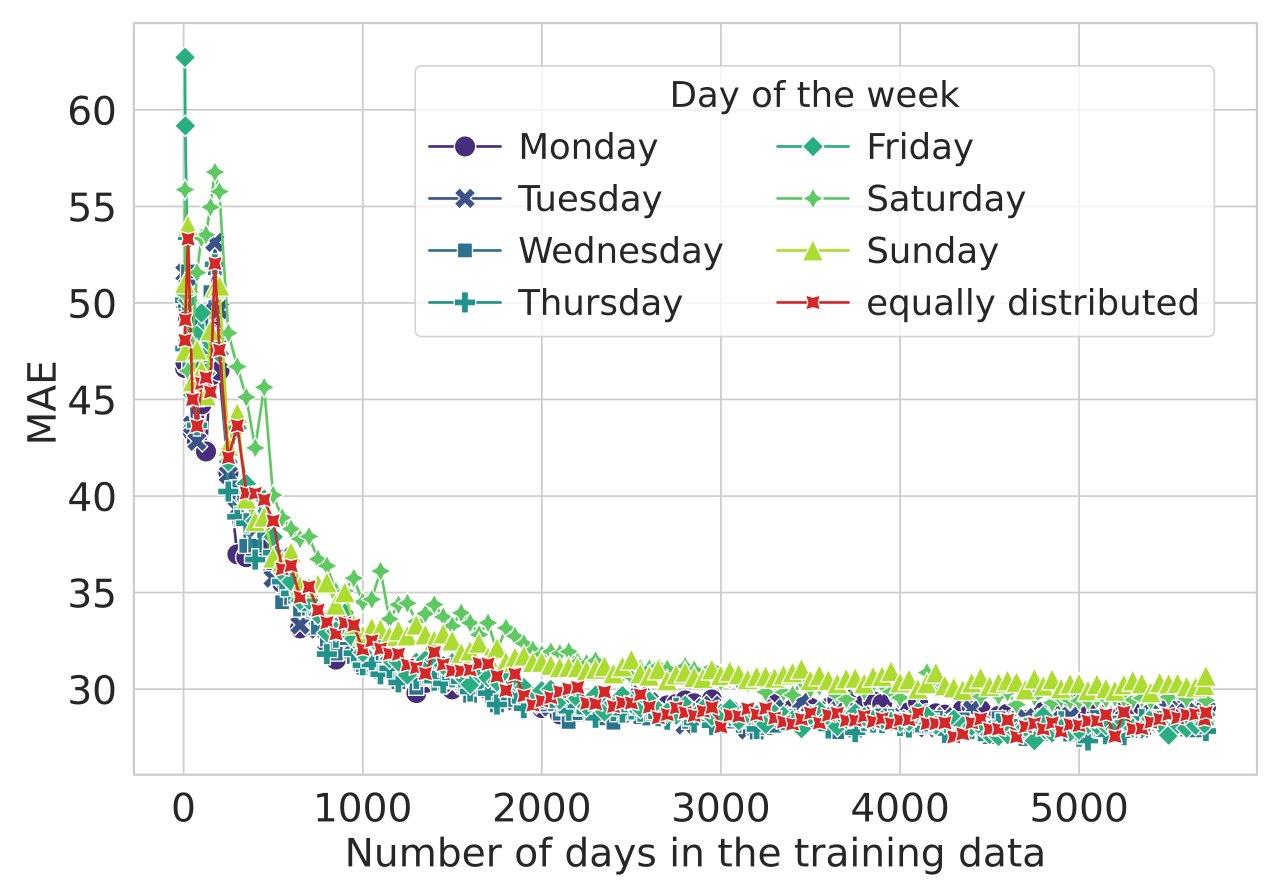}
        \caption{Manhattan — Day-of-week effects}
        \label{fig:newyork_week_mae_evaluation}
    \end{subfigure}
    \caption[Temporary placement considerations of temporary sensors (MAE)]{Temporary placement considerations of temporary sensors: The upper row compares deployment strategies that vary the number of days allocated to each location, effectively trading off between broader spatial coverage (more locations, fewer days each) and repeated measurement at fewer locations, while keeping the total number of observations constant. The lower row examines whether restricting temporary measurements to specific weekdays affects prediction accuracy. Results are shown for Berlin (left) and Manhattan (right). The reported error metric is \acrshort{mae}.}
    \label{fig:results3}
\end{figure}
  
This section examines how temporary deployment strategies influence interpolation performance, focusing on revisiting versus rotating sensor locations and the temporal distribution of sampled weekdays. The corresponding methodological details were explained in Section~\ref{sec:methods_sensor_placement_strategies_temporally}. Although the focus here is on temporal deployment, a spatial placement strategy must still be specified to determine which locations are sampled. As discussed in Section~\ref{sec:results_subsection1}, the three most favorable spatial placement strategies are spatial dispersion, Voronoi area inequality, and active learning. To maintain brevity, the main text reports results using spatial dispersion, while results for the other two strategies are presented in Appendix~\ref{apx:temporary_furtherstrategies}, as they lead to the same conclusions.

The effect of revisiting versus rotating locations is shown in Figures~\ref{fig:location_berlin_mae_evaluation} and ~\ref{fig:newyork_location_mae_evaluation} using \acrshort{mae} as the performance metric. Results are reported for increasing numbers of days in the training data. Because the total number of training days is fixed, strategies that revisit locations trade spatial coverage for repeated measurement: increasing the number of days per location proportionally reduces the number of distinct locations observed.

Across both cities, the MAE patterns indicate that rotating across many distinct locations yields better performance than repeatedly sampling the same locations. When the number of observations is small (fewer than approximately 1,000 days), results are volatile, and no strategy is consistently superior. As the training set grows, however, performance stabilizes, and a clear ordering emerges, with strategies that allocate fewer days per location consistently outperforming those that rely on more intensive revisits. In this stabilized regime, sampling more days per location only rarely performs comparably to, or marginally better than, broader-coverage strategies, and these isolated cases do not affect the overall pattern. Quantitatively, the average MAE across all numbers of days for Berlin is 0.8 lower when measuring one day per location instead of ten (a reduction of around 7\%); for Manhattan, the corresponding values are 8.3 and 21\%. Taken together, these results indicate that broader spatial coverage is generally more informative than repeated temporal measurement at the same locations. Accordingly, sampling one day per location emerges as the preferred strategy.
These findings are robust to the choice of error metric; the same qualitative patterns are observed when using \acrshort{rmse} instead of \acrshort{mae} (Appendix~\ref{apx:temporary_rmse}).

The influence of weekday selection is shown in Figures~\ref{fig:berlin_week_mae_evaluation} and ~\ref{fig:newyork_week_mae_evaluation}, comparing strategies that restrict sampling to a single weekday with a strategy distributing observations evenly across the week.

The weekday-based strategies differ little from one another, with the equally distributed strategy performing as a good average. Differences in MAE across weekdays are substantially smaller than those observed when varying the number of days per location, indicating that the specific day of the week is a comparatively minor factor. Solely Sunday tends to perform clearly worse, likely because movement patterns on that day differ largely from those on the other weekdays. Other than that, evenly distributing measurements across weekdays provides a robust average performance. In conclusion, evenly distributing sensor measurements across all weekdays is the most reliable choice, although the advantage over alternative weekday selections is modest. An analogous seasonal analysis for Berlin (Appendix~\ref{appx:results3_seasons}) further supports this conclusion, showing that evenly distributed deployments across seasons is also the most reliable choice. Such a seasonal analysis is not feasible for Manhattan due to limited temporal coverage.

\subsection{Comparison of Interpolation Performance under Optimal Permanent and Temporary Sensor Deployment \label{sec:results_subsection3}}

\begin{figure}[ht!]
    \centering
    \begin{subfigure}{0.49\linewidth}
        \includegraphics[width=\linewidth]{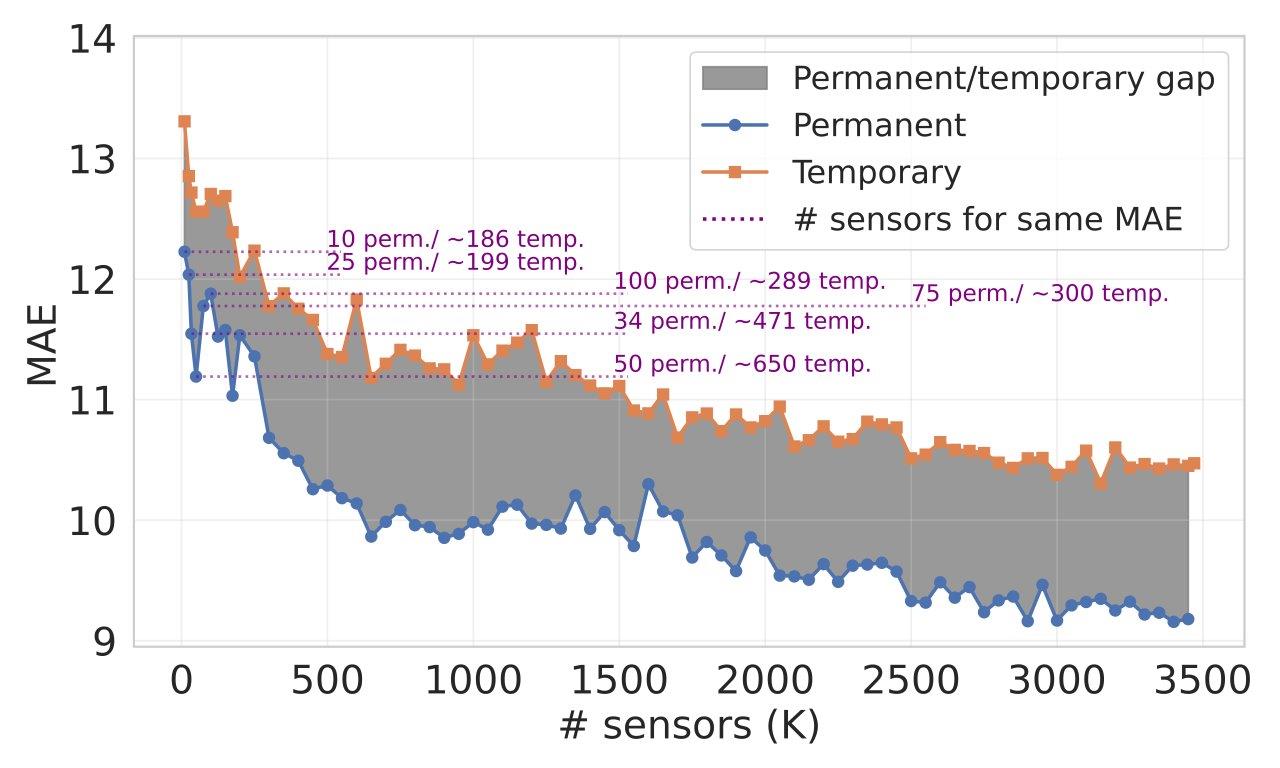}
        \caption{Berlin}
        \label{fig:temporary_berlin_mae}
    \end{subfigure}
    \hfill
        \begin{subfigure}{0.49\linewidth}
        \includegraphics[width=\linewidth]{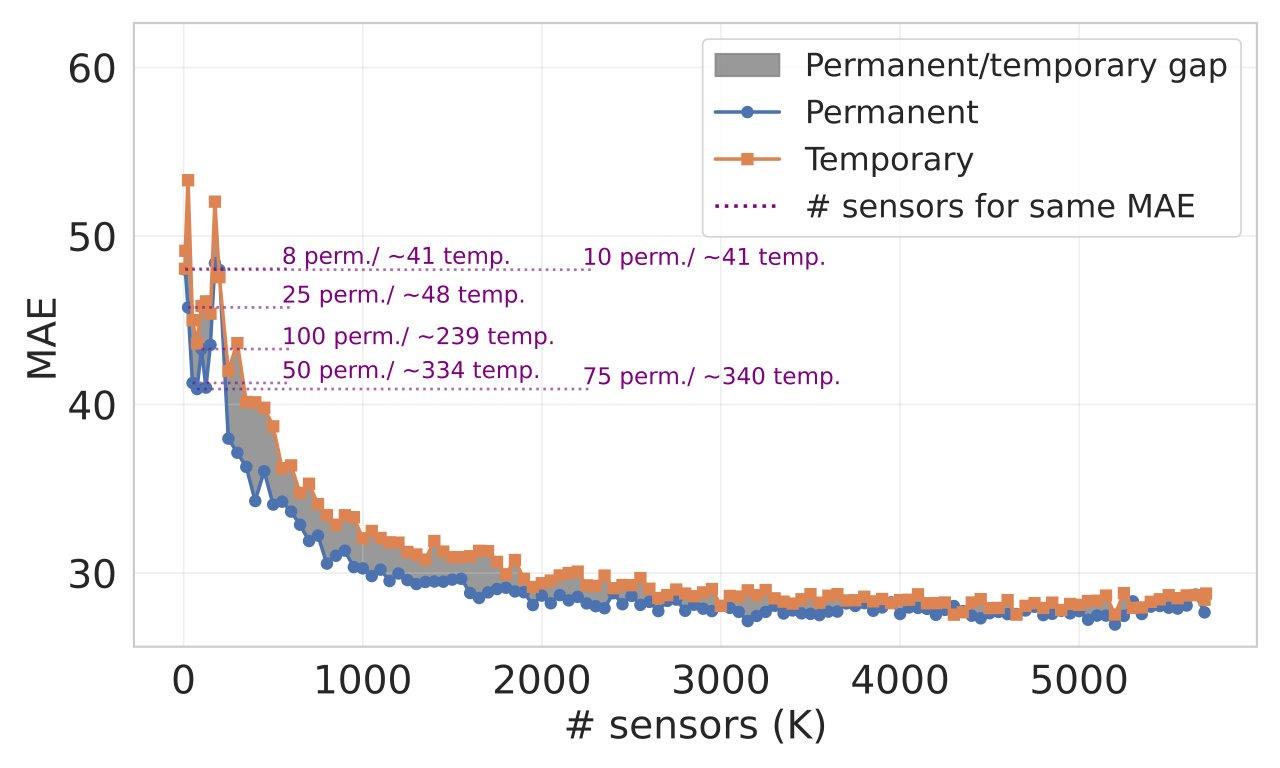}
        \caption{Manhattan}
        \label{fig:temporary_newyork_mae}
    \end{subfigure}
    \caption[Prediction performance under temporary and permanent deployment]{\acrshort{mae} performance under temporary and permanent sensor deployments for citywide traffic volume estimation, shown as a function of the sensor budget $K$. 
    \label{fig:temporary}}
\end{figure}

This section contrasts citywide interpolation performance between permanent and temporary sensor deployments. Methodological details are provided in Section~\ref{sec:methods_sensor_placement_strategies_temporally}. Figure~\ref{fig:temporary} reports the \acrshort{mae} achieved by optimally placed permanent sensors and by optimally placed and scheduled temporary sensors. Here, optimal refers to configurations guided by the findings of the previous sections: spatial placement follows the spatial dispersion criterion, and temporary sensors are deployed for one day per location and evenly distributed across weekdays. Results for Voronoi area inequality and active learning–based placement are reported in Appendix~\ref{appx:results3_otherlists}. 

Overall, temporary deployments yield higher prediction errors than permanent deployments, but the performance gap is consistently small. In Berlin, temporary sensors closely match the performance of permanent ones, with \acrshort{mae} being 0.5 to 1.7 higher across all sensor budgets $K$, corresponding on average to an 11.8\% increase. Manhattan exhibits greater variability, with \acrshort{mae} increases ranging from 0.0 to 7.6 (mean increase: 4.4\%). However, given the overall scale of the \acrshort{mae}, these increases remain small and relatively stable across sensor budgets in both cities.

To make the small performance gap between permanent and temporary deployments more interpretable, we translate prediction accuracy into an equivalent sensor requirement. For each permanent sensor budget considered in Section~\ref{sec:results_subsection1} (10, 25, 50, 75, 100, and the existing deployment), we identify how many temporary sensor observations are needed to attain the same MAE. Because temporary deployments are evaluated only in increments of 50 observations, the corresponding values are obtained by linear interpolation between adjacent points; when multiple temporary sensor counts yield comparable MAE, the smallest such count is reported.
This comparison confirms that the performance difference is minor. For example, in Manhattan, the MAE achieved with 50 permanently installed sensors can be matched with approximately 334 single-day observations from temporary sensors.

The small performance differences are particularly notable given the extreme imbalance in temporal data availability. For a fixed sensor budget $K$, both deployment strategies use identical spatial locations, but differ dramatically in temporal coverage. In Berlin, one permanent sensor provides approximately 4 years of daily observations per location, whereas one temporary sensor collects data for only 1 day; a roughly 1,460-fold difference in observations. In Manhattan, permanent sensors provide around two months of hourly data per location, compared to a single day for temporary deployment, corresponding to roughly a 60-fold difference.


\section{Discussion}

This discussion first interprets the empirical findings around spatially diverse and learning-based sensor placement strategies. It then examines which of these strategies is preferable given real-world conditions. The discussion subsequently situates the temporal placement results within existing research and considers the flexibility they imply regarding the choice between temporary and permanent sensor deployments. Finally, the findings are contrasted with current deployment practices before outlining key limitations.

Regarding the spatial placement strategies, the observed performance differences are consistent with established insights from spatial and data-efficient sampling. Spatially evenly distributed sensor placement improves model generalization by selecting locations that form a more representative subsample \citep{stevens_jr_spatially_2004}. Similarly, the strong performance of active learning is consistent with uncertainty-based sampling theory and prior empirical evidence showing its effectiveness in improving data acquisition efficiency \citep{krause_near-optimal_2008, settles_active_2009}. 
By contrast, centrality-based placement strategies consistently yield weaker interpolation performance. One plausible explanation is that, by prioritizing structurally important corridors, these strategies concentrate sensors in high-flow and strongly correlated locations. Feature-based strategies exhibit similarly reduced performance, which may be caused by an overemphasis on rare or extreme feature combinations.

Overall, this study recommends spatial dispersion to cities facing real-world operational constraints as a spatial placement strategy. When comparing spatial dispersion with Voronoi area inequality, the former performed more consistently and robustly across cities, transport modes, and sensor budgets, making it the more reliable choice.
When comparing spatial dispersion with active learning, spatial dispersion also entails substantially lower practical and organizational effort. Active learning requires an iterative placement process in which sensor locations are selected based on model uncertainty, which in turn presupposes the availability of sufficient data to train and repeatedly update an interpolation model. This implies that supplementary data must already be collected, processed, and modeled prior to placement decisions, and that placement and modeling must proceed in tandem. By contrast, spatial dispersion can be implemented as a one-time, model-free procedure based solely on network geometry. 
Beyond practical considerations, spatial dispersion is also a socially desirable placement strategy. Spatial sensor placement affects not only predictive accuracy but also whose mobility patterns are observed. Spatially diverse sensor placement can help mitigate equity concerns by reducing the underrepresentation of peripheral and socioeconomically disadvantaged neighborhoods in traffic monitoring \citep{gebru_datasheets_2021, robinson_optimizing_2022}. 

Regarding the temporal placement strategies, the observed patterns are consistent with work on diverse data assemblages. Distributing observations across many distinct locations proves more beneficial than repeatedly sampling the same sites, indicating that spatial coverage is more informative for citywide traffic estimation than long temporal histories at individual locations. This pattern is consistent with prior work showing that diversity in training data improves model performance by exposing learning algorithms to a broader range of conditions and patterns \citep{gong_diversity_2019}. In practice, temporally short but spatially expansive deployments increase the variety of observed traffic states and network contexts, enabling effective interpolation even when individual locations are observed only briefly.

The findings regarding temporary sensors highlight an important degree of flexibility in sensor deployment. When traffic is observed at a sufficiently large number of spatially diverse locations, interpolation accuracy remains largely comparable whether sensors are installed permanently or deployed temporarily. At the same time, both temporary and permanent sensors typically entail per-site costs, ranging from several hundred to a few thousand dollars \citep{minge_evaluation_2010, ozan_state---art_2021}. Given their comparable predictive performance, cities are therefore not locked into a single deployment model. Instead, they can select deployment strategies that best fit their financial, operational, and planning constraints.
Temporary sensors, in particular, offer substantial advantages in this regard. By allowing sensors to be rotated across locations, cities can expand spatial coverage, experiment with alternative monitoring strategies, or strategically complement existing permanent networks without committing to long-term installations. This flexibility lowers barriers to data-driven traffic monitoring and is especially valuable for cities facing tight budget constraints or evolving planning needs.

The practical relevance of these findings becomes evident when they are compared with current sensor deployment practices. Overall, the results indicate that spatial placement decisions are the primary driver of improvements in interpolation performance, far outweighing the gains achievable through temporal optimization. At the same time, politically motivated or visibility-driven placement decisions risk reinforcing existing spatial data inequalities by concentrating sensors in already well-observed, high-profile areas. Berlin provides a clear illustration of this tension. Bicycle counters are distributed according to administrative priorities: they are evenly allocated across districts for political notions of fairness and frequently installed at highly visible locations along major cycling routes. While such placements may serve symbolic or motivational purposes \citep{claes_bicycle_2016}, they are more likely to yield biased data and, as reflected in the estimation results, lead to suboptimal interpolation performance. The potential for improvement using spatial dispersion as a placement strategy is large.
Manhattan performs better in this regard, reflecting a greater alignment between deployment practices and data-driven placement principles. However, even there, performance could likely be further improved by deploying temporary sensors for shorter periods at each location and rotating them more frequently across the network.

While these results highlight substantial opportunities to improve sensor deployment, several limitations should be acknowledged, indicating further research directions.
A limitation is that the tested placement strategies do not account for real-world limitations in sensor placement. In practice, spatially optimal locations may not always be feasible due to requirements related to power supply, data transmission, accessibility, maintenance, or visibility \citep{leduc_road_2008, owais_traffic_2022}. Likewise, temporally optimal scheduling may conflict with operational realities, such as the availability of field staff, restrictions on weekend or holiday work, or adverse weather conditions during deployment periods. The proposed framework is flexible and could be extended to incorporate such constraints, enhancing applicability and realism. For example, feasibility masks, deployment calendars, or cost weights could be integrated directly into the optimization process. 
Furthermore, the analysis relies on proxy traffic data (Strava cycling volumes and taxi trips). While both correlate with observed traffic counts, they reflect the behavior of specific user groups and do not capture total population flows. As a result, absolute performance levels should be interpreted cautiously. However, the consistency of the main findings across both cities and transport modes supports the applicability of the qualitative conclusions to other urban contexts.
A further limitation is that all placement strategies are evaluated using a single interpolation model. While this model is computationally lightweight and performs well on tabular traffic data, future work should examine whether optimal sensor placements differ for more complex models, such as graph neural networks \citep{kaiser_spatio-temporal_2025}, which may offer performance gains but come with higher computational and scalability costs.
Beyond this, this study focuses on interpolation rather than forecasting. In practice, cities may seek to deploy sensors that support both purposes simultaneously. Future research could therefore explicitly investigate joint placement strategies that optimally balance the requirements of interpolation and forecasting.
Lastly, although the proposed placement strategies already achieve strong estimation performance, they are all conceptually simple and rely on relatively lightweight decision rules. It therefore remains an open question whether slightly more complex, yet still practically implementable strategies could yield further improvements. 
In this context, reinforcement learning is a promising approach to discover such more complex strategies. Reinforcement learning is not directly deployable as a placement strategy in any given city because it requires near-complete ground-truth data for training. 
However, reinforcement learning can be used on available proxy datasets to iteratively explore the space of possible sensor configurations, identify high-performing placement patterns, and encode them in a learned policy. These patterns can subsequently be analyzed using explainability techniques and distilled into new, interpretable placement strategies that can then be applied to arbitrary cities.
A corresponding conceptual framework is outlined in Appendix~\ref{appx:RL_model} or in \cite{gupta_inspire-gnn_2025}.


\section{Conclusion}
This study demonstrates that the spatial and temporal placement of traffic sensors is critical to the accuracy of citywide traffic volume interpolation. By systematically comparing a broad set of spatial and temporal placement strategies and by contrasting permanent and temporary deployments across two cities and transport modes, the analysis provides a novel, comprehensive empirical assessment of sensor placement strategies. 

Across both cities and transport modes, the results demonstrate that carefully designed spatial and temporal sensor placement substantially improves citywide traffic estimation performance. Spatial placement strategies, especially based on spatial dispersion, Voronoi area inequality, and active learning, consistently outperform alternative approaches, reducing the mean absolute error by over 60\% for Berlin and 70\% for Manhattan. For temporal placement, the findings indicate that temporary sensors are most effective when deployed for a single day per location and rotated across many distinct locations. Moreover, distributing these observations evenly across weekdays yields the best performance, reducing the error by an additional 7\% in Berlin and 21\% in Manhattan. When sensors are placed optimally, temporary deployments achieve prediction accuracy comparable to permanent installations, even though they rely on substantially fewer observations. That these patterns hold across two structurally different cities and transport modes underscores the robustness and transferability of the proposed placement strategies. From a policy perspective, these findings suggest that cities can substantially improve traffic volume interpolation by carefully considering both the spatial and temporal placement of sensors. In particular, depending on budgetary and operational constraints, cities can rely on temporary rather than permanent sensor deployments without substantial loss of interpolation performance.

Several avenues for future research remain. 
First, while the spatial and temporal placement strategies evaluated here already achieve substantial performance gains, the space of possible strategies is not exhaustive, and more advanced approaches may yield further improvements. 
Second, future work could extend the temporal dimension by applying active learning not only to spatial but also to temporal sensor placement. 
Third, practical decision-support tools would help translate these strategies into operational practice, for instance, through web-based applications that generate placement recommendations from basic network inputs. Such tools could explicitly incorporate deployment constraints and planning priorities, enabling practitioners to explore trade-offs between statistical performance, cost, and operational feasibility. 
Finally, extending the framework to jointly optimize sensor placement for both interpolation and forecasting, and to explicitly account for real-world constraints such as power supply, accessibility, and maintenance requirements, would further enhance its practical relevance.

\clearpage
\printglossary[type=\acronymtype, title={Abbreviations}, nonumberlist]

\vspace{1cm}
\paragraph{Funding Statement}
I am grateful the European Union’s Horizon Europe research and innovation program funded this project under Grant Agreement No 101057131, Climate Action To Advance HeaLthY Societies in Europe (CATALYSE). 

\paragraph{Acknowledgments}
 I thank E. Kolibacz, L. H. Kaack, C. L. Azevedo, and C. Sobral for their valuable feedback and comments.

\paragraph{Competing Interests}
The author declares no competing interests.

\newpage
\appendix
\section{Error metrices \label{app:error_metrices}}
 \acrshort{mae} and \acrshort{rmse} are defined as:
\begin{equation}
\text{MAE} = \frac{1}{n} \sum_{i=1}^{n} |y_i - \hat{y}_i|
\end{equation}

\begin{equation}
\text{RMSE} = \sqrt{\frac{1}{n} \sum_{i=1}^{n} (y_i - \hat{y}_i)^2}
\end{equation}

with $y$ the true and $\hat{y}$ the predicted values, and $n$ the number of observations.
\section{Permanent Sensor Placement - further deployment strategies \label{appx:permanent_further_strategies}}
This appendix presents additional results that complement Section~\ref{sec:results_subsection1}, showing performance for further feature-based spatial sensor placement strategies. The performance of feature-based sensor placement strategies depends on the specific feature sets used in computing them. While the main paper reports results based on all time-invariant features and a selected set of infrastructure-related features, this appendix explores further feature sets.

Specifically, three additional feature sets are considered: 
(1) a connectivity-based feature set (betweenness, degree, closeness, and clustering coefficient); 
(2) a full infrastructure feature set, including both built-environment and points-of-interest variables (for example, but not limited to: the maximum street, the number of lanes, the type of cycling lane, the type of road pavement, the number of shops, educational institutions, hospitals, or bus stops within the area); and 
(3) a points-of-interest-only feature set (for example, but not limited to: the number of shops, educational institutions, hospitals, or bus stops within the area). 
A complete list and detailed description of all features are provided in \citet{kaiser_data_2025} and \citet{kaiser_spatio-temporal_2025}.

Tables~\ref{Table:results_1_permanent_appendix} report MAE and RMSE prediction results for both Berlin and Manhattan under permanent sensor deployment for these additional feature configurations. The feature specifications presented here yield performance patterns that are consistent with the main results reported in Section~\ref{sec:results_subsection1}. Notably, the relative ranking of the placement strategies remains essentially unchanged, indicating that the main conclusions are robust to alternative feature selections.

\begin{table*}[ht]
\centering
\caption[Prediction performance for additional feature-based placement strategies]{MAE and RMSE prediction results for citywide traffic under permanent sensor deployment with additional feature-based placement strategies for Berlin and Manhattan.}
\label{Table:results_1_permanent_appendix}
\scriptsize
\begin{subtable}{\textwidth}
\centering
\caption{MAE}
\begin{tabular}{ l|
    rrrrrr|rrrrrr}
    \toprule
    City & \multicolumn{6}{c|}{Berlin} & \multicolumn{6}{c}{Manhattan}\\
    Sensor budget (K) &10 & 25 & 34$^{*}$ & 50 & 75 & 100  & 8$^{*}$ & 10 & 25 & 50 & 75 & 100  \\
    \midrule
Feature div. (connectivity) &18.1 & 12.5 & 12.9 & 12.0 & 11.9 & 12.3 & 66.9 & 64.0 & 51.6 & 43.6 & 42.9 & 43.8 \\
Feature div. (infrastructure) &17.4 & 14.4 & 13.2 & 13.9 & 13.7 & 13.2 & 66.4 & 53.4 & 44.7 & 52.7 & 51.2 & 55.3 \\
Feature div. (points of interest) &21.0 & 12.2 & 15.9 & 16.8 & 15.2 & 13.6 & 56.2 & 48.2 & 48.3 & 54.8 & 44.4 & 52.6 \\
Redundancy (connectivity) &14.0 & 12.7 & 13.2 & 12.0 & 12.4 & 12.4 & 43.2 & 43.2 & 42.4 & 40.2 & 40.5 & 41.7 \\
Redundancy (infrastructure) &32.8 & 18.5 & 19.2 & 16.3 & 18.7 & 15.2 & 45.7 & 46.0 & 49.4 & 46.8 & 44.7 & 43.1 \\
Redundancy (points of interest) &14.7 & 14.7 & 17.1 & 14.7 & 12.7 & 12.3 & 43.7 & 43.1 & 42.9 & 37.9 & 38.9 & 39.2 \\
Coverage (connectivity) &15.5 & 12.2 & 13.0 & 13.5 & 14.9 & 12.7 & 117.7 & 113.5 & 66.7 & 50.7 & 46.7 & 42.1 \\
Coverage (infrastructure) &19.4 & 12.7 & 14.2 & 15.3 & 13.6 & 13.2 & 80.9 & 51.1 & 43.5 & 49.0 & 51.0 & 52.7 \\
Coverage (points of interest) &18.6 & 15.4 & 14.7 & 13.8 & 13.8 & 19.3 & 53.8 & 54.6 & 43.9 & 53.5 & 50.8 & 54.0 \\
    \bottomrule
\end{tabular}
\end{subtable}

\vspace{1em} 

\begin{subtable}{\textwidth}
\centering
\caption{RMSE}
\begin{tabular}{ l|
    rrrrrr|rrrrrr}
    \toprule
    City & \multicolumn{6}{c|}{Berlin} & \multicolumn{6}{c}{Manhattan}\\
    Sensor budget (K) &10 & 25 & 34$^{*}$ & 50 & 75 & 100  & 8$^{*}$ & 10 & 25 & 50 & 75 & 100  \\
    \midrule
Feature div. (connectivity) &31.8 & 30.0 & 29.2 & 28.6 & 29.1 & 28.7 & 125.9 & 123.6 & 109.9 & 106.2 & 105.4 & 103.2 \\
Feature div. (infrastructure) &30.6 & 28.5 & 28.3 & 28.5 & 28.4 & 27.9 & 115.3 & 109.0 & 97.4 & 101.5 & 97.1 & 108.0 \\
Feature div. (points of interest) &32.4 & 28.2 & 29.9 & 29.3 & 28.8 & 28.5 & 108.6 & 105.8 & 95.5 & 101.0 & 97.3 & 102.1 \\
Redundancy (connectivity) &30.3 & 29.5 & 29.8 & 29.0 & 28.3 & 27.8 & 113.6 & 114.1 & 106.1 & 104.7 & 105.1 & 104.9 \\
Redundancy (infrastructure) &52.1 & 34.2 & 35.0 & 30.4 & 34.3 & 29.8 & 109.4 & 110.8 & 108.0 & 106.2 & 102.5 & 100.6 \\
Redundancy (points of interest) &31.8 & 28.9 & 30.6 & 28.7 & 27.9 & 27.2 & 111.5 & 112.0 & 91.1 & 90.0 & 94.3 & 95.9 \\
Coverage (connectivity) &30.1 & 29.3 & 29.3 & 29.0 & 29.5 & 28.3 & 168.7 & 158.2 & 111.2 & 109.6 & 108.7 & 106.3 \\
Coverage (infrastructure) &31.9 & 28.1 & 29.0 & 28.9 & 28.4 & 27.8 & 126.0 & 103.0 & 102.4 & 102.2 & 101.9 & 102.2 \\
Coverage (points of interest) &30.8 & 29.0 & 29.2 & 28.8 & 27.8 & 33.8 & 105.0 & 104.8 & 95.2 & 99.6 & 102.9 & 102.4 \\

\bottomrule
    \bottomrule
\end{tabular}
\end{subtable}

\vspace{1em} 

{\raggedright \footnotesize $^{*}$ For both cities, the tables include a column corresponding to the number of existing sensors: 34 in Berlin and 8 in Manhattan.\par}
\end{table*}

\section{Debiasing Data}\label{apx:debiasing_data}

Because the available citywide traffic data sources (Strava bicycle counts and taxi counts) represent biased subsamples of total traffic, this appendix examines whether an explicit debiasing step would be appropriate in the present setting. In this context, debiasing refers to estimating true cycling or motorized traffic volumes, as measured by physical traffic sensors, from the biased Strava and taxi data, in combination with additional covariates.

Such an approach would involve training a predictive model on locations equipped with physical sensors and applying it to infer corrected traffic volumes on unobserved street segments. Before such a correction can be meaningfully applied, its feasibility must be assessed. This is done following an established practice for debiasing sparse traffic sensor data \citep{kaiser_counting_2025}.
Rather than splitting the already sparse ground-truth data into separate training and test sets, a \acrlong{logo} (\acrshort{logo}) cross-validation scheme is implemented, treating each sensor location as one fold: each location is held out once as the test set while the model is trained on the remaining sensors. All auxiliary features and the biased traffic counts (Strava and taxi data, respectively) are included as predictors. An XGBoost regressor is used as the predictive model.

\begin{figure}[ht!]
    \centering
    \begin{subfigure}{0.49\linewidth}
        \includegraphics[width=\linewidth]{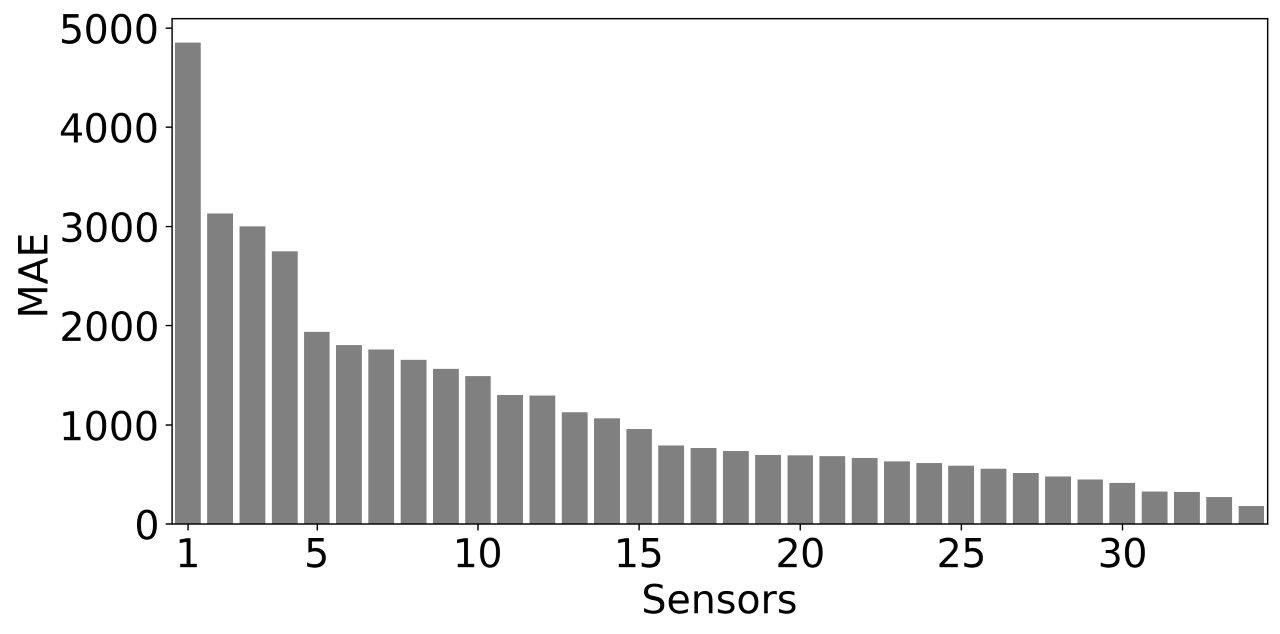}
        \caption{Berlin}
        \label{fig:debiasing_data_mae_berlin}
    \end{subfigure}
    \hfill
    \begin{subfigure}{0.49\linewidth}
        \includegraphics[width=\linewidth]{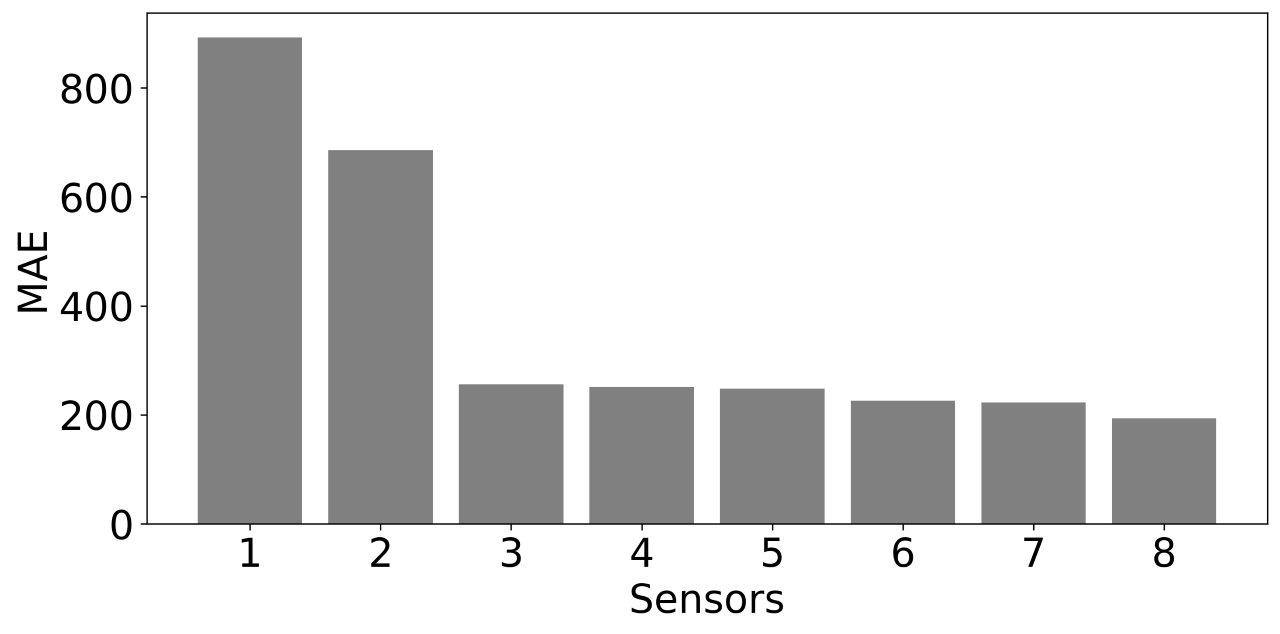}
        \caption{Manhattan}
        \label{fig:debiasing_data_mae_ny}
    \end{subfigure}
    \caption[Per-sensor \acrshort{mae} under \acrshort{logo} cross-validation]{Distribution of per-fold \acrshort{mae} values across sensor locations under \acrshort{logo} cross-validation.}
    \label{fig:debiasing_data}
\end{figure}

\begin{figure}[ht!]
    \centering
    \begin{subfigure}{0.49\linewidth}
        \includegraphics[width=\linewidth]{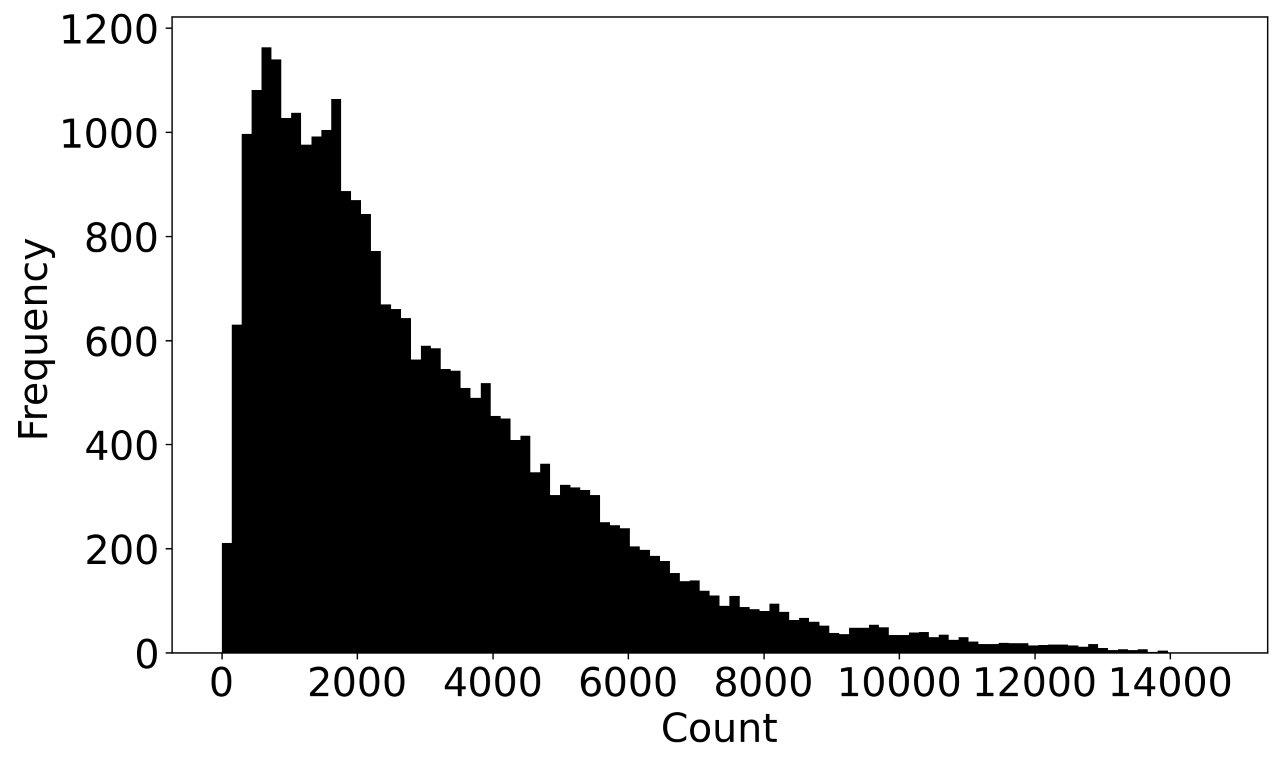}
        \caption{Berlin}
        \label{fig:debiasing_data_measruements_berlin}
    \end{subfigure}
    \hfill
        \begin{subfigure}{0.49\linewidth}
        \includegraphics[width=\linewidth]{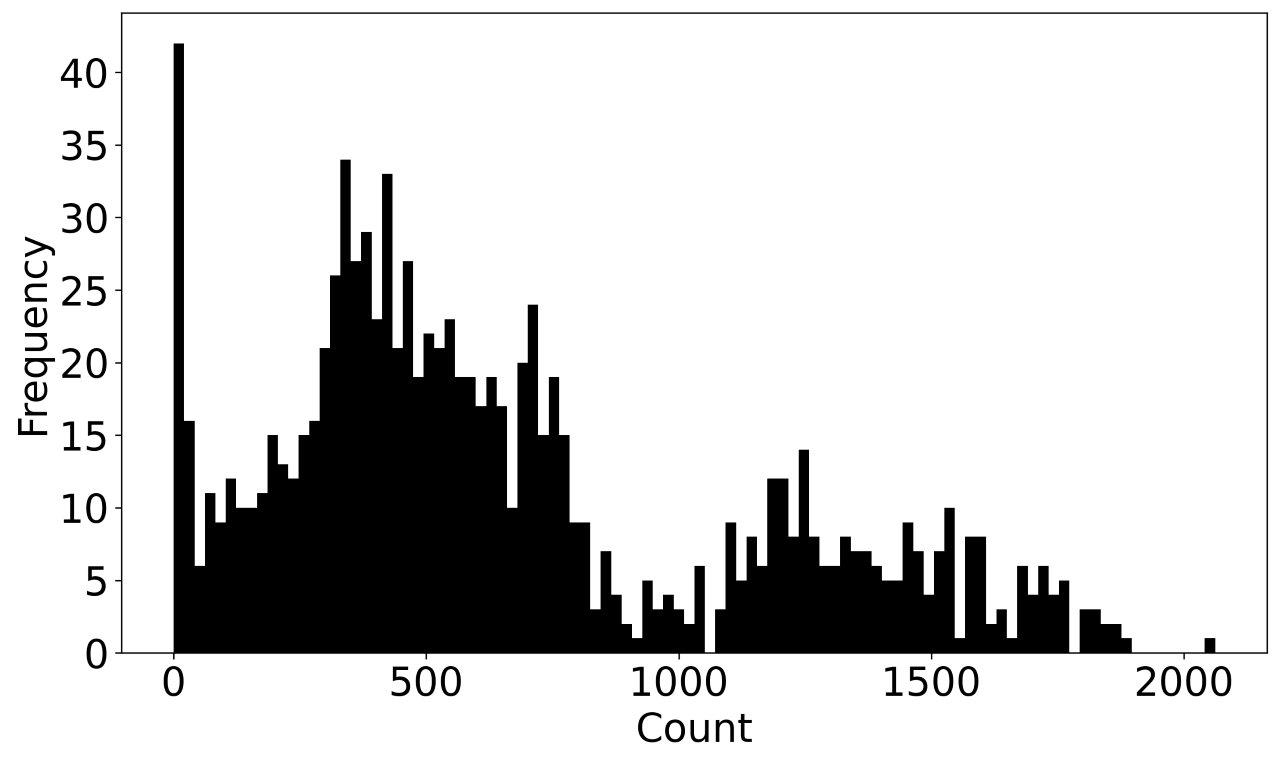}
        \caption{Manhattan}
        \label{fig:debiasing_data_measruements_ny}
    \end{subfigure}
    \caption[Distributions of ground-truth sensor measurements for both cities]{Distributions of ground-truth sensor measurements for both cities. Berlin measurements are recorded at the daily level, whereas Manhattan measurements are recorded at the hourly level.}
    \label{fig:debiasing_data_grountruth}
\end{figure}

Table~\ref{tab:debiasing_results} reports the mean test error across folds, while Figure~\ref{fig:debiasing_data} illustrates the distribution of per-fold \acrshort{mae} values across sensor locations for both cities. Figure \ref{fig:debiasing_data_grountruth} further illustrates the distributions of true sensor measurements.

\begin{table}[!ht]
\centering
\caption[Performance of debiasing model]{Performance of debiasing model under \acrshort{logo} cross-validation.}
\label{tab:debiasing_results}
\begin{tabular}{lccc }
\toprule
\textbf{Dataset} & \textbf{\acrshort{mae}}  & \textbf{\acrshort{rmse}} & \textbf{\acrshort{mape}}$^{*}$ \\
\midrule
Berlin (Strava) & 1177.34 & 1363.47 & 75.63\\
Manhattan (Taxi) & 371.97 & 412.03 & 225.92\\
\bottomrule

\end{tabular}\\
\vspace{0.2cm}
{\raggedright \footnotesize
$^{*}$ \acrlong{mape} (\acrshort{mape}) is computed as 
$\text{MAPE} = \frac{1}{n} \sum_{i=1}^{n} \left| \frac{y_i - \hat{y}_i}{y_i} \right| \times 100$. 
Since MAPE is undefined for zero ground-truth values, it is calculated only on observations with non-zero ground truth.\par}
\end{table}

The results indicate limited predictive power, given the large errors relative to the observed traffic volumes. For example, the \acrlong{mape} (\acrshort{mape}), computed only on non-zero ground-truth observations where it is well-defined, amounts to 75.63\% for Berlin and 225.92\% for Manhattan. This finding is corroborated by \acrshort{mae}, which reaches 1177.34 for Berlin and 371.97 for Manhattan. When put into context with the empirical distributions of the ground-truth sensor measurements (Figures~\ref{fig:debiasing_data_measruements_berlin} and~\ref{fig:debiasing_data_measruements_ny}), these absolute errors are substantial. Moreover, the \acrshort{logo} \acrshort{mae} distributions (Figures~\ref{fig:debiasing_data_mae_berlin} and~\ref{fig:debiasing_data_mae_ny}) reveal firm heterogeneity across sensor locations, indicating that predictive performance varies widely between different sites. This implies that while a small subset of locations may be reasonably well predicted, others exhibit very poor performance, making it impossible to identify where debiasing is reliable.

Debiasing would therefore not provide sufficiently reliable data for citywide analysis. The high average error levels, combined with the pronounced spatial variability in prediction quality, introduce substantial uncertainty into the reconstructed traffic volumes. In addition, the debiased estimates would be anchored to the small, potentially unrepresentative set of existing sensors, thereby limiting the model’s ability to generalize to unsensed locations. Since the objective of this work is to evaluate strategies for optimal sensor placement, such dependence on the existing sensor network would introduce an unquantifiable and potentially critical bias. For these reasons, the analysis refrains from using debiased data in the analysis.

\section{Permanent sensors - extending existing sensor deployment}\label{apx:extending_existing_sensor_network}
\begin{center}
\begin{table*}[h]
\centering
\caption[Prediction performance under different placement strategies accounting for existing sensors]{Prediction performance of citywide traffic interpolation under permanent deployment when extending existing sensor deployments in Berlin and Manhattan. For each column, the best value is shown in bold, followed by the second–, third–, and fourth–best values shaded from dark to light gray.}
\label{Table:results_1_withexisting}
\definecolor{rankSecond}{gray}{0.70}
\definecolor{rankThird}{gray}{0.82}
\definecolor{rankFourth}{gray}{0.92}
\scriptsize
\begin{subtable}{\textwidth}
\caption{MAE}
    \centering
\begin{tabular}{ l|
    rrrrrr|rrrrrr}

\toprule
    City & \multicolumn{6}{c|}{Berlin} & \multicolumn{6}{c}{Manhattan}\\
      &10 & 25 & 34$^{*}$ & 50 & 75 & 100  & 8$^{*}$ & 10 & 25 & 50 & 75 & 100  \\
\midrule
Betweenness &-- & -- & \textbf{21.7} & 21.7 & 25.5 & 22.5 & \textbf{44.0} & 66.8 & 52.9 & 53.7 & 55.0 & 47.0 \\
Closeness &-- & -- & \textbf{21.7} & 20.4 & 21.1 & 19.9 & \textbf{44.0} & 66.8 & 47.4 & 59.9 & 55.7 & 50.7 \\
\lightmidrule
Feature div. (all) &-- & -- & \textbf{21.7} & 36.8 & 37.2 & 26.5 & \textbf{44.0} & 78.8 & 100.0 & 95.9 & 105.4 & 103.8 \\
Feature div. (infr. sel.) &-- & -- & \textbf{21.7} & 18.8 & \cellcolor{rankFourth}{15.1} & 15.2 & \textbf{44.0} & 63.1 & 48.4 & 44.5 & \cellcolor{rankThird}{42.0} & 42.3 \\
Redundancy (all) &-- & -- & \textbf{21.7} & 19.8 & 15.2 & 14.4 & \textbf{44.0} & 43.3 & 48.7 & \cellcolor{rankFourth}{43.9} & 50.7 & 46.0 \\
Redundancy (infr. sel.) &-- & -- & \textbf{21.7} & 20.4 & 18.6 & 16.5 & \textbf{44.0} & 60.4 & 46.9 & 45.0 & \cellcolor{rankFourth}{42.2} & \cellcolor{rankThird}{41.3} \\
Coverage (all) &-- & -- & \textbf{21.7} & 35.2 & 37.7 & 30.8 & \textbf{44.0} & 65.0 & 112.9 & 89.2 & 87.6 & 75.0 \\
Coverage (infr. sel.) &-- & -- & \textbf{21.7} & 20.8 & 16.0 & 15.2 & \textbf{44.0} & 63.1 & \cellcolor{rankThird}{42.5} & 46.0 & 43.0 & 42.1 \\
\lightmidrule
Voronoi &-- & -- & \textbf{21.7} & \cellcolor{rankFourth}{18.1} & 15.7 & 15.7 & \textbf{44.0} & \cellcolor{rankFourth}{43.2} & \cellcolor{rankSecond}{41.5} & \cellcolor{rankSecond}{40.7} & \cellcolor{rankSecond}{39.7} & \cellcolor{rankSecond}{39.8} \\
Spatial dispersion &-- & -- & \textbf{21.7} & \cellcolor{rankSecond}{17.0} & \cellcolor{rankSecond}{14.5} & \cellcolor{rankSecond}{13.2} & \textbf{44.0} & \cellcolor{rankSecond}{42.4} & 47.4 & 52.0 & 51.8 & 44.1 \\
\lightmidrule
Active learning &-- & -- & \textbf{21.7} & 18.9 & \cellcolor{rankThird}{14.8} & \cellcolor{rankFourth}{14.1} & \textbf{44.0} & \cellcolor{rankThird}{43.0} & 56.2 & \cellcolor{rankThird}{43.3} & 43.6 & 43.3 \\
\midrule
Random (median) &-- & -- & \textbf{21.7} & \cellcolor{rankThird}{17.7} & 15.3 & \cellcolor{rankThird}{14.1} & \textbf{44.0} & 44.7 & \cellcolor{rankFourth}{46.3} & 44.4 & 42.8 & \cellcolor{rankFourth}{41.8} \\
Random (min) &-- & -- & \textbf{21.7} & \textbf{13.9} & \textbf{12.1} & \textbf{11.9} & \textbf{44.0} & \textbf{41.4} & \textbf{39.7} & \textbf{36.7} & \textbf{36.3} & \textbf{35.5} \\
Random (max) &-- & -- & \textbf{21.7} & 30.0 & 22.5 & 18.6 & \textbf{44.0} & 235.4 & 97.7 & 79.0 & 60.6 & 54.7 \\

    \bottomrule
\end{tabular}
\end{subtable}

\vspace{1em}

\begin{subtable}{\textwidth}
\caption{RMSE}
    \centering
\begin{tabular}{ l|
    rrrrrr|rrrrrr}
\toprule
    City & \multicolumn{6}{c|}{Berlin} & \multicolumn{6}{c}{Manhattan}\\
    Sensor budget (K) &10 & 25 & 34$^{*}$ & 50 & 75 & 100  & 8$^{*}$ & 10 & 25 & 50 & 75 & 100  \\
\midrule
Betweenness &-- & -- & \textbf{34.7} & 33.5 & 36.3 & 33.7 & \textbf{109.3} & 118.8 & 107.6 & 109.7 & 111.4 & 104.6 \\
Closeness &-- & -- & \textbf{34.7} & 32.5 & 32.1 & 31.3 & \textbf{109.3} & 118.8 & 109.9 & 107.8 & 113.7 & 104.7 \\
\lightmidrule
Feature div. (all) &-- & -- & \textbf{34.7} & 54.9 & 52.8 & 39.5 & \textbf{109.3} & 135.3 & 144.4 & 162.9 & 176.8 & 170.5 \\
Feature div. (infr. sel.) &-- & -- & \textbf{34.7} & 33.1 & 30.0 & 29.4 & \textbf{109.3} & 112.3 & 113.8 & 108.5 & 105.9 & 102.1 \\
Redundancy (all) &-- & -- & \textbf{34.7} & 32.6 & \cellcolor{rankThird}{28.8} & 28.7 & \textbf{109.3} & 110.7 & \cellcolor{rankSecond}{97.8} & \cellcolor{rankThird}{95.0} & \cellcolor{rankFourth}{101.3} & 97.1 \\
Redundancy (infr. sel.) &-- & -- & \textbf{34.7} & 34.7 & 33.1 & 30.5 & \textbf{109.3} & 115.5 & 111.7 & 106.7 & 106.6 & 103.8 \\
Coverage (all) &-- & -- & \textbf{34.7} & 51.6 & 51.1 & 46.0 & \textbf{109.3} & 120.5 & 171.6 & 166.0 & 162.6 & 137.1 \\
Coverage (infr. sel.) &-- & -- & \textbf{34.7} & 35.0 & 30.1 & 29.5 & \textbf{109.3} & 112.3 & 110.4 & 108.3 & 105.4 & 107.0 \\
\lightmidrule
Voronoi &-- & -- & \textbf{34.7} & 31.9 & 29.6 & 30.4 & \textbf{109.3} & 110.3 & 105.5 & 106.5 & 101.5 & 93.9 \\
Spatial dispersion &-- & -- & \textbf{34.7} & \cellcolor{rankSecond}{30.1} & \cellcolor{rankFourth}{29.0} & \cellcolor{rankSecond}{27.4} & \textbf{109.3} & \cellcolor{rankThird}{109.7} & \cellcolor{rankThird}{104.6} & 100.7 & 102.2 & \cellcolor{rankThird}{92.1} \\
\lightmidrule
Active learning &-- & -- & \textbf{34.7} & \cellcolor{rankFourth}{31.7} & \cellcolor{rankSecond}{28.2} & \cellcolor{rankThird}{28.2} & \textbf{109.3} & \cellcolor{rankSecond}{107.9} & 109.4 & \cellcolor{rankSecond}{93.5} & \textbf{78.7} & \textbf{78.2} \\
\midrule
Random (median) &-- & -- & \textbf{34.7} & \cellcolor{rankThird}{31.2} & 29.4 & \cellcolor{rankFourth}{28.5} & \textbf{109.3} & \cellcolor{rankFourth}{110.1} & \cellcolor{rankThird}{104.6} & \cellcolor{rankFourth}{99.0} & \cellcolor{rankThird}{95.5} & \cellcolor{rankFourth}{93.4} \\
Random (min) &-- & -- & \textbf{34.7} & \textbf{28.3} & \textbf{26.2} & \textbf{25.1} & \textbf{109.3} & \textbf{100.1} & \textbf{90.8} & \textbf{84.0} & \cellcolor{rankSecond}{80.2} & \cellcolor{rankSecond}{79.4} \\
Random (max) &-- & -- & \textbf{34.7} & 49.0 & 42.9 & 36.3 & \textbf{109.3} & 359.5 & 158.1 & 145.3 & 119.4 & 111.4 \\
\bottomrule
\end{tabular}
\end{subtable}
{\raggedright \footnotesize $^{*}$ For both cities, the tables include a column corresponding to the number of  existing sensors: 34 in Berlin and 8 in Manhattan.\par}
\end{table*}
\end{center}

This appendix reports results analogous to those in Section~\ref{sec:results_subsection1}, with the key distinction that here the placement strategies are applied to extend the existing sensor deployments. While the main text evaluates sensor placement from scratch, Table~\ref{Table:results_1_withexisting} presents prediction performance when additional sensors are added to the existing sensors using the respective placement strategies. The existing sensors are included in $S_{\text{selected}}$ at initialization and are fully incorporated into the computation of all placement strategies and baselines. For example, under feature-diversity-based placement, new sensor locations are selected to maximize diversity jointly with the existing sensors. The all-training-data baseline is not included, as it is identical to the corresponding benchmark in the main paper.

This setup constrains the feasible sensor budgets. In Berlin, where 34 sensors are already deployed, configurations with fewer than 34 sensors cannot be evaluated; likewise, in Manhattan, the minimum feasible budget is 8 sensors. Accordingly, the columns corresponding to 34 sensors in Berlin and 8 sensors in Manhattan are identical across all placement strategies and represent the prediction error of the existing deployments.

Across both cities, extending existing deployments yields consistent performance improvements as additional sensors are added, though the gains are generally smaller than in the from-scratch scenarios. The relative performance ranking of the placement strategies remains essentially unchanged, indicating that the main conclusions regarding the effectiveness of the strategies are robust in realistic network extension settings.

\section{Permanent Sensor Placement - RMSE \label{appx:permanent_rmse}}

This appendix reports results analogous to those presented in Section~\ref{sec:results_subsection1}, using \acrshort{rmse} instead of (\acrshort{mae}) as the evaluation metric. Overall, the results mirror the main findings: spatial dispersion, Voronoi area inequality, and active learning consistently outperform network-centrality and feature-based placement strategies across sensor budgets and cities.

\begin{table*}[ht]
\centering
\caption[Prediction results under different placement strategies - RMSE]{RMSE prediction results of city-wide traffic using spatial dispersion as the placement strategy for permanent deployment under different sensor budgets in Berlin and Manhattan. For each column, the best value is shown in bold, followed by the second–, third–, and fourth–best values shaded from dark to light gray.}
\label{Table:results_1_permanent_rmse}
\definecolor{rankSecond}{gray}{0.70}
\definecolor{rankThird}{gray}{0.82}
\definecolor{rankFourth}{gray}{0.92}
\scriptsize
\begin{subtable}{\textwidth}
\centering
\begin{tabular}{ l|
    rrrrrrr|rrrrrrr}
    \toprule
    City & \multicolumn{7}{c|}{Berlin} & \multicolumn{7}{c}{Manhattan}\\
    Sensor budget (K) &10 & 25 & 34$^{*}$ & 50 & 75 & 100 & all & 8$^{*}$ & 10 & 25 & 50 & 75 & 100 & all \\
    \midrule
Betweenness &38.4 & 30.9 & 41.5 & 45.5 & 46.0 & 61.2 & -- & 165.2 & 161.7 & 128.2 & 111.7 & 108.0 & 97.4 & -- \\
Closeness &34.3 & 43.4 & 51.2 & 34.9 & 36.8 & 39.1 & -- & 137.6 & 141.0 & 139.0 & 108.6 & 102.8 & 107.8 & -- \\
\lightmidrule
Feature div. (all) &52.6 & 91.0 & 48.8 & 51.9 & 44.2 & 42.5 & -- & 205.7 & 220.4 & 164.4 & 162.7 & 174.1 & 171.8 & -- \\
Feature div. (infr. sel.) &42.8 & 29.8 & 29.6 & 29.2 & 28.7 & 31.0 & -- & 113.4 & 113.1 & 108.3 & 107.0 & 104.5 & 108.2 & -- \\
Redundancy (all) &\cellcolor{rankFourth}{30.5} & 30.1 & 29.6 & 28.9 & \cellcolor{rankSecond}{27.3} & \cellcolor{rankSecond}{27.2} & -- & 143.8 & 131.0 & 105.4 & 100.1 & \cellcolor{rankThird}{94.7} & \cellcolor{rankFourth}{93.2} & -- \\
Redundancy (infr. sel.) &30.8 & 38.1 & 32.9 & 30.6 & 29.2 & 29.2 & -- & \cellcolor{rankSecond}{104.5} & 115.4 & 103.0 & 105.9 & 105.5 & 106.3 & -- \\
Coverage (all) &52.7 & 86.0 & 54.1 & 54.3 & 47.9 & 40.8 & -- & 288.8 & 174.8 & 204.4 & 162.6 & 141.1 & 120.8 & -- \\
Coverage (infr. sel.) &35.9 & 30.2 & 29.5 & 28.7 & 28.6 & 28.4 & -- & 112.5 & 114.0 & \cellcolor{rankFourth}{102.3} & 101.5 & 102.1 & 99.1 & -- \\
\lightmidrule
Voronoi &\cellcolor{rankSecond}{29.3} & \cellcolor{rankSecond}{28.6} & \cellcolor{rankSecond}{29.0} & \cellcolor{rankSecond}{28.1} & \cellcolor{rankThird}{27.7} & 30.5 & -- & 120.7 & 115.1 & \cellcolor{rankThird}{100.6} & \cellcolor{rankFourth}{98.5} & 97.2 & 98.2 & -- \\
Spatial dispersion &\cellcolor{rankThird}{29.9} & 29.6 & 29.5 & 28.7 & 29.0 & 29.1 & -- & \cellcolor{rankThird}{104.9} & \cellcolor{rankThird}{105.7} & 103.4 & 103.0 & 99.9 & 97.1 & -- \\
\lightmidrule
Active learning &30.7 & \cellcolor{rankThird}{29.3} & \cellcolor{rankFourth}{29.3} & \cellcolor{rankFourth}{28.7} & 28.9 & \cellcolor{rankFourth}{28.3} & -- & 111.8 & \cellcolor{rankSecond}{104.1} & \cellcolor{rankSecond}{98.6} & \cellcolor{rankSecond}{91.4} & \cellcolor{rankSecond}{92.1} & \cellcolor{rankSecond}{86.3} & -- \\
\midrule
Random (median) &30.5 & \cellcolor{rankFourth}{29.4} & \cellcolor{rankThird}{29.1} & \cellcolor{rankThird}{28.6} & \cellcolor{rankFourth}{28.2} & \cellcolor{rankThird}{27.9} & -- & 109.8 & \cellcolor{rankFourth}{108.7} & 103.1 & \cellcolor{rankThird}{98.0} & \cellcolor{rankFourth}{94.9} & \cellcolor{rankThird}{92.6} & -- \\
Random (min) &\textbf{27.7} & \textbf{27.3} & \textbf{26.2} & \textbf{26.0} & \textbf{25.9} & \textbf{25.4} & -- & \textbf{96.0} & \textbf{92.9} & \textbf{88.0} & \textbf{82.1} & \textbf{80.6} & \textbf{81.6} & -- \\
Random (max) &72.2 & 61.0 & 43.9 & 38.5 & 37.4 & 45.3 & -- & 280.3 & 360.4 & 204.6 & 149.0 & 131.1 & 122.1 & -- \\
\midrule
All training data &-- & -- & -- & -- & -- & -- & \textbf{23.4} & -- & -- & -- & -- & -- & -- & \textbf{77.8} \\
Existing &-- & -- & 34.7 & -- & -- & -- & -- & \cellcolor{rankFourth}{109.3} & -- & -- & -- & -- & -- & -- \\
    \bottomrule
\end{tabular}
\end{subtable}

\vspace{1em} 

{\raggedright \footnotesize $^{*}$ For both cities, the table includes a column corresponding to the number of existing sensors: 34 in Berlin and 8 in Manhattan.\par}
\end{table*}

\section{Temporary sensors - further placement strategies \label{apx:temporary_furtherstrategies}}

\begin{figure}[ht!]
    \centering
    \begin{subfigure}{0.49\linewidth}
        \includegraphics[width=\linewidth]{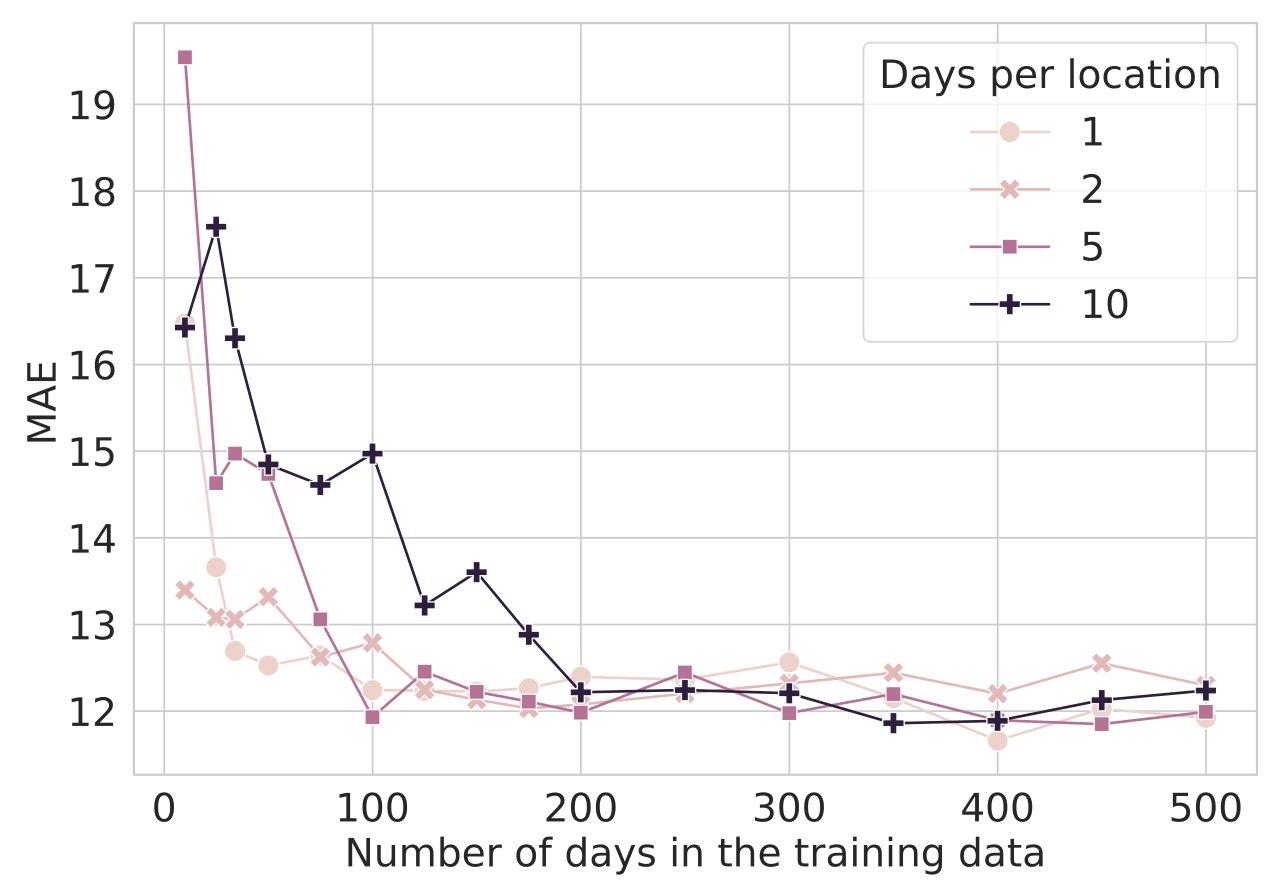}
        \caption{Berlin — Days per location}
        \label{fig:location_berlin_mae_evaluation_voronoi}
    \end{subfigure}
    \hfill
            \begin{subfigure}{0.49\linewidth}
        \includegraphics[width=\linewidth]{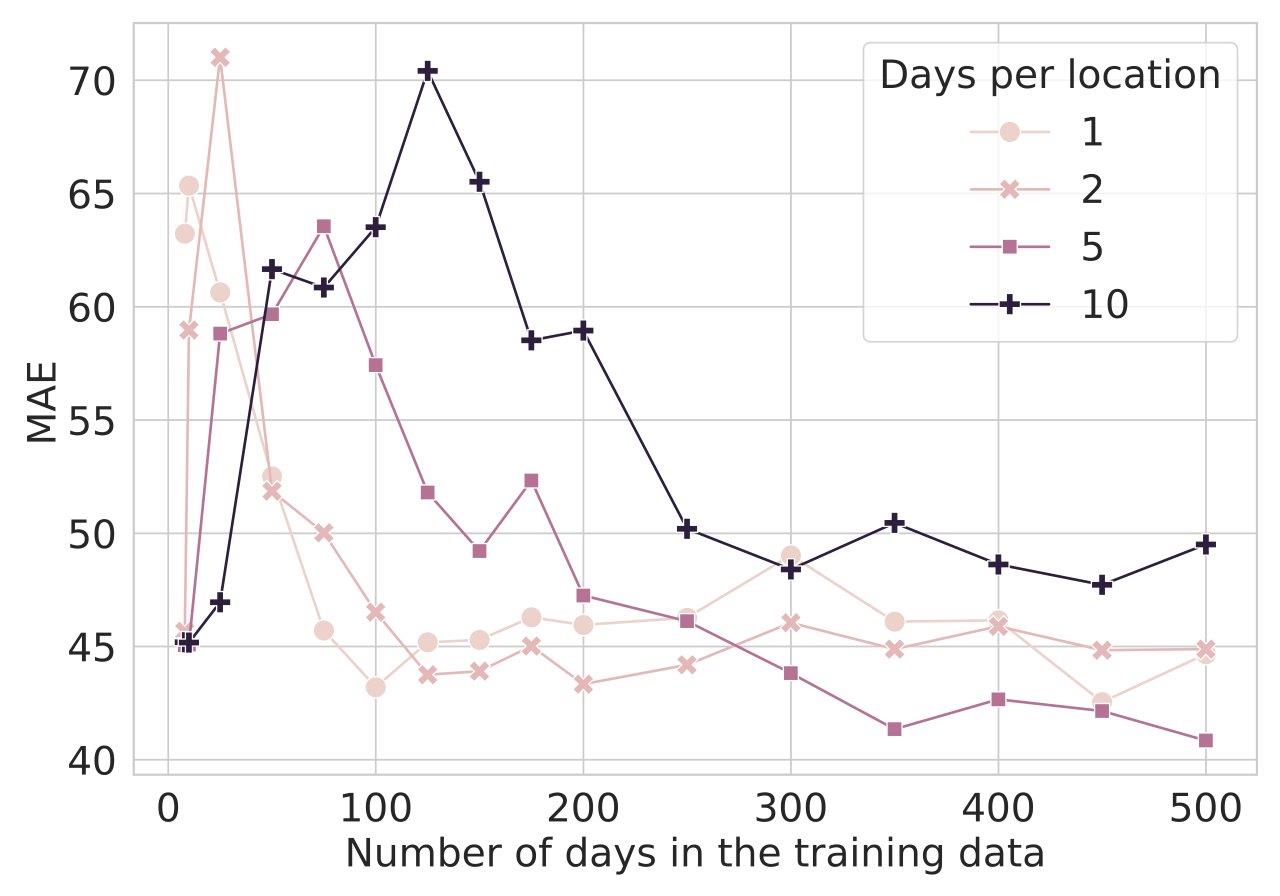}
        \caption{Manhattan — Days per location}
        \label{fig:newyork_location_mae_evaluation_voronoi}
    \end{subfigure}

    \vspace{0.5cm}
    
        \begin{subfigure}{0.49\linewidth}
        \includegraphics[width=\linewidth]{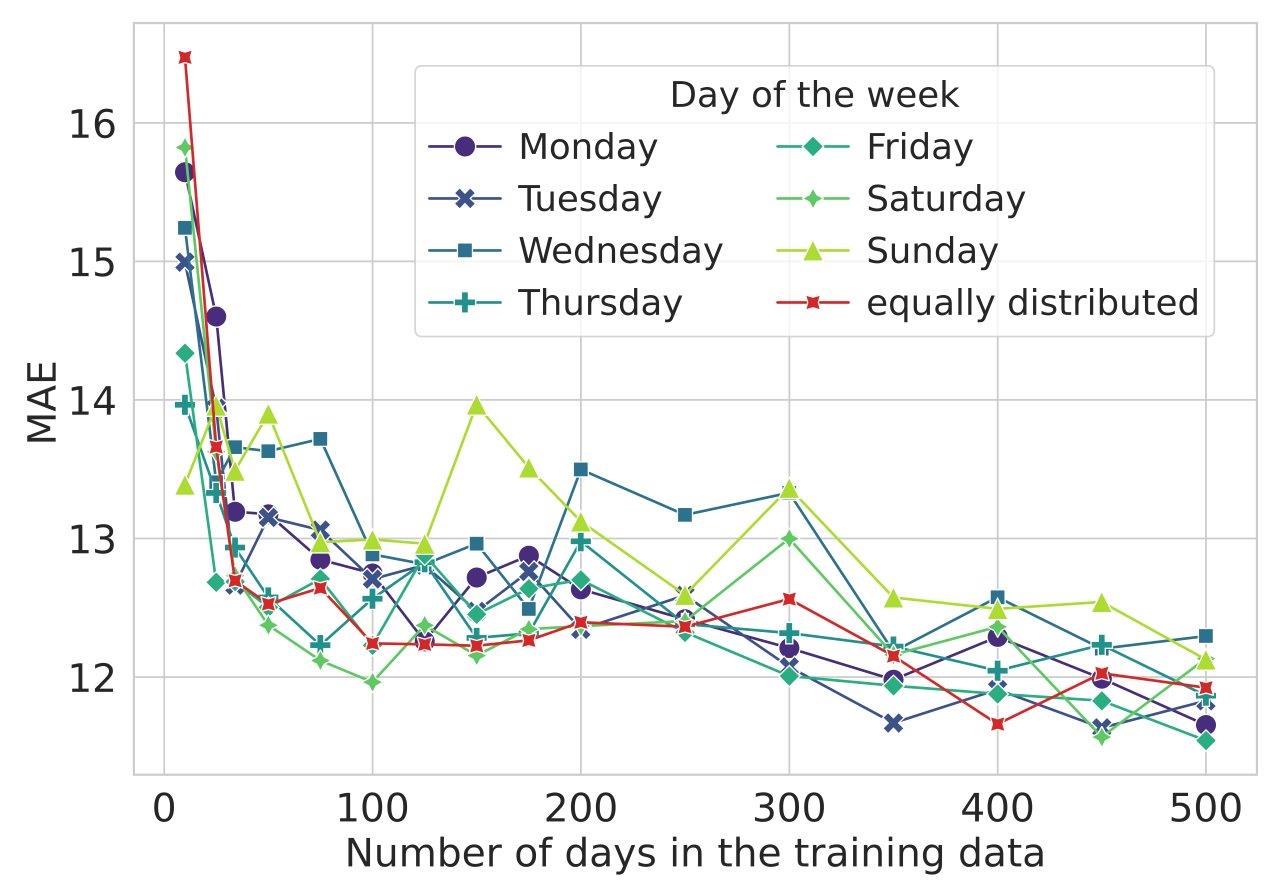}
        \caption{Berlin — Day-of-week effects}
        \label{fig:berlin_week_mae_evaluation_voronoi}
    \end{subfigure}
    \hfill
    \begin{subfigure}{0.49\linewidth}
        \includegraphics[width=\linewidth]{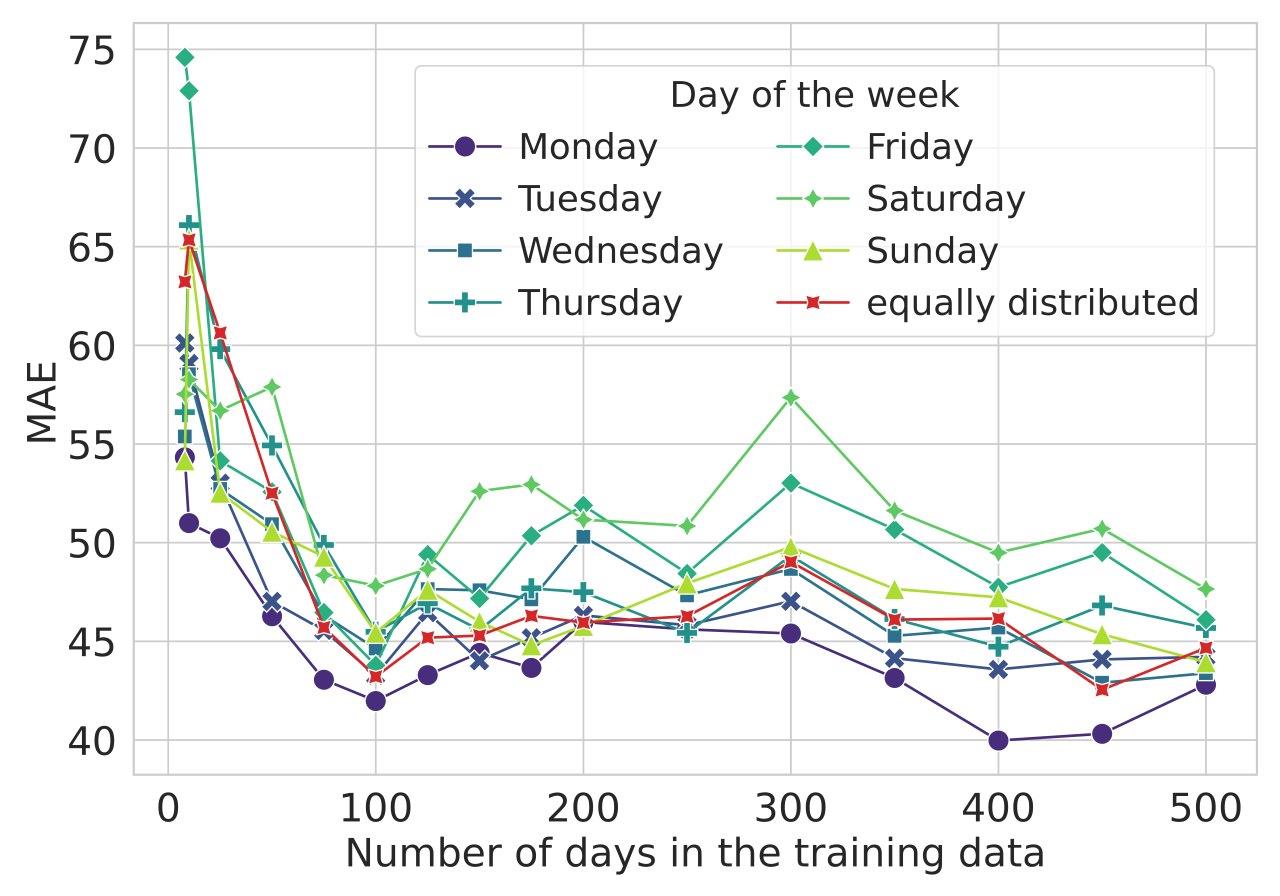}
        \caption{Manhattan — Day-of-week effects}
        \label{fig:newyork_week_mae_evaluation_voronoi}
    \end{subfigure}
    \caption[Temporary placement considerations of temporary sensors (MAE) - Voronoi area inequality]{Temporary placement considerations of temporary sensors using Voronoi area inequality as spatial placement strategy: The upper row compares deployment strategies that vary the number of days allocated to each location, effectively trading off between broader spatial coverage (more locations, fewer days each) and repeated measurement at fewer locations, while keeping the total number of observations constant. The lower row examines whether restricting temporary measurements to specific weekdays affects prediction accuracy. Results are shown for Berlin (left) and Manhattan (right). The reported error metric is \acrshort{mae}.}
    \label{fig:results3_mae_voronoi}
\end{figure}

\begin{figure}[ht!]
    \centering
    \begin{subfigure}{0.49\linewidth}
        \includegraphics[width=\linewidth]{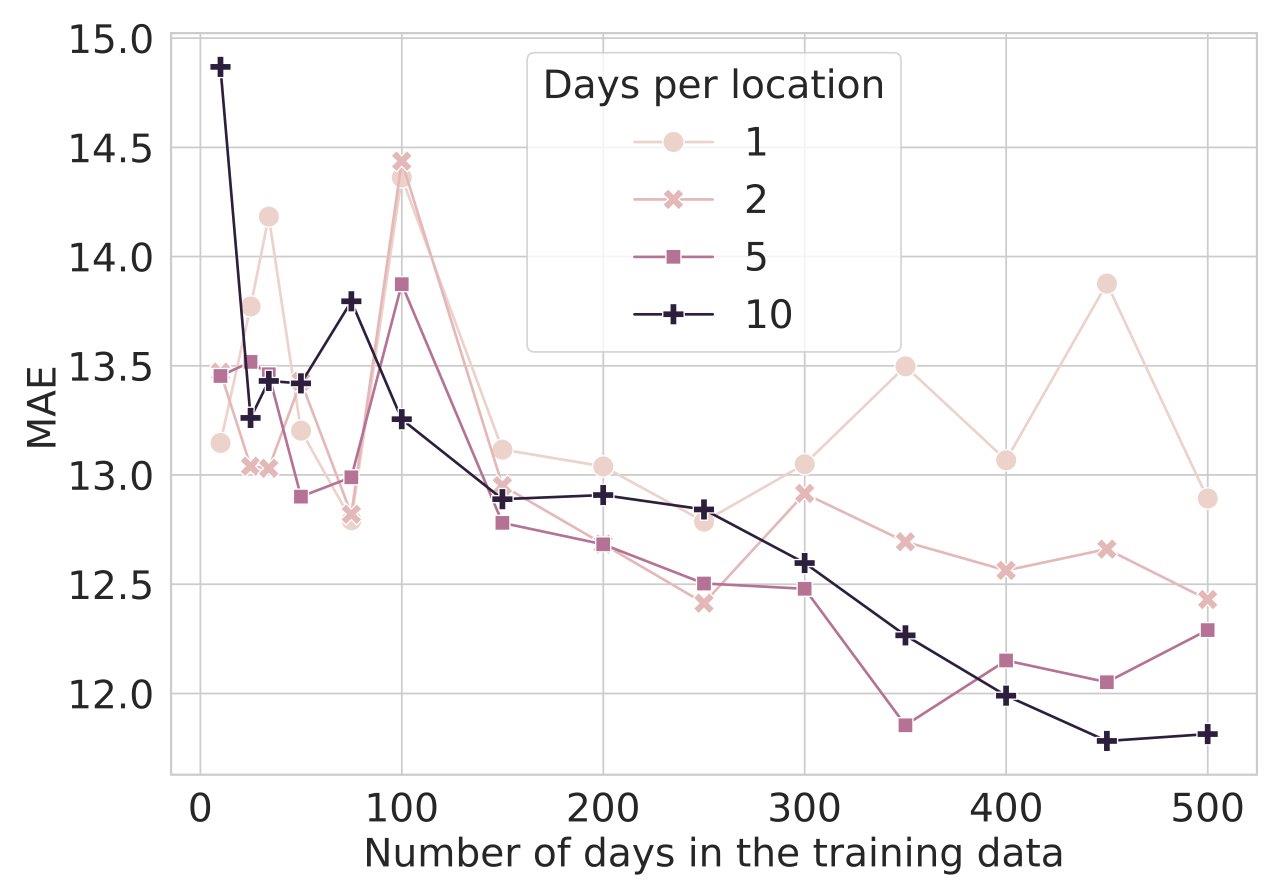}
        \caption{Berlin — Days per location}
        \label{fig:location_berlin_mae_evaluation_active}
    \end{subfigure}
    \hfill
            \begin{subfigure}{0.49\linewidth}
        \includegraphics[width=\linewidth]{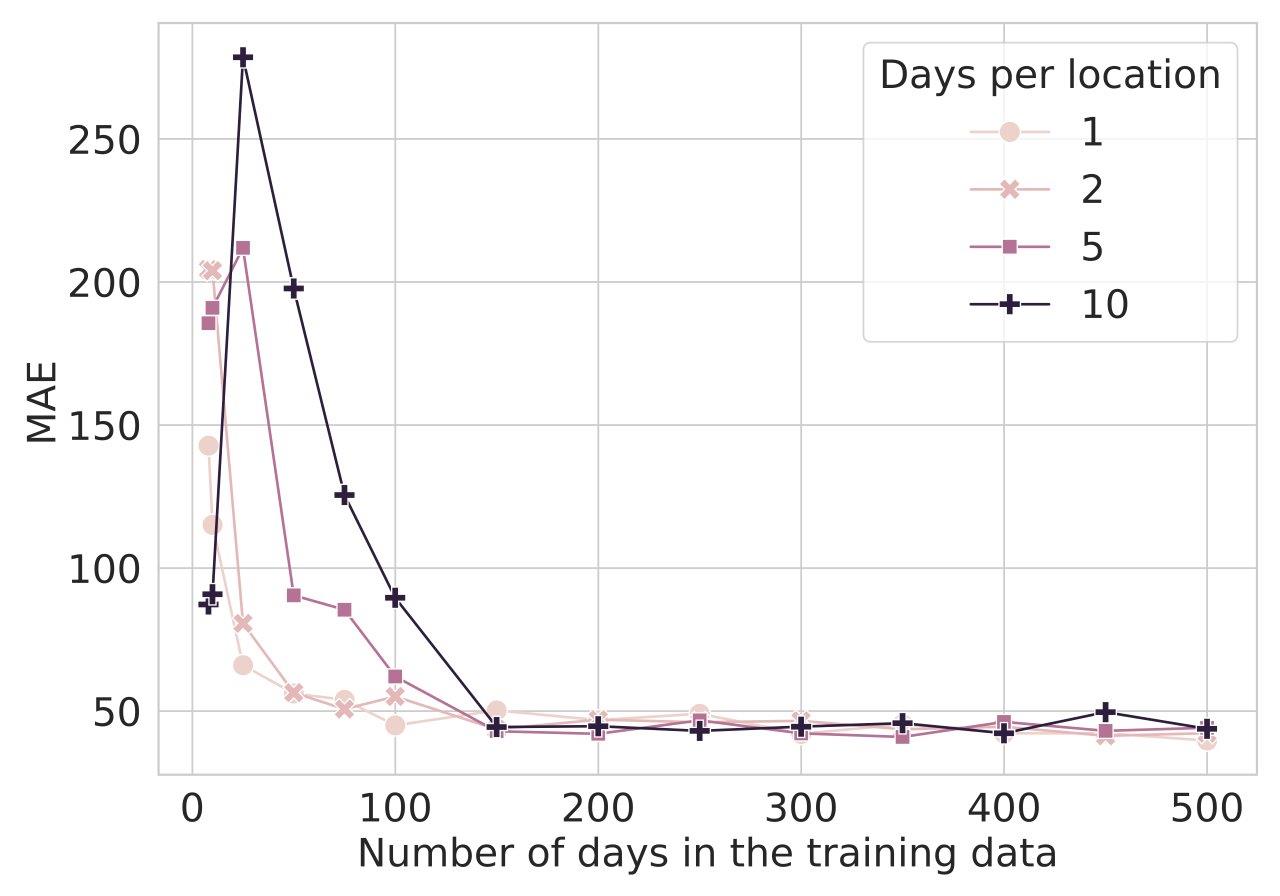}
        \caption{Manhattan — Days per location}
        \label{fig:newyork_location_mae_evaluation_active}
    \end{subfigure}

    \vspace{0.5cm}
    
        \begin{subfigure}{0.49\textwidth}
        \includegraphics[width=\textwidth]{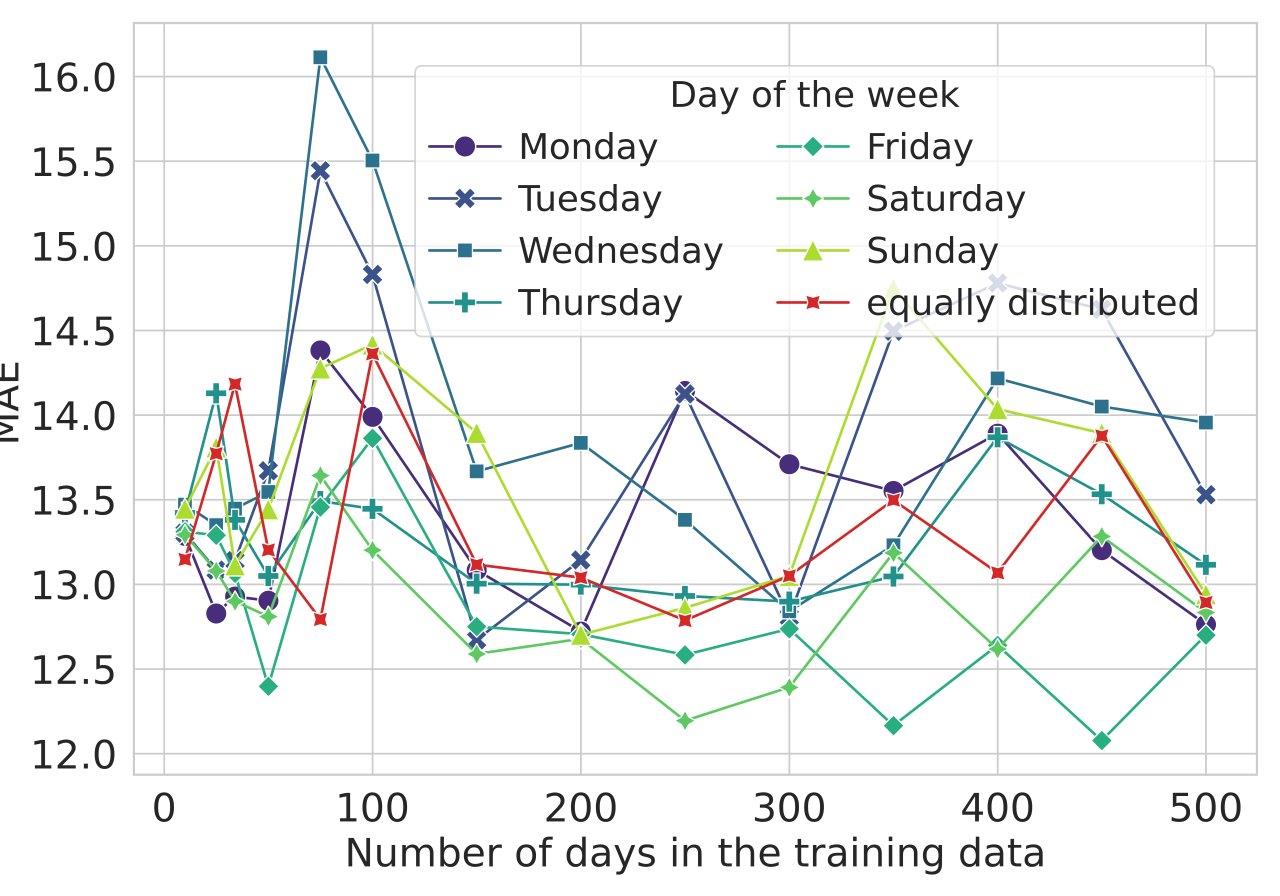}
        \caption{Berlin — Day-of-week effects}
        \label{fig:berlin_week_mae_evaluation_active}
    \end{subfigure}
    \hfill
    \begin{subfigure}{0.49\textwidth}
        \includegraphics[width=\textwidth]{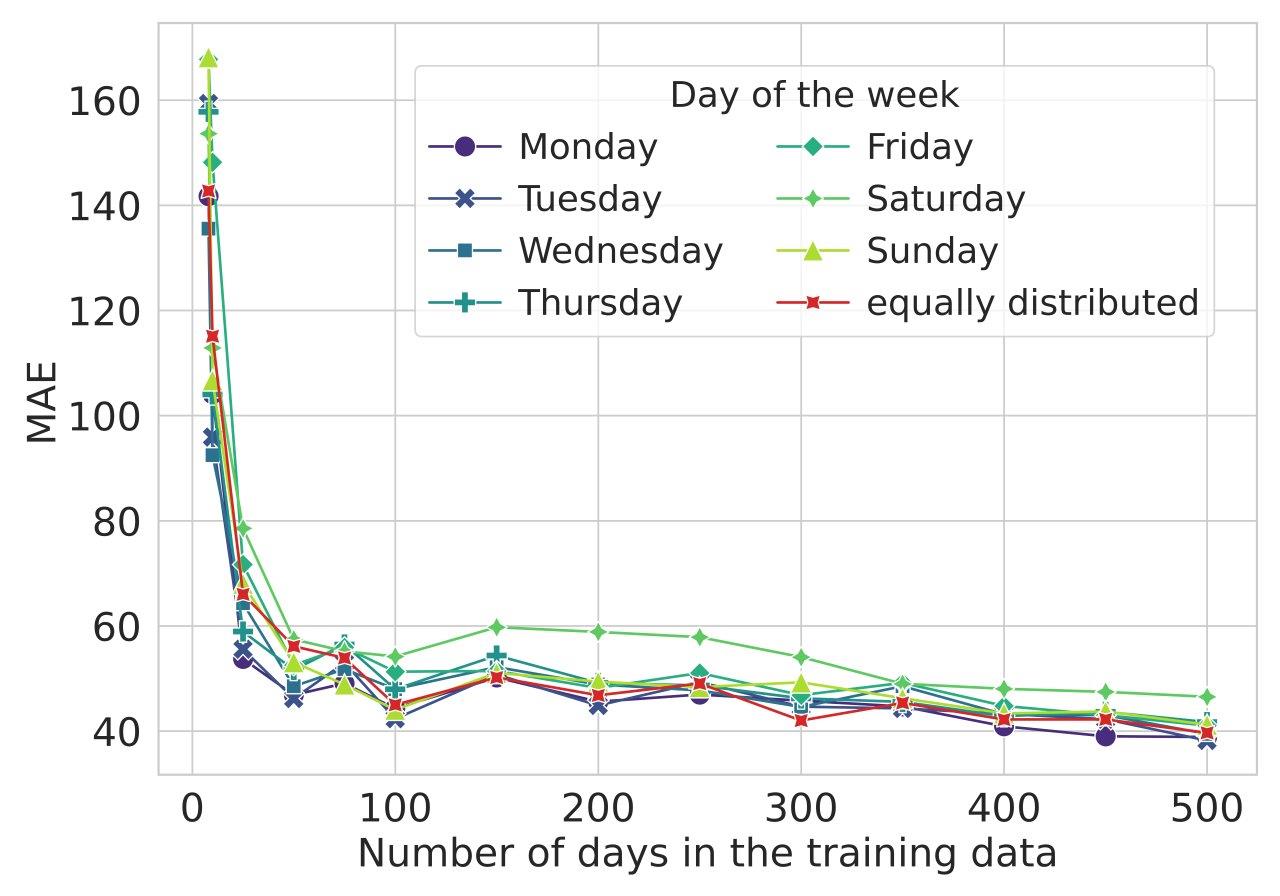}
        \caption{Manhattan — Day-of-week effects}
        \label{fig:newyork_week_mae_evaluation_active}
    \end{subfigure}
    \caption[Temporary placement considerations of temporary sensors (MAE)]{Temporary placement considerations of temporary sensors using the active learning spatial placement strategy: The upper row compares deployment strategies that vary the number of days allocated to each location, effectively trading off between broader spatial coverage (more locations, fewer days each) and repeated measurement at fewer locations, while keeping the total number of observations constant. The lower row examines whether restricting temporary measurements to specific weekdays affects prediction accuracy. Results are shown for Berlin (left) and Manhattan (right). The reported error metric is \acrshort{mae}.}
    \label{fig:results3_mae_active}
\end{figure}

This appendix reports results analogous to those presented in Section~\ref{sec:results_subsection2}, evaluating rotating versus revisiting locations and the temporal distribution of sampled days. Specifically, Voronoi area inequality and active learning are considered in place of spatial dispersion. For computational reasons, the analysis is restricted to a training horizon of up to 500 days. Results are presented in Figures~\ref{fig:results3_mae_voronoi} and \ref{fig:results3_mae_active}.

Across both cities, the qualitative patterns for rotating versus revisiting locations are consistent with the main results. The curves shown in this appendix exhibit higher volatility overall, which can be attributed to two main factors. First, as also observed in Section~\ref{sec:results_subsection2}, performance is inherently more unstable at small training horizons, where the limited amount of available data amplifies random variation. Because the present analysis focuses on the early stages of data accumulation, this volatility is more pronounced. Importantly, despite these fluctuations, the relative ordering of the deployment strategies follows the same trends as in the main analysis, with broader spatial coverage generally outperforming repeated sampling of fewer locations.
Second, the increased volatility is particularly pronounced for active learning–based placement. Active learning explicitly prioritizes locations with high predictive uncertainty, which often correspond to segments with more volatile or heterogeneous traffic patterns. When such locations are observed for only a single day, the resulting measurements may be unrepresentative, thereby increasing noise in model training. In this setting, strategies that revisit the same locations multiple times (e.g., the 10-day strategy) can perform comparatively better by averaging over short-term fluctuations and stabilizing the learned signal. As the number of observations grows, however, this effect is expected to diminish, and performance should stabilize more quickly once sufficient data are available across locations.

The qualitative findings regarding the temporal distribution of sampled days also align with the main results. No single weekday consistently outperforms others, and differences between weekday-specific strategies remain small. The strategy that distributes observations evenly across weekdays continues to provide a robust, low-risk average performance, avoiding sensitivity to idiosyncratic day-specific traffic patterns.

\clearpage
\section{Temporary sensors - RMSE \label{apx:temporary_rmse}}

\begin{figure}[ht!]
    \centering
    \begin{subfigure}{0.49\linewidth}
        \includegraphics[width=\linewidth]{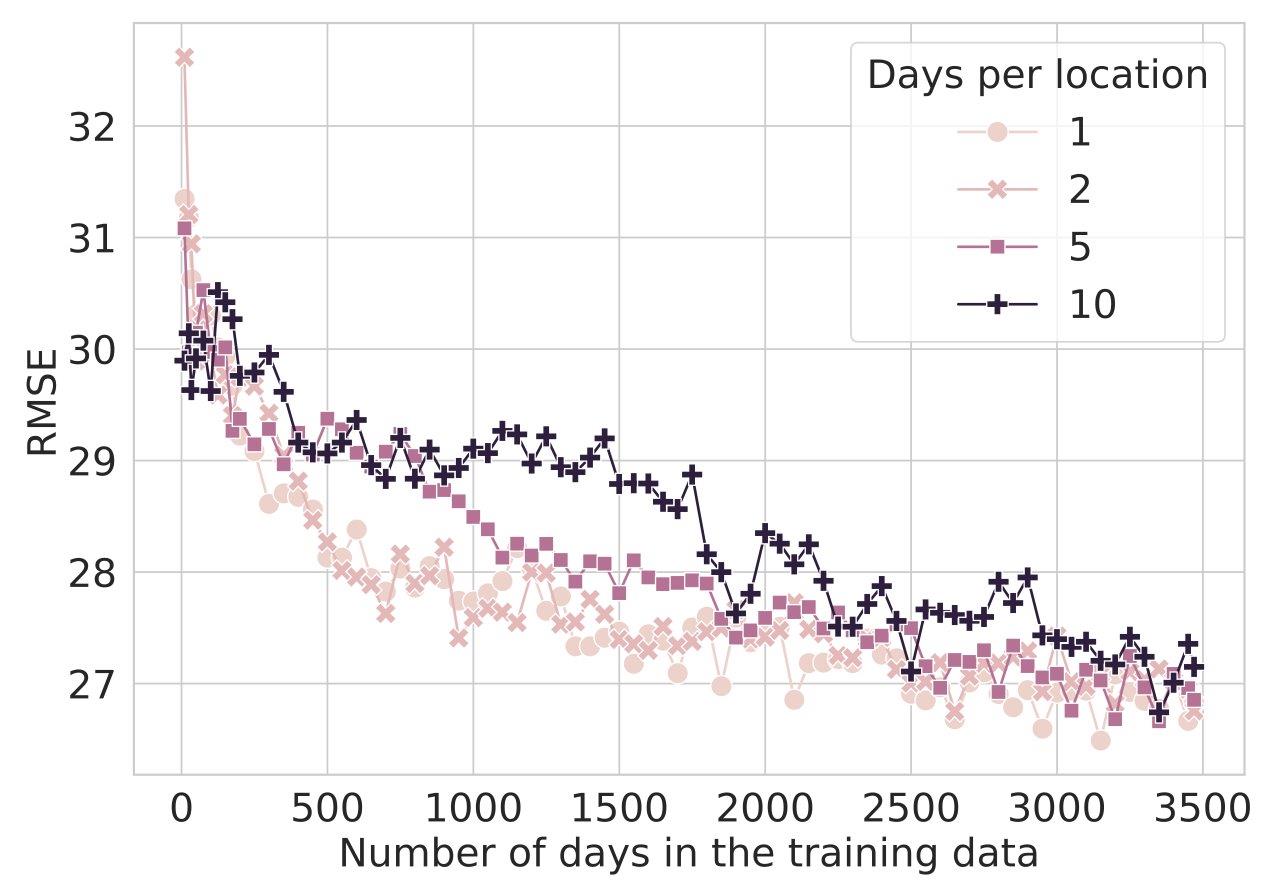}
        \caption{Berlin — Days per location}
        \label{fig:location_berlin_rmse_evaluation}
    \end{subfigure}
    \hfill
            \begin{subfigure}{0.49\linewidth}
        \includegraphics[width=\linewidth]{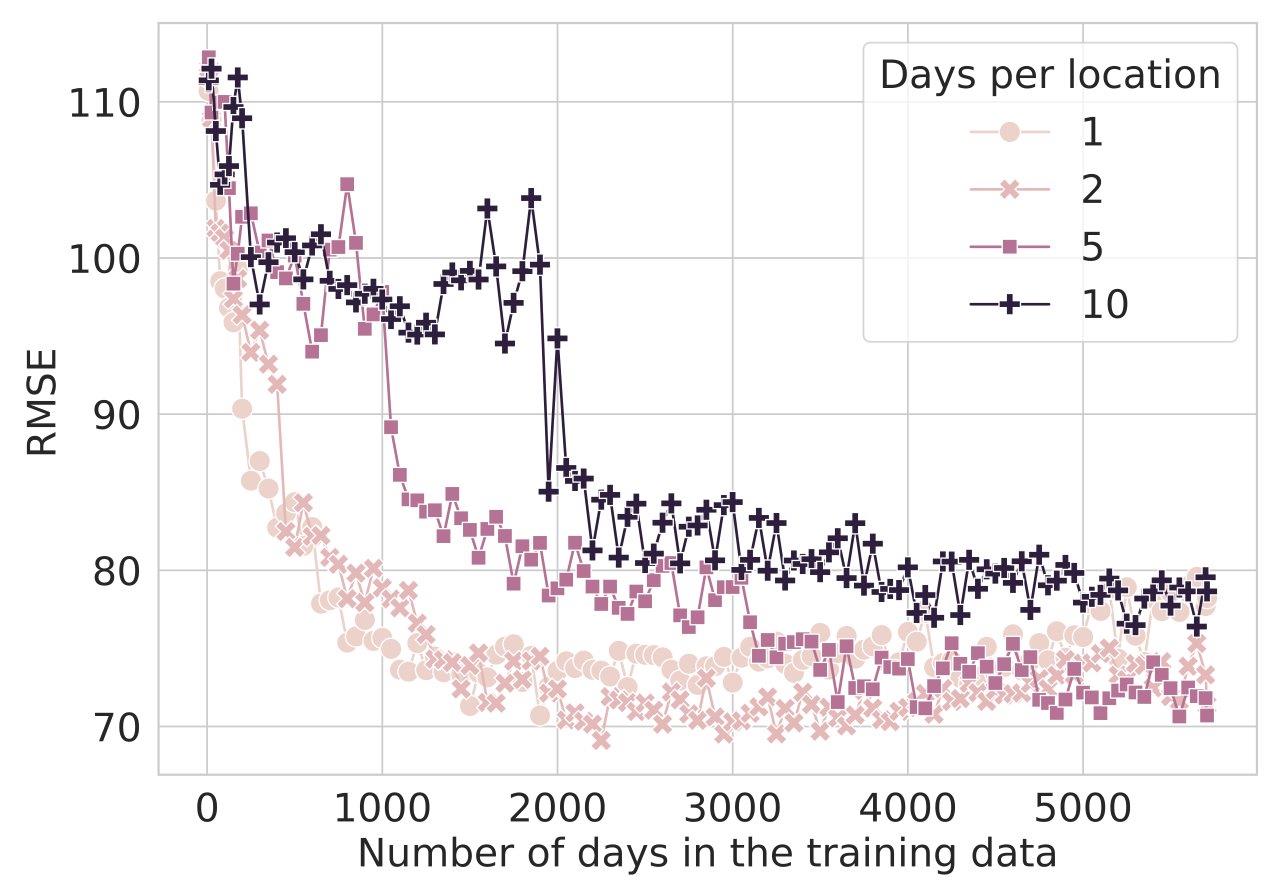}
        \caption{Manhattan — Days per location}
        \label{fig:newyork_location_rmse_evaluation}
    \end{subfigure}

    \vspace{0.5cm}
    
        \begin{subfigure}{0.49\linewidth}
        \includegraphics[width=\linewidth]{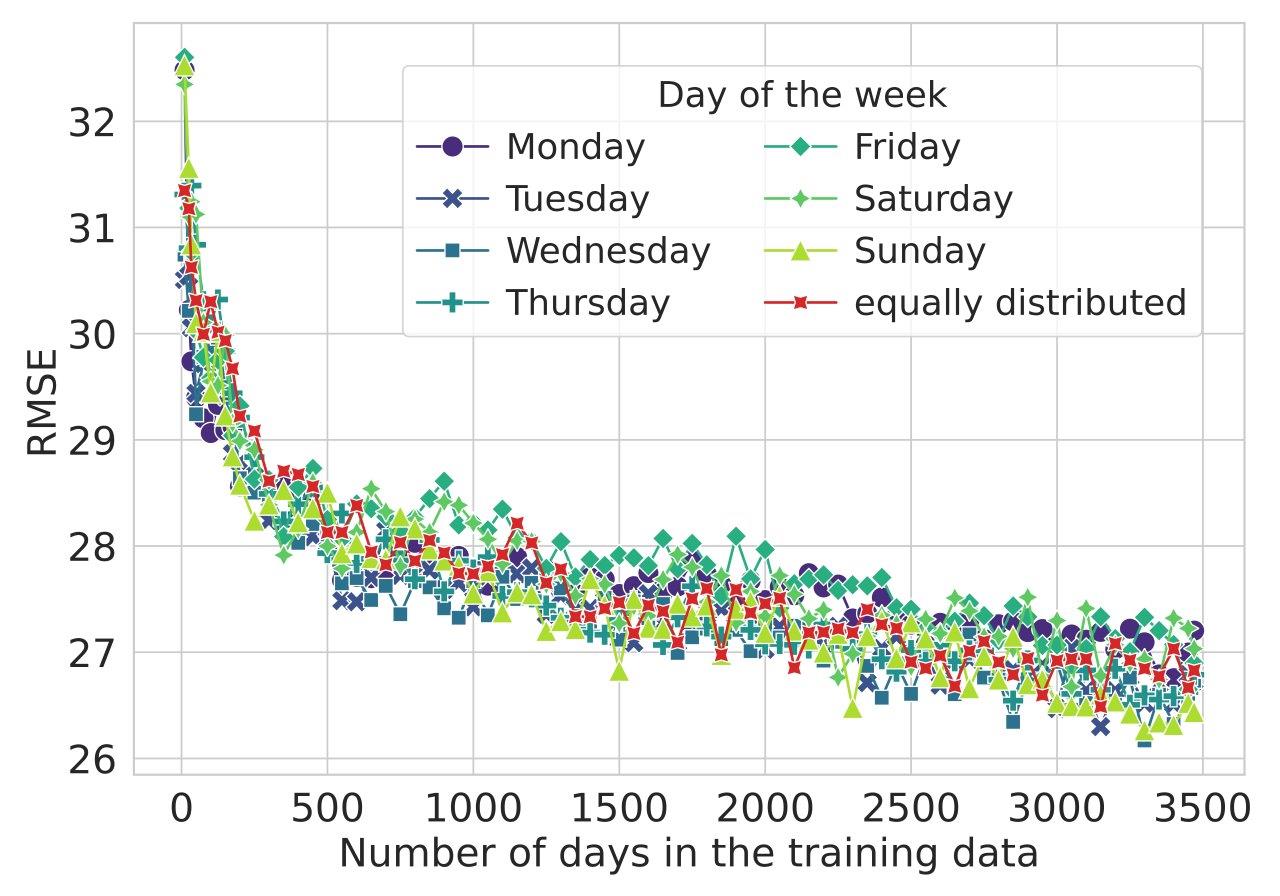}
        \caption{Berlin — Day-of-week effects}
        \label{fig:berlin_week_rmse_evaluation}
    \end{subfigure}
    \hfill
    \begin{subfigure}{0.49\linewidth}
        \includegraphics[width=\linewidth]{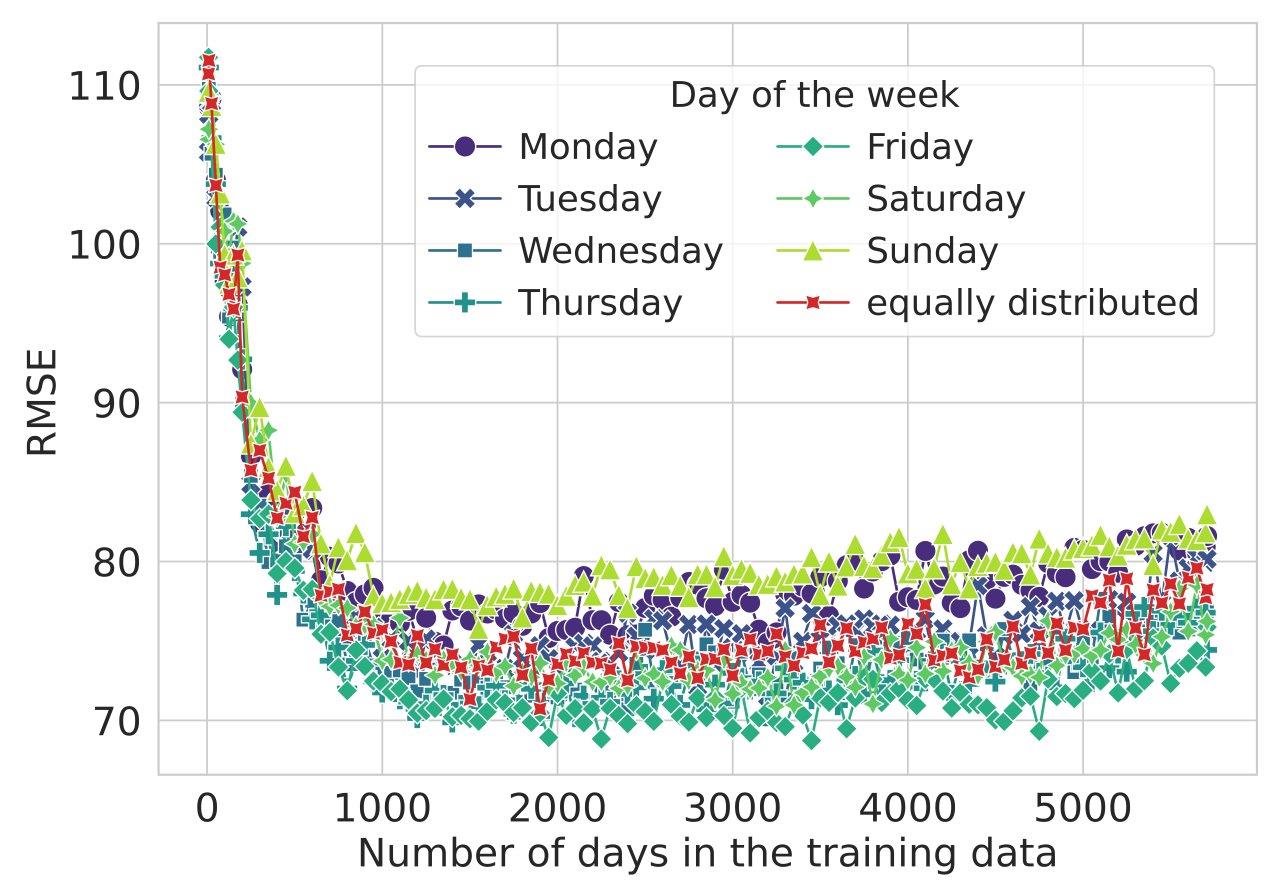}
        \caption{Manhattan — Day-of-week effects}
        \label{fig:newyork_week_rmse_evaluation}
    \end{subfigure}
    \caption[Temporary placement considerations of temporary sensors (RMSE)]{Temporary placement considerations of temporary sensors: The upper row compares deployment strategies that vary the number of days allocated to each location, effectively trading off between broader spatial coverage (more locations, fewer days each) and repeated measurement at fewer locations, while keeping the total number of observations constant. The lower row examines whether restricting temporary measurements to specific weekdays affects prediction accuracy. Results are shown for Berlin (left) and Manhattan (right). The reported error metric is \acrshort{rmse}.}
    \label{fig:results3_rmse}

\end{figure}

This appendix reports results analogous to those presented in Section~\ref{sec:results_subsection2}, which evaluates the performance of temporal sensor deployment strategies, using \acrshort{rmse} instead of \acrshort{mae} as the evaluation metric. The corresponding results are shown in Figure~\ref{fig:results3_rmse}.

Overall, the qualitative patterns remain consistent with the main findings. In both cities, deployment strategies that allocate fewer days per location, thereby increasing spatial coverage, tend to achieve lower errors than strategies relying on repeated measurement at fewer locations. Compared to the \acrshort{mae}-based results presented in the main text, the ordering of strategies is less pronounced at larger training horizons, particularly beyond approximately 2,000 days in Berlin and 3,000 in Manhattan. This reduced separation may reflect that broader spatial sampling occasionally includes atypical observations, whose influence is amplified under a squared-error metric. However, given that the primary objective of this study is to minimize typical absolute interpolation errors rather than squared deviations, these differences do not alter the qualitative conclusions.

The analysis of weekday effects yields conclusions under \acrshort{rmse} that are consistent with those in the main text. In both cities, evenly distributing temporary measurements across weekdays provides robust average performance. In Manhattan, \acrshort{rmse} exhibits a slight increase at the largest training horizons. This behavior likely reflects the limited temporal coverage of the underlying dataset, which amplifies the influence of a small number of extreme errors under a squared-error metric. However, because the primary objective of this study is to minimize typical absolute interpolation errors, this behavior does not alter the qualitative conclusions.

\clearpage
\section{Temporal placement of sensors - across seasons \label{appx:results3_seasons}}

In addition to weekday-based scheduling, the seasonal timing of temporary sensor deployments may also affect interpolation performance. Traffic volumes exhibit pronounced seasonal variability, as is well documented for cycling activity \citep{fournier_sinusoidal_2017,nankervis_effect_1999}. Analogous to within-week variation, this raises the question of how temporary sensors should be scheduled across seasons to ensure representative training data and robust model performance.
To investigate this effect, the analysis defines seasons according to standard Northern Hemisphere conventions: spring (March–May), summer (June–August), fall (September–November), and winter (December–February). Deployment strategies that restrict sampling to a single season are compared against a strategy that distributes observations evenly across all seasons. 

\begin{figure}[ht!]
    \centering
    \begin{subfigure}{0.49\linewidth}
        \includegraphics[width=\linewidth]{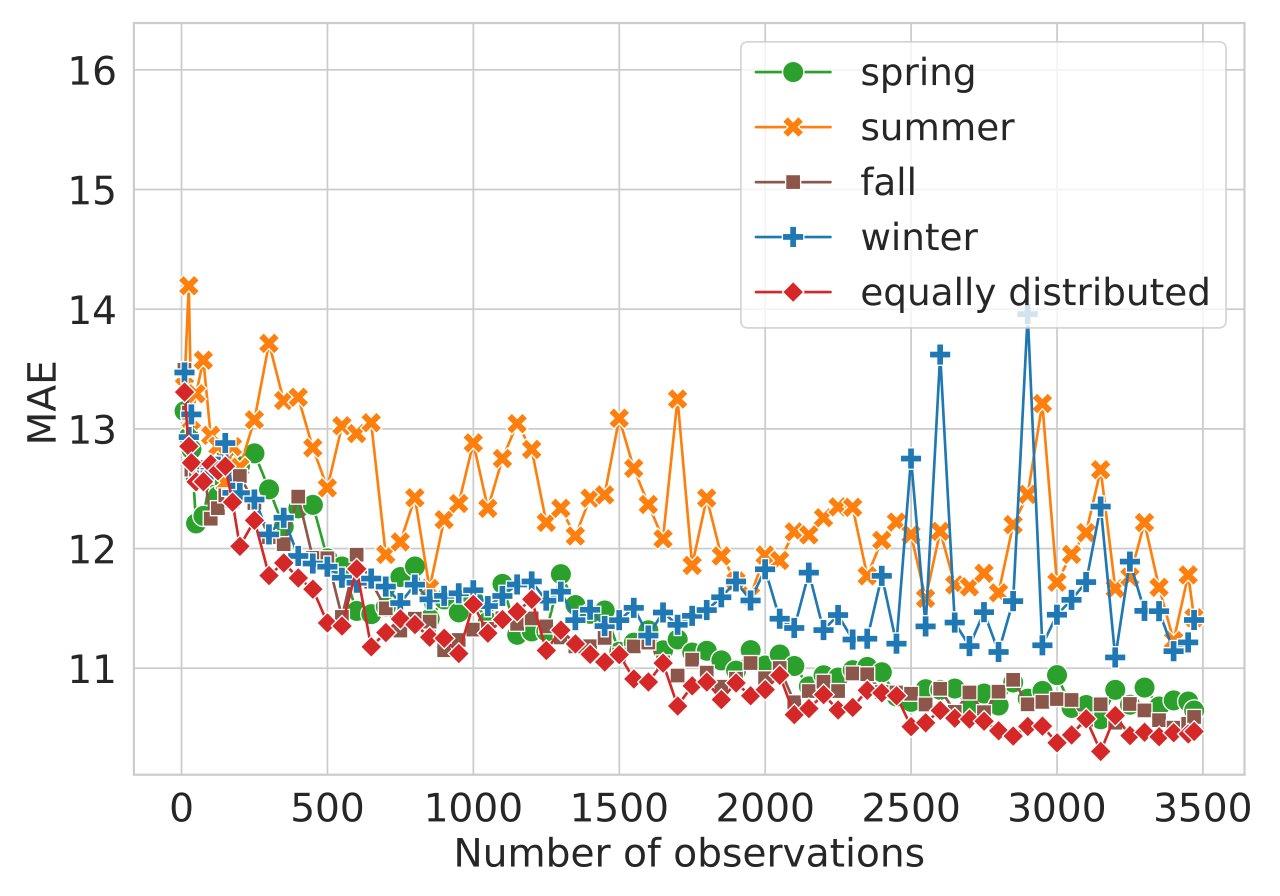}
        \caption{Berlin - MAE}
        \label{fig:location_berlin_mae_evaluation_appendix}
    \end{subfigure}
    \hfill
    \begin{subfigure}{0.49\linewidth}
        \includegraphics[width=\linewidth]{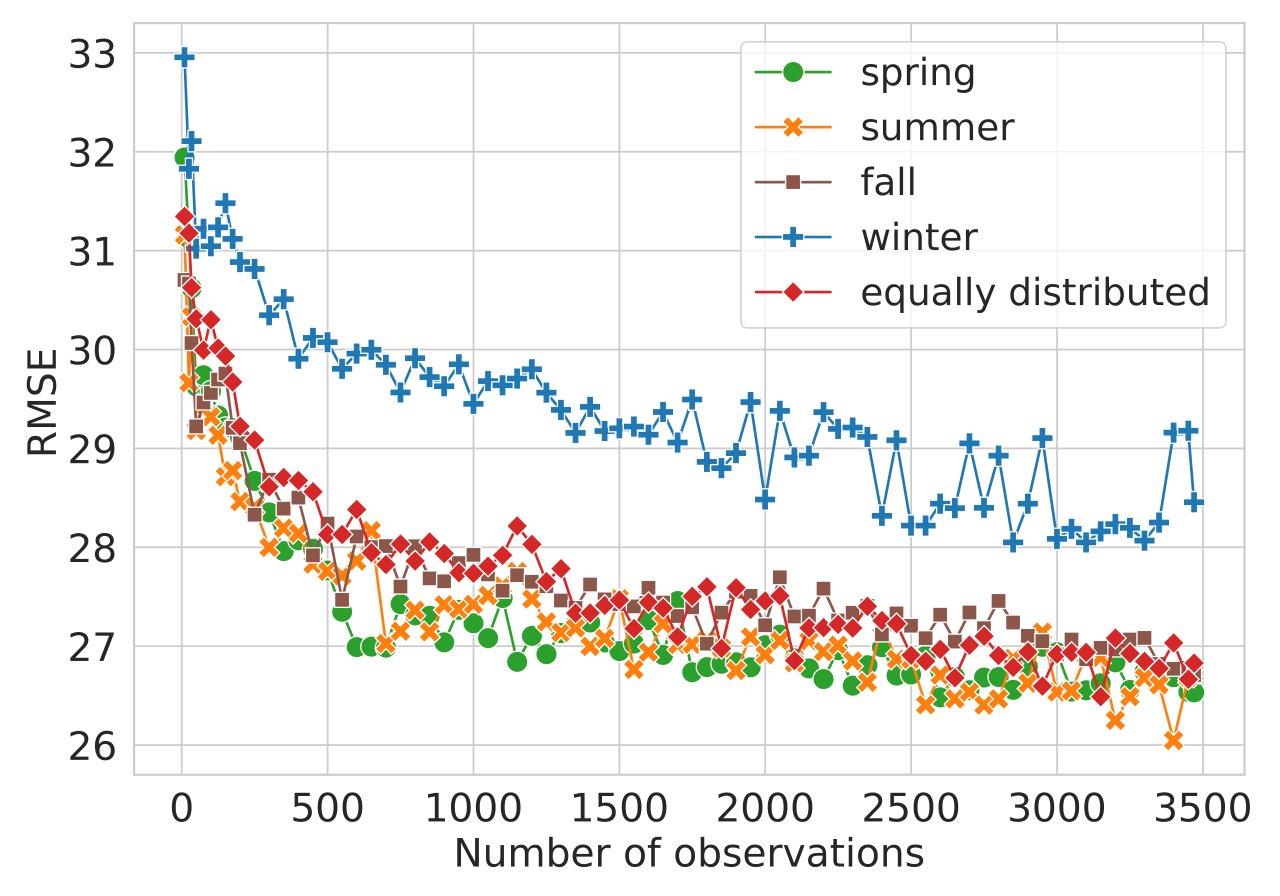}
        \caption{Berlin - RMSE}
        \label{fig:location_berlin_rmse_evaluation_appendix}
    \end{subfigure}

    \caption[Seasonal placement of temporary sensors]{Interpolation performance resulting from seasonal placement of temporary sensors in Berlin. Interpolation performance is evaluated for strategies that restrict temporary measurements to individual seasons and for a strategy that distributes observations evenly across the year. Results are reported using both \acrshort{mae} (left) and \acrshort{rmse} (right). }
    \label{fig:temporal_evaluation_month}
\end{figure}

Figure~\ref{fig:temporal_evaluation_month} reports the resulting interpolation performance for Berlin, evaluated using both \acrshort{mae} and \acrshort{rmse}. The analysis adopts the best-performing design choices identified in the main text: spatial diversity as the spatial placement strategy, single-day deployments per site, and an even distribution of observations across weekdays. The results show that seasonal restrictions in temporary sensor deployment lead to systematic differences in interpolation performance. Across all sample sizes, deployments restricted to summer consistently perform among the worst strategies. A likely explanation is that summer observations capture disproportionately high cycling volumes, thereby biasing the training data toward elevated traffic levels and reducing generalization to lower-volume periods.
Winter-only deployments exhibit a mixed pattern: they perform relatively well when evaluated with \acrshort{mae} but substantially worse under \acrshort{rmse}. This suggests that models trained predominantly on low-volume winter data tend to systematically underpredict traffic volumes, including occasional high-volume observations arising from the right-skewed distribution of traffic volumes. While such underprediction is only weakly penalized by \acrshort{mae}, it is strongly amplified under \acrshort{rmse} due to its quadratic error structure.
In contrast, the strategy that distributes temporary measurements evenly across all seasons consistently achieves the best or second-best performance once more than ten sensors are deployed. This mirrors the weekday results and indicates that temporal diversity in the training data is more desirable.

\clearpage
\section{Comparing interpolation performance using temporary and permanent sensors - further placement strategies \label{appx:results3_otherlists}}

\begin{figure}[ht!]
    \centering
    \begin{subfigure}{0.49\linewidth}
        \includegraphics[width=\linewidth]{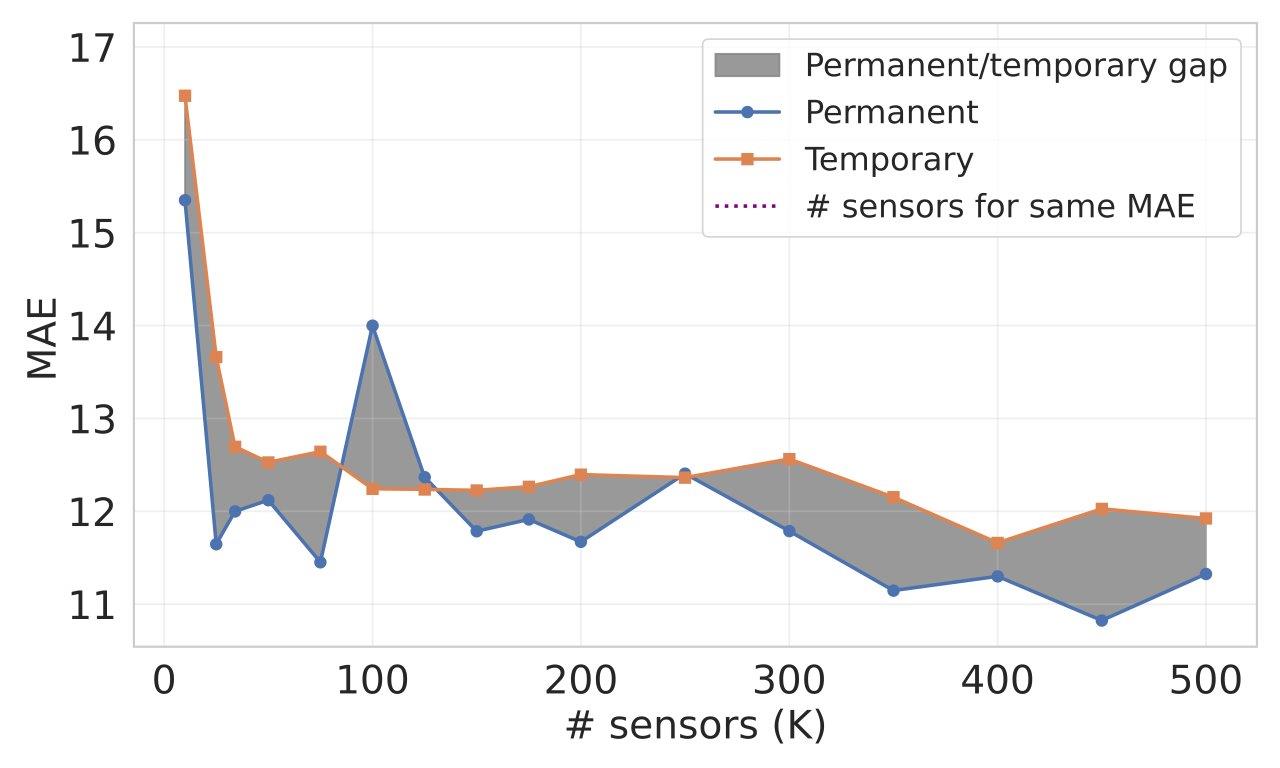}
        \caption{Berlin}
        \label{fig:temporary_berlin_mae_voronoi}
    \end{subfigure}
    \hfill
        \begin{subfigure}{0.49\linewidth}
        \includegraphics[width=\linewidth]{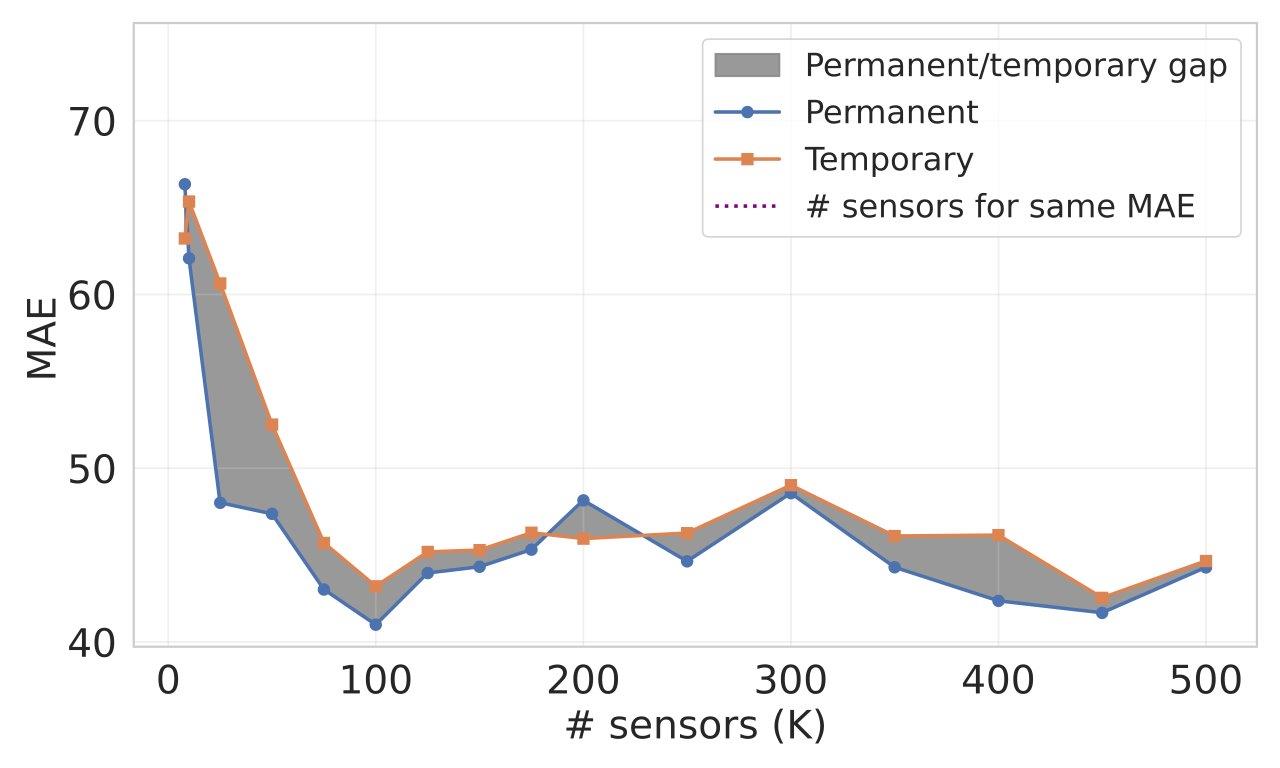}
        \caption{Manhattan}
        \label{fig:temporary_newyork_mae_voronoi}
    \end{subfigure}
    \caption[Prediction performance under temporary and permanent deployment using Voronoi area inequality]{\acrshort{mae} performance under temporary and permanent sensor deployments for citywide traffic volume estimation, shown as a function of the sensor budget $K$, using Voronoi area inequality as the spatial placement strategy.
    \label{fig:results3_appendix_voronoi}}
\end{figure}

\begin{figure}[ht!]
    \centering
    \begin{subfigure}{0.49\linewidth}
        \includegraphics[width=\linewidth]{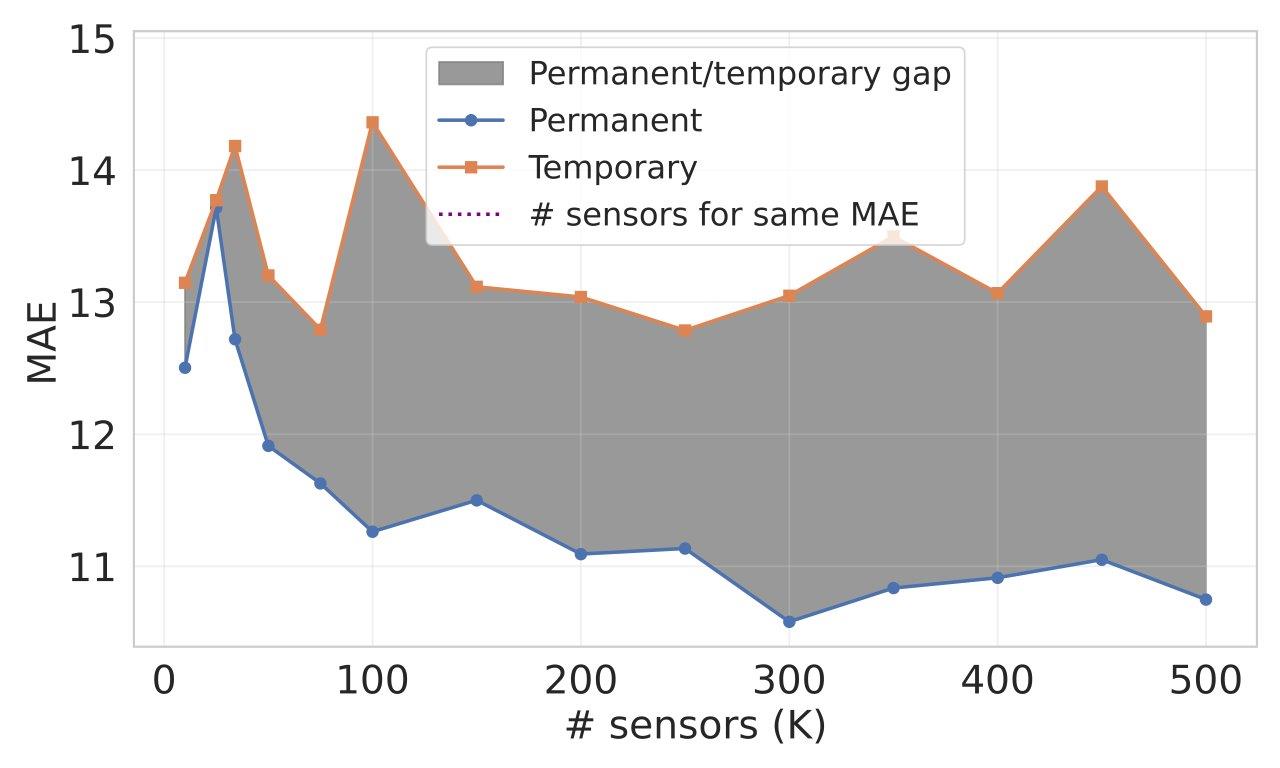}
        \caption{Berlin}
        \label{fig:temporary_berlin_mae_al}
    \end{subfigure}
    \hfill
        \begin{subfigure}{0.49\linewidth}
        \includegraphics[width=\linewidth]{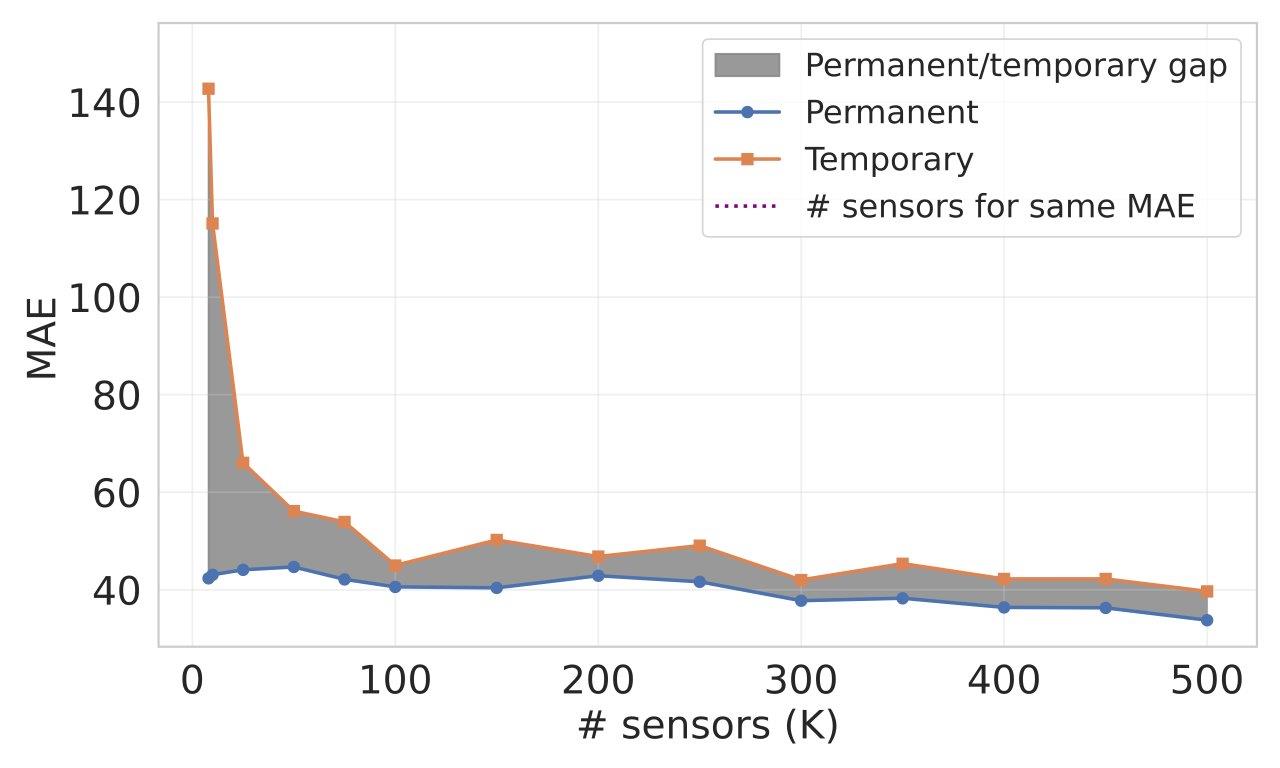}
        \caption{Manhattan}
        \label{fig:temporary_newyork_mae_al}
    \end{subfigure}
    \caption[Prediction performance under temporary and permanent deployment using active learning]{\acrshort{mae} performance under temporary and permanent sensor deployments for citywide traffic volume estimation, shown as a function of the sensor budget $K$, using active learning as the spatial placement strategy.
    \label{fig:results3_appendix_al}}
\end{figure}

This appendix reports results analogous to those presented in Section~\ref{sec:results_subsection3}, which compares the interpolation performance between temporary and permanent sensors, but considers alternative spatial placement strategies. Specifically, Voronoi area inequality and active learning are evaluated in place of spatial dispersion. For computational reasons, results are shown only up to 500 sensors. The corresponding results are depicted in Figures~\ref{fig:results3_appendix_voronoi} and \ref{fig:results3_appendix_al}.

Across both cities, the qualitative patterns closely mirror those observed in the main analysis. As in Section~\ref{sec:results_subsection3}, performance differences between permanent and temporary deployments remain relatively small across the displayed sensor budgets. However, when using active learning, the performance gap between permanent and temporary sensors is more pronounced than for Voronoi area inequality or spatial dispersion. A plausible explanation is that active learning deliberately selects locations with high predictive uncertainty, which may reflect greater temporal variability or volatility in traffic patterns. When such locations are observed for only a single day—as in the temporary deployment setting—the resulting observations may be unrepresentative or introduce additional noise, thereby reducing their usefulness for model training.

These results suggest an important direction for future research. In particular, it may be beneficial to extend active learning beyond spatial placement to also inform temporal placement decisions, i.e., by identifying not only uncertain locations but also uncertain time periods.

Although results are shown only up to 500 sensors, the same trends are expected to hold at larger sensor budgets. As demonstrated in Section~\ref{sec:results_subsection1}, performance differences between spatial placement strategies systematically diminish as sensor coverage becomes denser, indicating that marginal placement decisions become less consequential. Because the present analysis already covers the range in which this convergence begins to emerge, similar or smaller differences are expected beyond the reported sensor budgets.

\clearpage
\section{Reinforcement learning proposal \label{appx:RL_model}}
\acrlong{rl} (\acrshort{rl}) may offer a data-driven approach to discover sensor configurations that improve interpolation performance through iterative feedback, and coupling this approach with explainability techniques could help derive new, interpretable placement heuristics. This appendix outlines (A) an \acrshort{rl} framework for sensor placement and (B) a complementary explainability approach. For brevity, here the study outlines such a model for permanent sensor placement only; the model could be extended to temporary sensor placement. 

\paragraph{A) \acrshort{rl} Set Up}\par
\textcolor{white}{.}
\vspace{0.5em}

Sensor placement can be formulated as a sequential decision-making problem, as illustrated in Figure \ref{appx:RL_model}. In this setting, an \acrshort{rl} agent sequentially selects $K$ sensor locations over steps $t \in \{1,2,\dots,K\}$, with the objective of minimizing city-wide traffic-volume interpolation error. After each placement, the current sensor configuration is evaluated using an interpolation model, and the resulting performance feedback is used to guide subsequent placement decisions. Through repeated interaction with this evaluation process across multiple episodes, the agent learns a placement strategy that progressively improves overall interpolation performance. The individual components of this framework are described in more detail below.

\textbf{Evaluation environment.} The evaluation environment includes the relevant city data, the evaluator (which is the interpolation model), and what is needed to compute the reward (see the right of Figure \ref{appx:RL_model}). 
The city data contains the feature matrix of all street segment $X\in \mathbb{R}^{N \times d}$, the feature matrix of time invariant features $X'\in \mathbb{R}^{N \times d'}$, the set of candidate locations $S_{candidate}^{(t)}$ at step t, the selected sensor locations $S_{selected}^{(t)}$ and the validation set $S_{val}$.

The evaluator is an interpolation model that assesses how well traffic volume can be predicted given the selected sensor locations. This appendix proposes using the same interpolation model as in the main analysis: Beyond the advantages discussed in the main paper regarding its strong performance on tabular traffic data, XGBoost is also well-suited for this role due to its relatively low training time, which is essential for the repeated model updates required in a \acrshort{rl} loop, and because it supports efficient fine-tuning with additional data without requiring full retraining at every step. 
In this \acrshort{rl} setup, at each step, the interpolation would be trained on all observations from the $S_{selected}^{(t)}$ (as simulation of permanent deployment, see main paper for more details) and return the loss $L_{\text{val}}^{(t)}$ on the validation set $S_{val}$. 
To balance computational efficiency and model performance, a hybrid training strategy can be employed in which the model is fully retrained at the beginning of each episode and at regular intervals during sensor placement, while intermediate updates are performed incrementally using a reduced learning rate and fewer estimators. This maintains efficiency while mitigating catastrophic forgetting.

A natural choice for the reward signal is the improvement in interpolation performance induced by each newly placed sensor. In line with the evaluation strategy used in the main paper, the reward at step $t$ could be defined as the reduction in validation \acrlong{mae} (\acrshort{mae}):
\begin{equation}
r_t = L_{\text{val}}^{(t-1)} - L_{\text{val}}^{(t)},
\end{equation}
where $L_{\text{val}}^{(t)}$ denotes the \acrshort{mae} on the $S_{val}$after placing the $t$-th sensor. This reward formulation encourages the agent to maximize cumulative reward over the entire episode by selecting sensors that progressively improve city-wide traffic volume estimation performance.

The state $s_t$ could be defined as a compact representation of the current sensor placement and the static characteristics of all candidate locations. The binary selection vector $\mathbf{b}_t \in \{0,1\}^N$ encodes the current placement status, where $b_{t,i}=1$ if sensor $i$ is selected and $b_{t,i}=0$ otherwise.
In addition, a feature matrix $\mathbf{X'} $ contains the $d'$-dimensional, time-invariant feature vectors of all locations. Only time-invariant features are considered, since sensor locations are selected independently of time; time-varying attributes therefore enter only at the interpolation stage and not into the placement decision (cf. Section \ref{sec:methods}). All numerical features are standardized, and categorical features are one-hot encoded.
The resulting state vector can then be expressed as
\begin{equation}
s_t = [\mathbf{b}_t, \text{vec}(\mathbf{X'})],
\end{equation}
where $\text{vec}(\cdot)$ denotes vectorization. This formulation yields a state space of dimensionality $N + Nd'$.%

\textbf{Learning agent.} 
A natural choice for the learning agent in this setting is a \acrlong{dqn} (\acrshort{dqn}), which approximates the action–value function for state–action pairs and is depicted on the left of Figure \ref{appx:RL_model}. The \acrshort{dqn} estimates the action–value function $Q(s_t, a_t; \theta)$ using a neural network parameterized by $\theta$. The action $a_t$ corresponds to selecting a single sensor location from the available candidate set $S_{\text{candidate}}^{(t)}$ at step $t$, where $S_{\text{candidate}}^{(t)}$ is dynamically updated to exclude all sensors that have been placed in previous steps, ensuring each sensor is selected at most once per episode. he selected location is then added to the current sensor configuration $S_{\text{selected}}$. The network architecture comprises an input layer that takes the state $s_t$ as input, followed by two fully connected hidden layers with ReLU activations and dropout, and a final output layer that outputs a single Q-value for each state–action pair.

At each step, the agent evaluates Q-values for all available candidate actions and follows an $\epsilon$-greedy policy:
\begin{equation}
a_t =
\begin{cases}
\text{random choice from } S_{\text{candidate}}^{(t)} & \text{with probability } \epsilon, \\
\arg\max_{a \in S_{\text{candidate}}^{(t)}} Q(s_t, a; \theta) & \text{with probability } 1-\epsilon .
\end{cases}
\end{equation}
Random exploration with probability $\epsilon$ prevents the agent from getting stuck in local optima, while exploitation with probability $1-\epsilon$ allows it to leverage learned Q-values for near-optimal decision-making. 

The agent is trained over several episodes. Each episode consists of $K$ sequential sensor placement decisions. The goal is to learn parameters $\theta$ that maximize the expected cumulative reward:
\begin{equation}
\max_{\theta} \mathbb{E}\left[\sum_{t=1}^{K} r_t\right]
\end{equation}
The DQN is trained by minimizing the \acrlong{td} (\acrshort{td}) loss with the target value:
\begin{equation}
y_t = r_t + \gamma \max_{a'} Q(s_{t+1}, a'; \theta^{-})
\end{equation}
where $\gamma$ is the discount factor and $\theta^{-}$ are the parameters of the target network. The TD loss uses Huber loss for robustness:
\begin{equation}
L_{\text{DQN}} = \mathbb{E}{(s_t, a_t, r_t, s_{t+1})} \left[ L_{\text{Huber}}(y_t - Q(s_t, a_t; \theta)) \right]
\end{equation}

\begin{figure}
    \centering
    \includegraphics[width=0.8\linewidth]{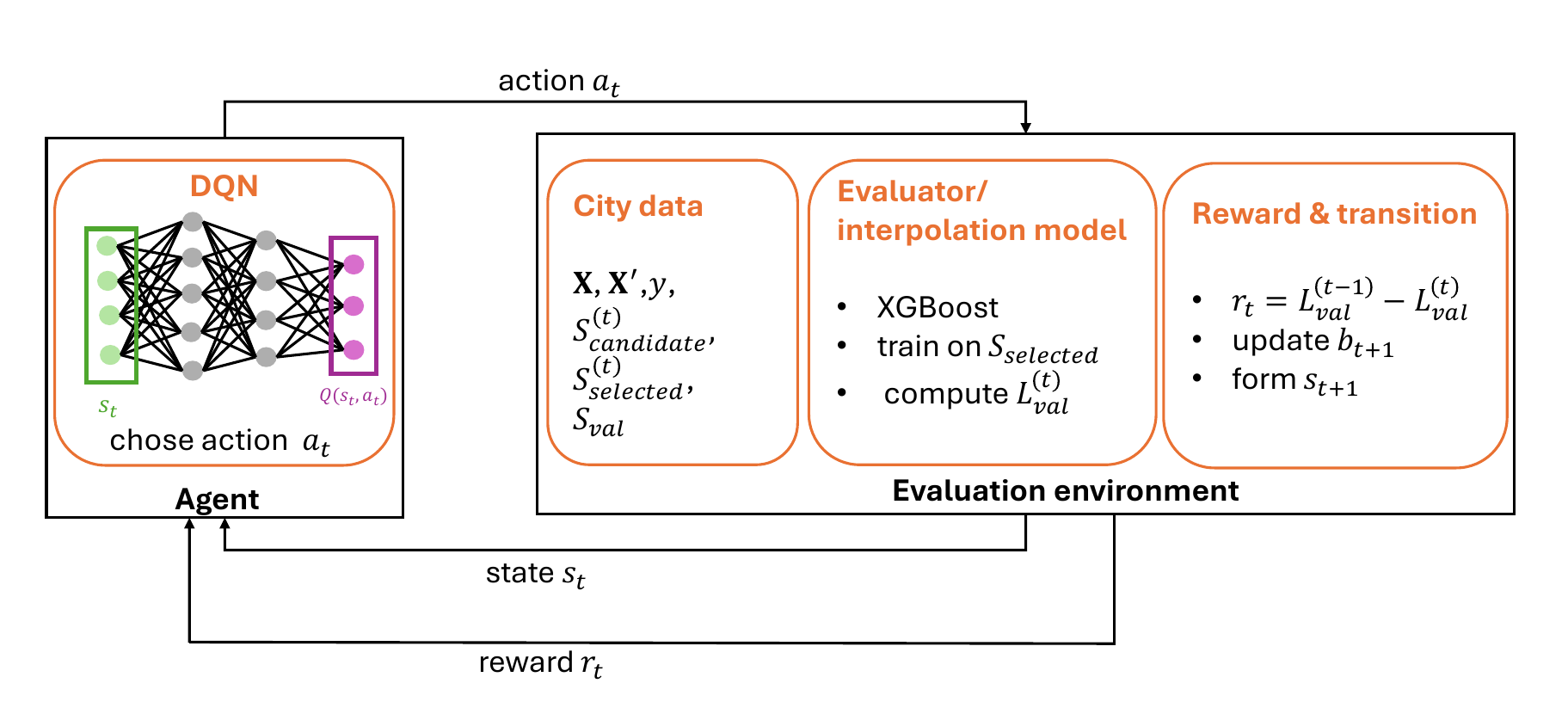}
    \caption{Graphical explanation of the RL model.}
    \label{fig:RL_model_Explanation}
\end{figure}


\paragraph{B) Explainability}\par
\textcolor{white}{.}
\vspace{0.5em}

Once an \acrshort{rl}–based placement policy has been trained using fully observed traffic data, it can identify sensor configurations with strong interpolation performance. However, to make these results transferable to cities where no ground-truth data are available, a systematic explainability analysis is required to extract interpretable placement principles from the learned policy.

Since sensor selection in the proposed framework is driven by location-specific feature values ($X'$), understanding which features influence the agent’s decisions, and how this influence evolves over the $K$ placement steps, is therefore of central importance. A range of post hoc explainability methods could be used for this purpose; here, this appendix focus on \acrlong{ig} (\acrshort{ig}) \citep{sundararajan_axiomatic_2017} as one suitable example. \acrshort{ig} is a gradient-based attribution method that quantifies the contribution of each input feature by integrating the model gradients along a straight-line path between a baseline input and the actual input. For a given state vector $s_t$, the attribution for the $i$-th feature can be expressed as

\begin{equation}
    \mathrm{IG}_i(s_t) = (s_{t,i} - s_{t',i}) \int_{0}^{1} \frac{\partial DQN\big(s_t' + \alpha (s_t - s_t')\big)}{\partial s_{t,i}} \, d\alpha ,
\end{equation}

where $DQN$ denotes the trained Q-network, $s_t'$ is a chosen baseline state and $\alpha \in [0,1]$ parametrizes the straight-line interpolation path between the baseline state $s'_t$ and the actual state $s_t$. In the given context, a zero vector could serve as a natural baseline, representing a neutral configuration in which no sensors are selected, and all standardized and one-hot-encoded features take their reference values.
By computing integrated gradients over all $K$ selection steps, the analysis tracks how the importance of different state features evolves as the sensor deployment grows. Since the state representation contains features for all $N$ locations, the resulting attributions can be aggregated across sensors to yield interpretable feature-type importance scores. For a feature $\ell$, an aggregated attribution measure can be defined as:

\begin{equation}
    \mathrm{IG}_{\text{feature},\ell} 
    = \sum_{i \in \mathcal{I}_\ell} \big| \mathrm{IG}_i(s_t) \big|,
    \qquad \ell = 1,\dots,d',
\end{equation}
where $\mathcal{I}_\ell$ denotes the index set of state components corresponding to 
feature dimension $\ell$ across all $N$ locations. The absolute values are used to capture both positive and negative contributions. This aggregation highlights which features systematically drive the placement decisions.

This analysis can be complemented by examining Q-values to assess the relative preference of the model for different candidate actions, with larger gaps between the selected action and alternatives indicating a stronger expected return. In combination with the integrated gradients, this enables the identification of interpretable and transferable sensor placement heuristics for cities without ground-truth coverage.

\newpage
\bibliographystyle{plainnat}
\bibliography{references}

\end{document}